\documentclass{aastex62}

\usepackage{amsmath}
\usepackage{amssymb}
\usepackage{subfig}
\usepackage{placeins}

\received{June 9, 2017}
\revised{October 26, 2017}
\accepted{October 31, 2017}
\published{December 19, 2017}
\submitjournal{The Astrophysical Journal}

\shorttitle{Viscous Overstability in Saturn's Rings}
\shortauthors{Lehmann et al.}

\begin{document}

\title{Viscous Overstability in Saturn's Rings: Influence of Collective Self-gravity}

\correspondingauthor{Marius Lehmann}
\email{marius.lehmann@oulu.fi}

\author[0000-0002-0496-3539]{Marius Lehmann}
\affil{Astronomy Research Unit, University of Oulu, Finland}

\author{J\"urgen Schmidt}
\affil{Astronomy Research Unit, University of Oulu, Finland}

\author[0000-0002-4400-042X]{Heikki Salo}
\affil{Astronomy Research Unit, University of Oulu, Finland}

%
%
%
%



\begin{abstract}

 We investigate the influence of collective self-gravity forces on the nonlinear, large-scale evolution of the viscous overstability in Saturn's rings.
 We numerically solve the axisymmetric nonlinear hydrodynamic equations in the isothermal and non-isothermal approximation, including radial 
 self-gravity and employing transport coefficients derived by \citet{salo2001}. We assume optical depths
 $\tau=1.5-2$ to model Saturn's dense rings.
 Furthermore, local N-body simulations, incorporating vertical and radial collective self-gravity are performed. Vertical self-gravity is mimicked through an 
 increased 
 frequency of vertical oscillations, while radial self-gravity is approximated by solving the Poisson equation for an axisymmetric thin disk with a Fourier 
 method. 
 Direct 
 particle-particle forces 
 are omitted, which prevents small-scale gravitational instabilities (self-gravity wakes) from forming, an approximation that allows us to study long 
 radial scales and to compare directly the hydrodynamic model and the N-body simulations.
 Our isothermal and non-isothermal hydrodynamic model results with vanishing self-gravity compare very well with results of \citet{latter2010} 
 and \citet{latter2013}, respectively.
 In contrast, for rings with radial self-gravity we find that the wavelengths of saturated overstable waves settle close to the 
 frequency minimum of the nonlinear dispersion relation, i.e. close to a state of vanishing group velocities of the waves.
 Good agreement is found between non-isothermal hydrodynamics and N-body simulations for moderate and strong radial self-gravity,
 while the largest deviations occur for weak self-gravity.
 The resulting saturation wavelengths of viscous overstability for moderate and strong self-gravity ($\lambda\sim 100-300\text{m}$) agree reasonably 
 well with the length scales of axisymmetric periodic micro-structure in Saturn's inner A-ring and the B-ring, as found by Cassini.

 \end{abstract}


%

\section{Introduction}\label{sec:intro}

Observational evidence for the presence of axisymmetric periodic 
micro-structure on length scales of $100 \,\text{m}-200 \,\text{m}$ in Saturn’s A and B 
rings was revealed by several instruments onboard the Cassini mission 
to Saturn. The structure was seen in radio occultations performed by the 
Radio Science Subsystem (RSS) (\citet{Thomson2007}) and stellar 
occultations carried out with the Ultraviolet Imaging Spectrograph 
(UVIS) (\citet{colwell2007};~ \citet{sremcevic2009}).
The axisymmetric nature of oscillations was demonstrated by the Visual and Infrared Mapping 
Spectrometer (VIMS) occultations analysed by \citet{hedman2014a}, indicating azimuthal coherence of the wave trains over length scales of thousands of 
kilometers.
To date, this micro-structure is best 
explained by axisymmetric waves induced in the rings by viscous overstability.

Since the work of \citet{schmit1995,schmit1999} an increasing amount of effort has been devoted to theoretical as well as simulational 
studies of the spontaneous viscous 
overstability in Saturn's rings. \citet{schmit1995} performed a detailed linear stability analysis of an isothermal 
hydrodynamic model of Saturn's B-ring by using transport coefficients estimated from the results of steady state particle simulations by \citet{wisdom1988}.
They concluded that Saturn's B-ring is most likely subject to viscous overstability, which arises as a spontaneous oscillatory instability of 
the ring flow if certain conditions are met.
In the hydrodynamic model of \citet{schmit1995} this condition is that the viscosity of the ring is a sufficiently steep function of the 
surface mass density, expressed in terms of a powerlaw dependence with an exponent $\beta\gtrsim 0\text{-}0.5$.
A steep dependence of viscosity on density, fulfilling the above condition by a significant margin, was found in studies of \citet{araki1986} and 
\citet{wisdom1988}.
\citet{schmit1999} followed the linear growth of overstable waves into the nonlinear regime by numerical 
solution of the isothermal hydrodynamic thin disk equations including radial self-gravity forces. They found that an initially disordered wave 
state evolves into a more ordered state, with a narrow band of preferred lengths scales.
They proposed the viscous overstability as structure forming mechanism in Saturn's B-ring, manifesting in form of nonlinear wave patterns with 
wavelengths corresponding to a few times the Jeans-wavelength. 
They concluded that it is the radial collective self-gravity force which sets this length scale.
At that time, no high resolution data was available to confirm the existence of such small scale structures in Saturn's rings.
Also, no signs of such overstable oscillations had been seen in any N-body simulations conducted so far, even though the condition derived by 
\citet{schmit1995}, $\beta\gtrsim 0\text{-}0.5$, should have been fulfilled. However, there were indications, based on idealized 2D simulations, that systems 
with even larger $\beta$ might become overstable (\citet{salo2001b}). 

The paper by \citet{salo2001} was the first study that demonstrated viscous overstability in a realistic N-body simulation of a 3D self-gravitating 
particulate ring. Furthermore, it was shown that axisymmetric overstable oscillations can co-exist with non-axisymmetric gravitational wake structures, which 
emerge for a wide range of parameters when 
particle-particle gravity is taken into account (\citet{salo1992a}). The condition found from the simulations for the onset of overstability, $\beta \gtrsim 
1$, was more 
stringent
than predicted by the isothermal model of \citet{schmit1995}. It was also found that the ring's vertical self-gravity is crucial in promoting overstable 
behavior at optical depths around unity.
Indeed, a basically similar overstable behavior as seen in fully self-gravitating systems, is obtained in non-gravitating systems, provided that the vertical  
component of the planet's gravity is artificially increased by using an enhanced frequency of vertical oscillations, a method devised by \citet{wisdom1988}. 
This treatment also has the advantage of possessing a uniform ground state, which makes it possible to measure transport coefficients, and other hydrodynamic 
quantities of 
interest. This is done by using simulations whose radial scale is smaller than the smallest unstable wavelength.

The linear stability criterion for viscous overstability found in \citet{salo2001} turned out to agree well with the non-isothermal linear model of 
\citet{schmidt2001b}, based on the transport coefficients measured from simulations. This model extended the hydrodynamic 
description of \citet{schmit1995} by including the thermal balance equation to the hydrodynamic model (see also \citet{spahn2000a}). The analysis of the 
non-isothermal model indicated that 
thermal variations mitigate overstability, shifting the stability boundary to higher values of $\beta$, corresponding to higher values of optical depth.

Later, \citet{schmidt2003} (hereafter SS2003) formulated a weakly nonlinear model for the viscous overstability in terms of coupled Landau-type 
amplitude equations for nonlinear waves. For this isothermal model they used the transport coefficients obtained by \citet{salo2001}, modified such that they - 
effectively - included thermal effects. The resulting stability boundary and the growth rates of of overstable modes agreed with those of a non-isothermal 
model based on the original transport 
coefficients. With the modified weakly nonlinear isothermal model SS2003 showed that the viscous overstability can saturate in form of nonlinear traveling 
waves and that the weakly nonlinear description is in qualitative agreement with N-body simulations, at least in the limit without self-gravity. 

\citet{latter2006} performed a detailed analysis of a linearized kinetic second-order moment description of a vertically averaged dilute ring.
Although they found that viscous overstability does not occur in a dilute ring, the authors addressed two interesting issues which are not assessable with 
hydrodynamics. 
One is the anisotropy of the velocity dispersion tensor. Compared with hydrodynamics, the kinetic treatment brings about additional deformation-modes of the 
velocity ellipsoid when considering small disturbances to the ring's ground state. 
In order to assess how these anisotropic perturbations influence the rings susceptibility to viscous 
overstability they compared a linear stability analysis incorporating a Krook collision term, previously introduced in the context of planetary rings by 
\citet{shu1985c}, with a stability analysis where 
particle collisions are modeled with a tri-axial Gaussian (\citet{goldreich1978a}) for the velocity-distribution function. 
The latter treatment should account for the effects of anisotropy of the velocity ellipsoid in a more realistic manner. The resulting stability boundaries for 
viscous overstability were found to differ for both treatments, though not by large amounts.
The second aspect which \citet{latter2006} assessed is the (collisional) relaxation of the pressure tensor components as it occurs in their kinetic treatment. 
They discovered that the (long) relaxation time of the stress components in a dilute ring destroys the synchronization with the density 
oscillations, which is crucial for the viscous overstability mechanism and which is assumed a priori in hydrodynamics.
In a following paper \citet{latter2008} investigated a dense ring within a kinetic treatment based on an Enskog-equation.
A linear stability analysis of the ground state, including vertical self-gravity, revealed similar threshold values for the optical depth to instigate 
viscous overstability, as had been found earlier in the N-Body simulations of \citet{salo2001}.
Within the linearized treatment \citet{latter2008} also found that, while the vertical component of self-gravity lowers the critical optical depth for the 
onset of viscous overstability, the radial component strengthens unstable behavior on intermediate length scales.

In subsequent studies, \citet{latter2009} (hereafter LO2009), \citet{latter2010} (hereafter LO2010), as well as  \citet{latter2013} (hereafter RL2013) 
investigated the \emph{large-scale} nonlinear evolution of the viscous overstability.
LO2009 performed a nonlinear stability analysis of periodic nonlinear density wave-trains in an isothermal hydrodynamic model. They found stable wave solutions 
on which the ring flow can settle. They suggested that the overstable state might be best described by an interplay of individually stable 
waves with different wavelengths undergoing modulations 
in phase and amplitude.

LO2010 solved the isothermal hydrodynamical model numerically and confirmed many of the results derived in 
their previous paper.
The main result was that the viscous overstability saturates in nonlinear traveling waves with wavelengths directly related to
the viscous parameters of the underlying model, in reasonable agreement with the wavelengths observed with UVIS and RSS.
However, it was also shown that during the process of initial, linear growth towards final saturation, a disordered state with counter-propagating 
waves, separated by sink and source structures, occurs. LO2010 pointed out that this intermediate state might even be more relevant to Saturn's rings
than the final state, since the latter is strongly influenced by the boundary conditions of the integration, which would in the rings vary themselves on larger 
timescales due to the effect of external perturbations and the evolution of the rings in response to existing gradients in the system. This hypothesis was 
further substantiated by the 
results of simulations with different boundary conditions. Both LO2009 and LO2010 omitted self-gravity forces in their considerations.

RL2013 presented the results of non self-gravitating N-body simulations. The results on the nonlinear evolution of overstability 
are qualitatively similar to those of LO2010, but the dynamics in the early stages of the simulation are different. These 
now include the occurrence of complicated standing wave 
patterns, which later progress into the source/sink states, discovered by LO2010.
The hydrodynamical integrations of LO2010 and the particle simulations of RL2013 are the most detailed large-scale studies of the viscous 
overstability in Saturn's rings to 
date. However, both studies omit the effect of self-gravity and there are indications that the inclusion of self-gravity may significantly reduce the 
wavelength range of oscillations \citep{schmit1999,salo2018}.
The only published nonlinear hydrodynamical study of the viscous overstability in Saturn's rings 
which includes the planar components of self-gravity is that of \citet{schmit1999}. However, the results of that study contradict to some extent those of 
LO2010 in the limit without self-gravity, motivating a re-assessment of a large-scale hydrodynamic model, including the effect of self-gravity.
Furthermore, \citet{schmit1999} considered only one fixed set of parameters, that is based on the particle simulations of \citet{wisdom1988}.

For a quantitative investigation of the nonlinear saturation of viscous overstability in Saturn's rings a study including the effect of full 
self-gravity would be ideal. However, this poses great challenges to theory and heavy computational demand in N-body simulations. For this reason we study in 
this 
paper the effects of a collective, axisymmetric self-gravity force on the nonlinear saturation of viscous 
overstability in Saturn's rings. To this extent we perform a series of N-body simulations, accompanied by corresponding hydrodynamical computations.
In the hydrodynamic model we distinguish between the isothermal and the non-isothermal approximation. Their comparison allows to bring out 
the influence of the temperature equation, which directly derives from a kinetic treatment of the ring flow and whose neglect is 
therefore in general not justified. On the other hand, due to the very high collision frequencies of the systems studied here, one can expect that 
effects of anisotropy are less important so that the assumption of a Newtonian stress tensor is still a valid approximation for the hydrodynamical model.

Section \ref{sec:theo} of the paper summarizes the hydrodynamic model for a dense ring. We also present a brief linear stability calculation and basic 
theoretical aspects, which are needed to describe the results of the simulations and model calculations. In Section \ref{sec:hydropar} we provide different 
sets of numerical values for the parameters and transport coefficients of the hydrodynamic model. In Section \ref{sec:numerics} we explain our 
numerical scheme to solve the 
hydrodynamic equations, presenting also numerical tests to verify accuracy and stability. We further outline our N-Body simulation method, and determine 
growth rates and occillation frequencies in the linear regime. 
In Section \ref{sec:results} we first present the results of our hydrodynamical computations without axisymmetric self-gravity, establishing a connection to the
results of LO2010 and RL2013. Then we turn to a description of the hydrodynamical solutions with a radial 
self-gravity force and the results of our N-Body runs. In Section \ref{sec:lambdasat} we present a critical and comparative 
discussion of the main results of both approaches, and infer properties of the nonlinearly saturated, final wave state. The section closes with a 
brief comparison with previous studies. Finally, in Section \ref{sec:conc} we summarize our main results and point out prospects for future work.

\section{Hydrodynamic Theory}\label{sec:theo}

We adopt a non-isothermal, axisymmetric hydrodynamic model for a dense planetary ring.
The nonlinear hydrodynamic equations, formulated in the shearing sheet approximation (\citet{goldreich1965}) at constant distance $r$ from Saturn read (cf.\ 
\citet{stewart1984};~\citet{schmidt2009})
 \begin{align}\label{eq:nleq}
\begin{split}
\partial_{t} \sigma & = -u \partial_{x} \sigma - \sigma \partial_{x} u \\[0.1cm]
 \partial_{t} u  &  = -u\partial_{x} u + 2\Omega  v  -\partial_{x}\phi - \frac{1}{\sigma} \partial_{x} \hat{P}_{xx} 
\\[0.1cm] 
\partial_{t} v & = - u \partial_{x}v -\frac{1}{2}\Omega u  - \frac{1}{\sigma} \partial_{x} \hat{P}_{xy}
\\[0.1cm]
\partial_{t} T & = -u \partial_{x}T -\frac{2}{3 \sigma} \left[\hat{P}:\hat{S} + \partial_{x} F_{\kappa} + \Gamma \right].
\end{split}
\end{align}
In these equations $x$ and $y$ denote the radial and azimuthal coordinate, respectively, in a frame rotating with local Keplerian 
frequency $\Omega = \sqrt{GM_{S} /r^3}$, where Saturn's mass is denoted by $M_{S}$ 
and $G$ is the gravitational constant.  
The quantity $\sigma$ denotes the surface mass density and $u$, $v$  stand for the radial and the azimuthal components of the velocity $\mathbf{u}$.
Furthermore, $T$, $\hat{P}$, $F_{\kappa}$ and $\Gamma$ are the granular temperature, the pressure tensor, the heat flux and the cooling function 
(more on these quantities follows below).
The central planet is assumed to be spherical so that we have equality between the orbital frequency $\Omega(r)$ and epicyclic frequency $\kappa(r)$. Note that 
the rings' ground state which describes the balance of central gravity and centrifugal force is subtracted from above equations, thereby also neglecting the 
secular viscous evolution which occurs on timescales much longer than those investigated here.
Equations (\ref{eq:nleq}), together with Poisson's equation for a thin axisymmetric disk
\begin{equation}\label{eq:poissoneq}
 (\partial_{x}^2 +\partial_{z}^2) \phi=4 \pi G \sigma \delta(z), 
\end{equation}
relating the self-gravity potential $\phi$ to the surface density $\sigma$, form a closed set, once we provide constitutive relations for  $\hat{P}$, 
$F_{\kappa}$ and $\Gamma$. These equations can be applied to describe the evolution of axisymmetric 
structures induced by 
intrinsic (no external forcing) instability mechanisms of the disk on length scales much smaller than the radial extent of the disk (neglect of curvature 
terms).  
Equations (\ref{eq:nleq}) can in principle be obtained from a vertical integration of the kinetic moment equations for the particle number density, the 
momentum 
density and the pressure tensor components, in the limit of high collision frequency, and under the neglect of vertical deformations of the disk. In the limit 
of high collision frequency, i.e.\ in the hydrodynamic limit, the viscous pressure tensor $\hat{P}$ is local in time and can be 
assumed to be of Newtonian form, so that its individual components need not be solved for from additional partial differential equations (\citet{shu1985c};~ 
\citet{latter2006};~\citet{latter2008}).

Self-gravity wakes are omitted in our study. The wakes would imply an inhomogeneous ground state and provide dominant 
contributions to the (angular) momentum transport (\citet{daisaka2001}). The vertical averaging is justified as long as the studied phenomena vary on 
radial length scales much larger than 
the vertical extent of the disk. 
Our description relies additionally on the assumption of hydrostatic equilibrium in $z$-direction within the disk.
While this assumption is adequate in near equilibrium states, it may be violated in strongly perturbed regions, as the compressed phase of nonlinear 
overstable oscillations, where the ring particles undergo a vertical splashing.

In this paper we also investigate the influence of the temperature equation [last of Equations (\ref{eq:nleq})] on the long term nonlinear evolution of 
the viscous overstability in Saturn's 
rings.
This equation corresponds to the trace of the (vertically averaged) kinetic equations for the velocity dispersion tensor $\hat{C}_{ij}=\langle w_{i} 
w_{j}\rangle$, with $\mathbf{w}$ being the peculiar velocity of ring particles, i.e.\ their velocity relative to the mean velocity field 
$\mathbf{u}$.
The temperature is defined by
\begin{equation}\label{eq:temp}
T=\frac{1}{3} \sum\limits_{i,j}^{} \delta_{ij} \hat{C}_{ij} \hspace{0.5cm} (i,j=x,y,z)
\end{equation}
and relates to the local isotropic pressure through
\begin{equation}\label{eq:ploc}
p^{l}=\sigma T
\end{equation}
which arises from the particle random motions.

The meaning of the remaining terms in Equations (\ref{eq:nleq}) is as follows. The cooling function $\Gamma$, which derives from the collisional 
relaxation of 
the 
diagonal components of the 
pressure tensor $\hat{P}$ (see below), describes the cooling due to inelastic particle collisions,
while $\hat{P}:\hat{S}$ is the rate at which the momentum flux (mainly the nonlocal contribution in the range of parameters 
addressed in this study) converts kinetic energy in form of systematic particle motions $\mathbf{u}$ into thermal energy in form of random motions. The term 
containing the heat flux 
\begin{equation}\label{eq:heatflux}
F_{\kappa} = \kappa_{D} \partial_{x} T
\end{equation}
describes thermal diffusion due to particle random motions (the local contribution) and energy transfer over one particle diameter during collisions (the 
nonlocal contribution). Both contributions are contained in the dynamic heat conductivity $\kappa_{D}$ 
(\citet{salo2001}).

The vertically integrated Newtonian pressure tensor reads
\begin{equation}\label{eq:pten}
\begin{split}
\begin{array}{@{}*{22}{l@{}}}
\hat{P} &=  \begin{pmatrix}  P_{xx}  \hspace{0.2 cm} & P_{xy} \\[0.18cm] 
P_{yx}  \hspace{0.2 cm}  & P_{yy}  \end{pmatrix}\\[0.5cm]
\quad & =\begin{pmatrix}   p -\eta\left(\frac{4}{3}+\gamma\right)\partial_{x}u   \hspace{0.2 cm} & -\eta\left(-\frac{3}{2}\Omega + 
\partial_{x}v \right) 
\\[0.15 cm] 
-\eta\left(-\frac{3}{2}\Omega +\partial_{x}v \right)  \hspace{0.2 cm}  & p + \eta \left(\frac{2}{3}-\gamma\right)\partial_{x} u  \end{pmatrix} 
\end{array}
\end{split}
\end{equation}
and is thus completely described by the velocities $u$, $v$, the dynamic shear viscosity $\eta$ and the total isotropic pressure $p$ (see below).
The ratio of the bulk and shear viscosity is denoted by $\gamma$, which is assumed to be constant (\citet{schmit1995}).
Furthermore,
\begin{equation}\label{eq:du}
\begin{array}{@{}*{22}{l@{}}}
\hat{S} = \begin{pmatrix}   \partial_{x}u   \hspace{0.2 cm} & \frac{1}{2}\partial_{x}v -\frac{3}{4}\Omega \\[0.15 cm] 
\frac{1}{2}\partial_{x}v -\frac{3}{4}\Omega  \hspace{0.2 cm}  & 0 \end{pmatrix} 
\end{array}
\end{equation}
is the rate of strain tensor.

The equation of state, the transport coefficients, and the cooling function are parameterized as
\begin{align}
 p&= p_{0} \left(\frac{\sigma}{\sigma_{0}}\right)^{p_{s}} \left(\frac{T}{T_{0}}\right)^{p_{T}}\label{eq:pres}\\[0.1cm]
 \eta&=\nu_{0} \sigma_{0} \left( \frac{\sigma}{\sigma_{0}}\right)^{\beta+1} \left( \frac{T}{T_{0}}\right)^{n_{T}}\label{eq:shearvis} \\[0.1cm]
 \kappa_{D} &= \kappa_{0} \sigma_{0} \left( \frac{\sigma}{\sigma_{0}}\right)^{\beta+1} \left( \frac{T}{T_{0}}\right)^{n_{T}}\label{eq:heatcon} \\[0.1cm]
 \Gamma &= \Gamma_{0} \left( \frac{\sigma}{\sigma_{0}}\right)^{G_{s}} \left( \frac{T}{T_{0}}\right)^{G_{T}}\label{eq:coolfunc}.
\end{align}
The ground state of the idealized disk is characterized by $\Gamma_{0}=\frac{9}{4} \nu_{0} \Omega^2 \sigma_{0}$ with $\sigma_{0}=\text{const.}$, $u_{0}=0$, 
$v_{0}=0$ and $T_{0}=\text{const.}$, together with the 
parameters in the above definition 
of the transport coefficients.

The ground state pressure $p_{0}$ in (\ref{eq:pres}) is the total isotropic ground state pressure 
\begin{equation}\label{eq:ptot}
 p_{0}= p_{0}^{l} + p_{0}^{nl} \equiv \sigma_{0} c_{0}^2
\end{equation}
containing local [Equation (\ref{eq:ploc})] and non-local contributions, the 
latter arising from transfer of momentum between particles over one particle diameter, during a collision.
With Equation (\ref{eq:ptot}) we define the \emph{effective} ground state velocity dispersion $c_{0}$, which effectively includes nonlocal pressure.
Note that also $\nu_{0}$ and $\kappa_{0}$ contain local and nonlocal contributions.
For later use we additionally define the hydrodynamic ground state Toomre-parameter as
\begin{equation}
 Q_{0}=\frac{\Omega c_{0}}{\pi G \sigma_{0}}.
\end{equation}

Values for these parameters were derived from small-scale steady state and mildly perturbed non-steady-state simulations in \citet{salo2001}. A similar 
theoretical approach as the one 
adopted here showed (\citet{schmidt2001b}) that these parameters reproduce the stability boundary and the growth rates of overstable modes, found 
in N-body simulations which had sufficient radial extent for perturbations to grow. However, these comparisons did not include axisymmetric gravity, which is 
the topic of the current study.

In the following we summarize basic results from linear theory relevant for this study.
We add small axisymmetric oscillatory disturbances to the homogeneous ground state 
\begin{equation}\label{eq:linper}
 \begin{pmatrix} \sigma   \\ u\\ v  \\ T \end{pmatrix} = \begin{pmatrix} \sigma_{0}   \\ 0\\ 0  \\ T_{0} \end{pmatrix} + \begin{pmatrix} 
\hat{\sigma}   
\\ \hat{u} \\ \hat{v}  \\ \hat{T} \end{pmatrix} \exp\left(\omega t + i k x\right),
\end{equation}
with complex oscillation frequency $\omega=\omega_{R} + i \, \omega_{I}$ and real-valued wavenumber $k>0$.
The solution of Poisson's equation provides the relation
\begin{equation}\label{eq:wkbsg}
 \hat{\phi} = -\frac{2 \pi G }{k}\hat{\sigma},
\end{equation}
for the perturbation in the self-gravitational potential generated by a single axisymmetric mode (\citet{binney1987}).

In the remainder of this section we apply the dimensional scalings as listed in Table \ref{tab:scalings} and drop the hat from the perturbation amplitudes. 
Inserting (\ref{eq:linper}) and (\ref{eq:wkbsg}) into (\ref{eq:nleq}) and linearizing with respect to the perturbations, results in an eigenvalue 
problem 
\begin{equation}\label{eq:evp}
 Det[\hat{M}]=0
\end{equation}
where 
\begin{equation}\label{eq:matni}
\begin{array}{@{}*{22}{l@{}}}
\hat{M} = \begin{pmatrix}   -\omega  \hspace{0.2 cm} & -ik \hspace{0.2 cm} & 0 \hspace{0.2 cm}  & 0 \\[0.15 cm] 
i \left(\frac{2}{Q_{0}} - k \,  p_{s}\right) \hspace{0.2 cm}  & -(\frac{4}{3} + \gamma) k^2 \nu_{0} -\omega  \hspace{0.2 cm}  & 2 \hspace{0.2 cm}  & -\frac{ i 
k \,  p_{T}}{T_{0}}\\[0.15 cm] 
-\frac{3}{2} i \left(1+\beta\right)k\nu_{0}  \hspace{0.2 cm} & -\frac{1}{2} \hspace{0.2 cm} & -k^2 \nu_{0} -\omega \hspace{0.2 cm}  & -\frac{3 i k \, n_{T} 
\nu_{0}}{2 T_{0}} \\[0.15cm] 
\frac{3}{2}\left(1+\beta-G_{s}\right)\nu_{0} \hspace{0.2 cm}  & -\frac{2}{3}i k  \hspace{0.2 cm}  & -2 i k \nu_{0} \hspace{0.2 cm}  & -\frac{9  
\nu_{0}\left(G_{T}- 
n_{T}\right)  + 4 k^2 \kappa_{0} T_{0} }{6 T_{0}}-\omega
\end{pmatrix}. 
\end{array}
\end{equation}
Later, we will calculate numerically $\omega_{R}$ and $\omega_{I}$ from Equation (\ref{eq:evp}).
Here, we solve (\ref{eq:evp}) perturbatively by inserting
\begin{equation}
 \omega = \omega^{(0)} + k \, \omega^{(1)} + k^2\, \omega^{(2)} +\ldots
\end{equation}
and solving for each order of $k$ separately.
Following this procedure we end up with four approximate eigenfrequencies, correct to order $k^2$
\begin{subequations}
 \begin{align}
\omega_{1} & =-\frac{3\nu_{0} \left(G_{T} - n_{T} \right)}{2 T_{0}} + k^2 \left(-\frac{2}{3}\kappa_{0} + F_{1} \right)\label{eq:emode},\\
\omega_{2} & = i - k\,   \frac{i}{Q_{0}} + k^2 \left[ \frac{1}{6} \nu_{0}\left(2 + 9 \beta - 3 \gamma \right) + \frac{1}{2}i \left(p_{s}-\frac{1}{Q_{0}^2} 
\right) +\frac{1}{2}\left(i\, F_{2} +F_{3}\right) \right]\label{eq:osmode1},\\
\omega_{3} & = \omega_{2}^{*}\label{eq:osmode2},\\
\omega_{4} & = k^2 \left[ - 3 \nu_{0} \left(1+\beta\right) -3\nu_{0} \frac{n_{T}}{G_{T}-n_{T}}\left(1+\beta - G_{s}\right)\right].
  \end{align}
\end{subequations}
The higher orders contain long expressions, providing little insight, and are therefore omitted here. 
The first mode ($\omega_{1}$) is the energy mode, describing the thermal relaxation of local disturbances of the thermal equilibrium. Generally $G_{T}-n_{T}>0$ 
in a thermally stable disk.
The isothermal limit is recovered if $G_{T}\to \infty$, corresponding to an infinitely fast decay of any temperature perturbation.
The last mode ($\omega_{4}$) is associated with the viscous instability (\citet{lukkari1981,lin1981,ward1981,schmit1995,salo2009}).
The second and third mode are the oscillatory modes of interest in this study,
the linear viscous overstability modes. The expressions $F_{1}$-$F_{3}$, arising from the temperature equation, contain a large number of 
terms and will not be displayed here (cf.\ \citet{schmidt2001b}). Numerical values of $F_{2}$ and $F_{3}$ for all parameter sets used in this paper are 
listed in Table \ref{tab:hydropar}.

Within the isothermal model for the nonlinear saturation of viscous overstability considered in this paper, we solve only the first three equations 
(\ref{eq:nleq}), adopting a constant temperature $T_{0}$. We use the \emph{ideal gas relation} 
for pressure
\begin{equation}\label{eq:ideal}
 p= p_{0} \frac{\sigma}{\sigma_{0}},
\end{equation}
allowing a comparison of our results to \citet{schmit1999} and LO2010. This implies $p_{s}=1$ in Equation (\ref{eq:pres}). But we will also investigate 
isothermal models with $p_{s}>1$. The only other quantity needed in the isothermal case is 
$\eta\left(T=T_{0}\right)$. 

Furthermore, for later use we define 
the $k^2$ approximations of the linear growth rate and the linear oscillation frequency from Equation (\ref{eq:osmode1}) by
\begin{equation}\label{eq:osgrate}
\omega_{R} =  \left[ \frac{1}{6} \nu_{0}\left(2 + 9 \beta - 3 \gamma \right)  + \frac{1}{2} F_{3}\right] k^2
\end{equation}
and 
\begin{equation}\label{eq:osfreq}
 \omega_{I} = 1-\frac{1}{Q_{0}}k + \frac{1}{2} \left(- \frac{1}{Q_{0}^2} + p_{s}  +  F_{2}\right) k^2,
\end{equation}
respectively.
From the condition that the real part of $\omega_{2}$ or $\omega_{3}$ vanishes one can define a critical value $\beta_{c}$
for the exponent of the density dependence of the viscosity (\ref{eq:shearvis}) so that for $\beta>\beta_{c}$ the 
system exhibits linear viscous overstability. An approximation for $\beta_{c}$ derives from (\ref{eq:osgrate}) and reads
$$\beta_{c}=\frac{1}{3} \, (\gamma-\frac{2}{3} - \frac{F_{3}}{\nu_{0}}).$$
If we set $\gamma=1$  and ignore thermal effects ($F_{3}=0$) we recover the value given in 
\citet{schmit1995}.

\begin{figure}[ht!]
\centering
\includegraphics[width = 0.4 \textwidth]{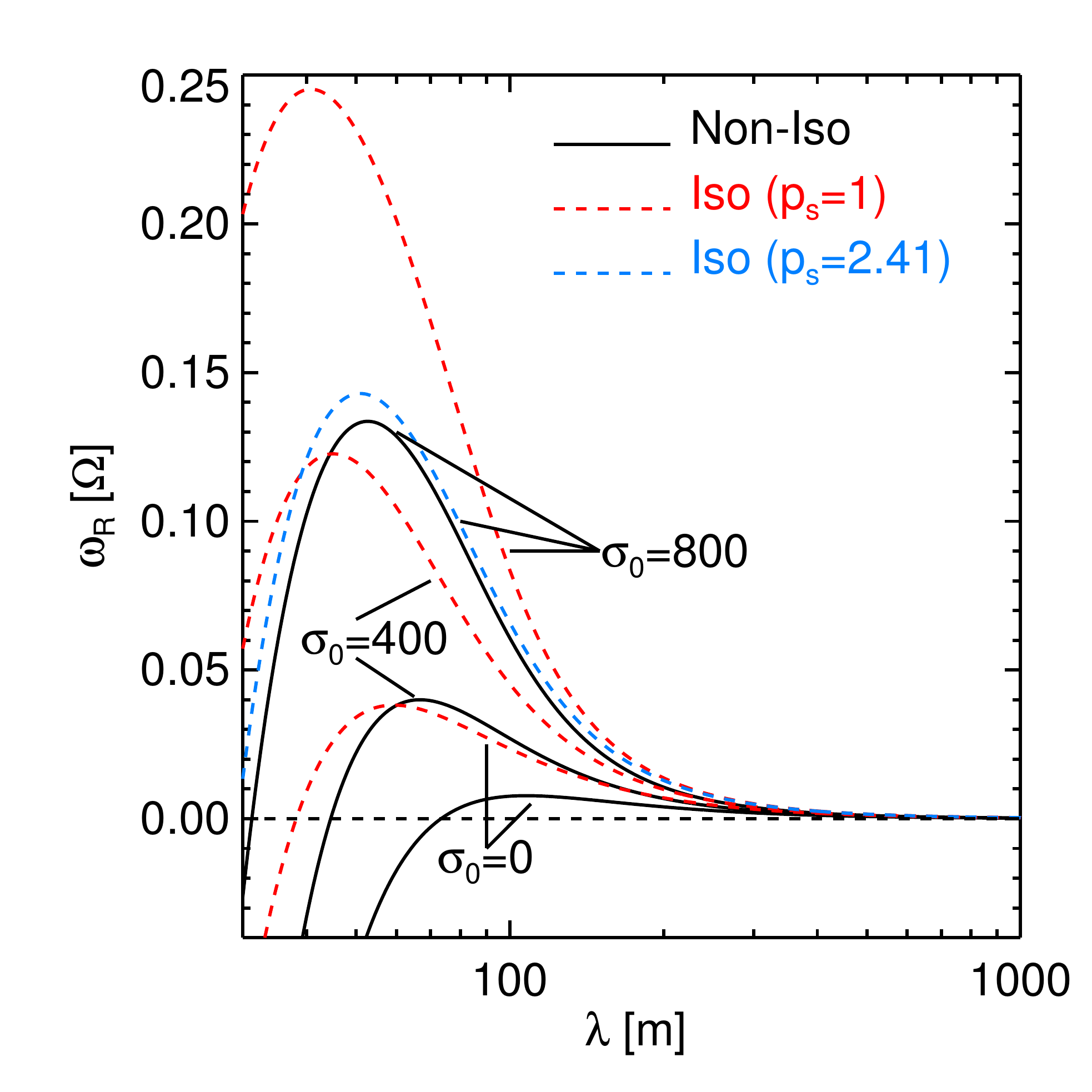}
\includegraphics[width =0.4 \textwidth]{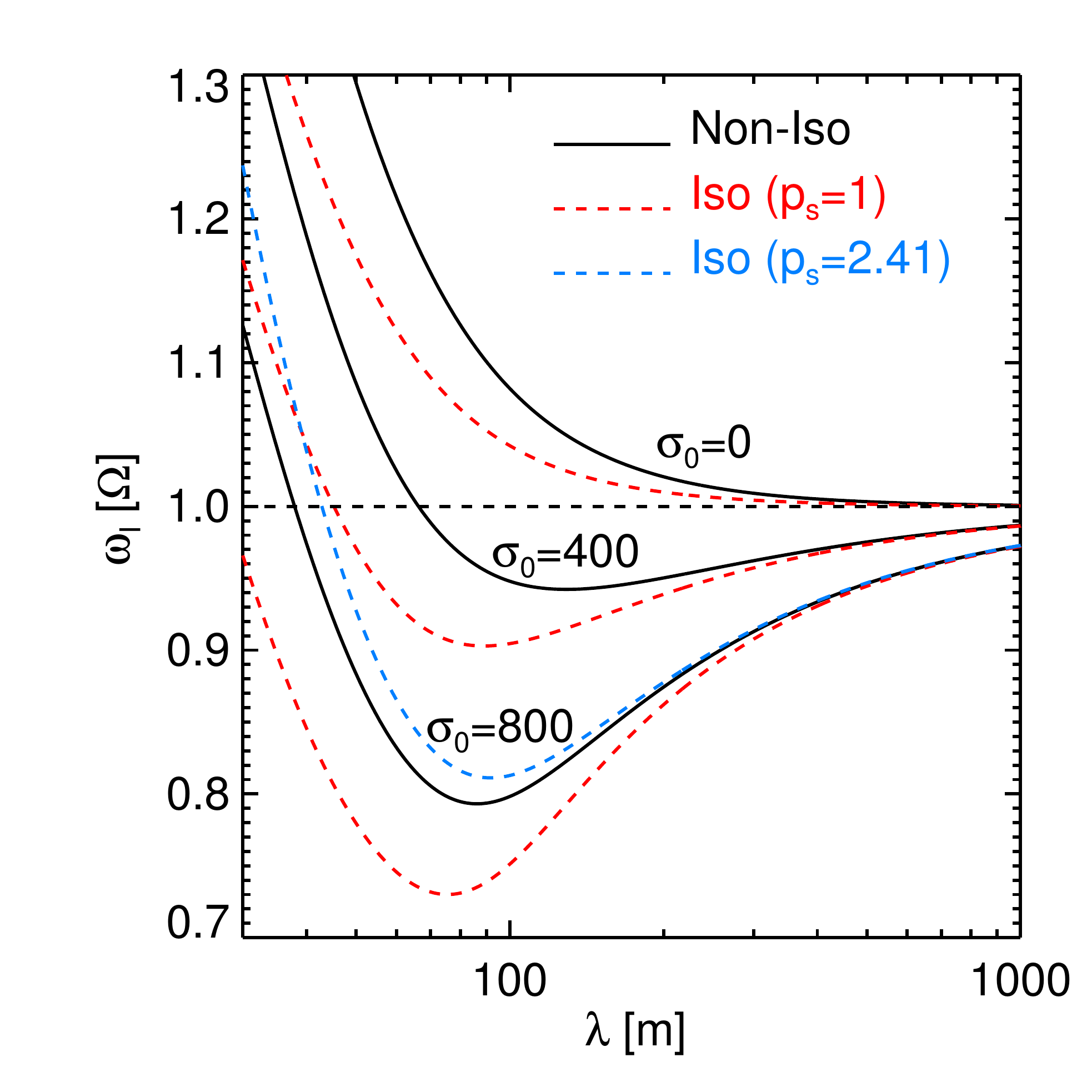}
\caption{Theoretical linear growth rates (left) and oscillation frequencies (right) of overstable waves for ground state 
surface mass densities $\sigma_{0}=0,400$ and $800\,\text{kg}\,\text{m}^{-2}$. Parameters correspond to an optical depth of $\tau=1.5$ and $\Omega_{z}=3.6$ 
(see Section \ref{sec:hydropar}). All curves are calculated numerically from the exact Equation  (\ref{eq:evp}). The red isothermal curves are obtained in 
the limit $G_{T}\to \infty$ and $p_{s}=1$. For the blue isothermal curve (for $\sigma_{0}=800\,\text{kg}\,\text{m}^{-2}$) in each panel the value $p_{s}=2.41$ 
is used.}
\label{fig:omega}
\end{figure}

Figure \ref{fig:omega} displays linear growth rates and oscillation frequencies of overstable waves, following from the isothermal and non-isothermal model for 
different surface densities $\sigma_{0}$, employing
a set of parameters that corresponds to an optical depth of $\tau=1.5$ and a vertical frequency enhanced by a factor 3.6 (see Section 
\ref{sec:hydropar} for details on 
the parameter sets). 
The black solid curves represent the non-isothermal model. The red dashed curves correspond to the isothermal model (with $G_{T}\to \infty$ and $p_{s}=1$).
Overall it is the (larger) pressure coefficient $p_{s}=2.41$ of the non-isothermal model, which causes the main difference from the isothermal 
model. 
To bring out the deviations caused by thermal effects alone, i.e. those arising from the temperature equation, we show for comparison the oscillation 
frequencies resulting from an isothermal model with a pressure coefficient $p_{s}=2.41$ (blue dashed curve for the case 
$\sigma_{0}=800\,\text{kg}\,\text{m}^{-2}$).
The differences between the blue curve and the black curve for $\sigma_{0}=800\,\text{kg}\,\text{m}^{-2}$ reveal that for the used parameter set thermal 
effects (mildly) reduce both growth rates and oscillation frequencies.
This behavior is also predicted by the approximations (\ref{eq:osgrate}) and (\ref{eq:osfreq}), using the corresponding values for $F_{2}$ and $F_{3}$ 
from Table \ref{tab:hydropar}.

\begin{figure}[h!]
\centering
\includegraphics[width = 0.4 \textwidth]{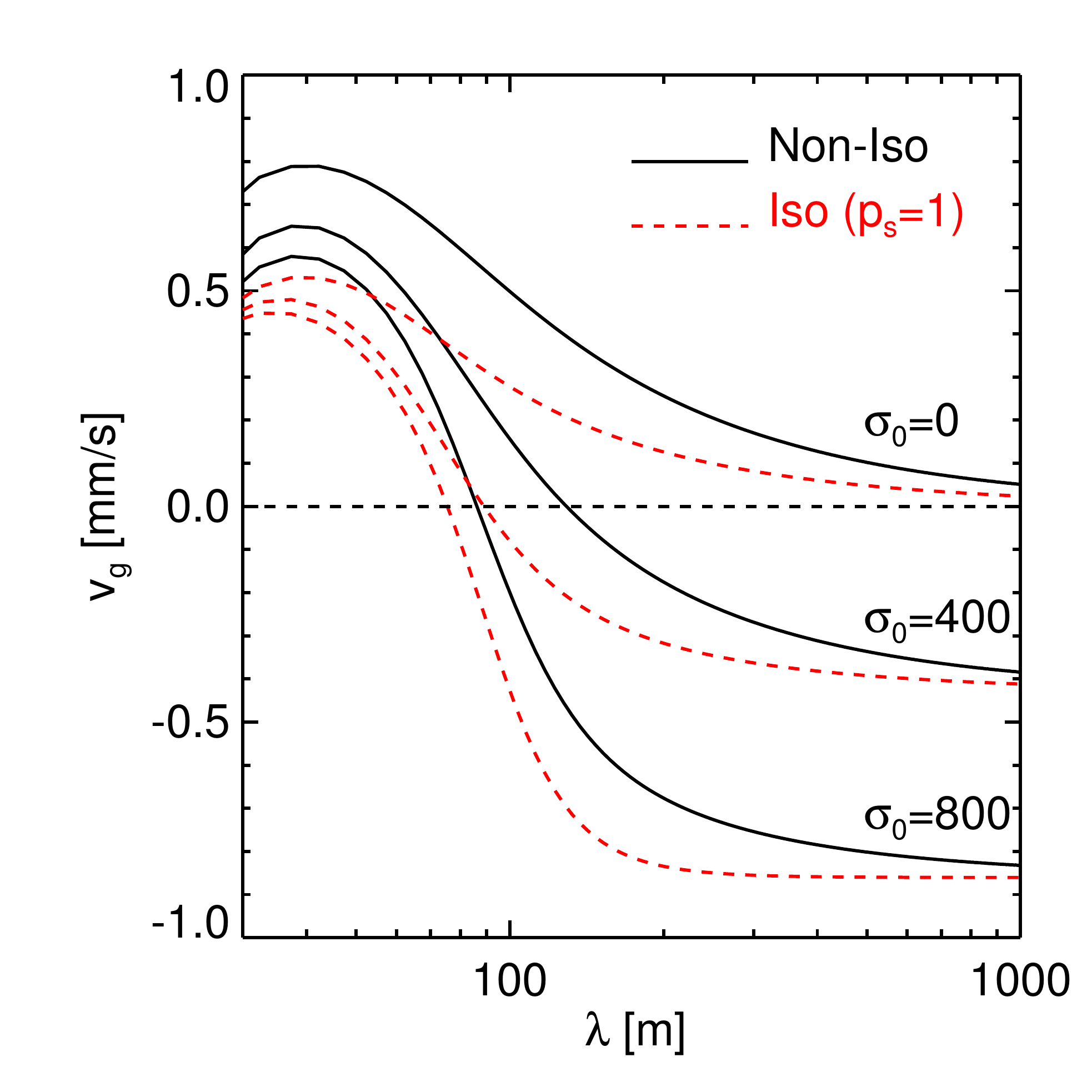}
\includegraphics[width =0.4 \textwidth]{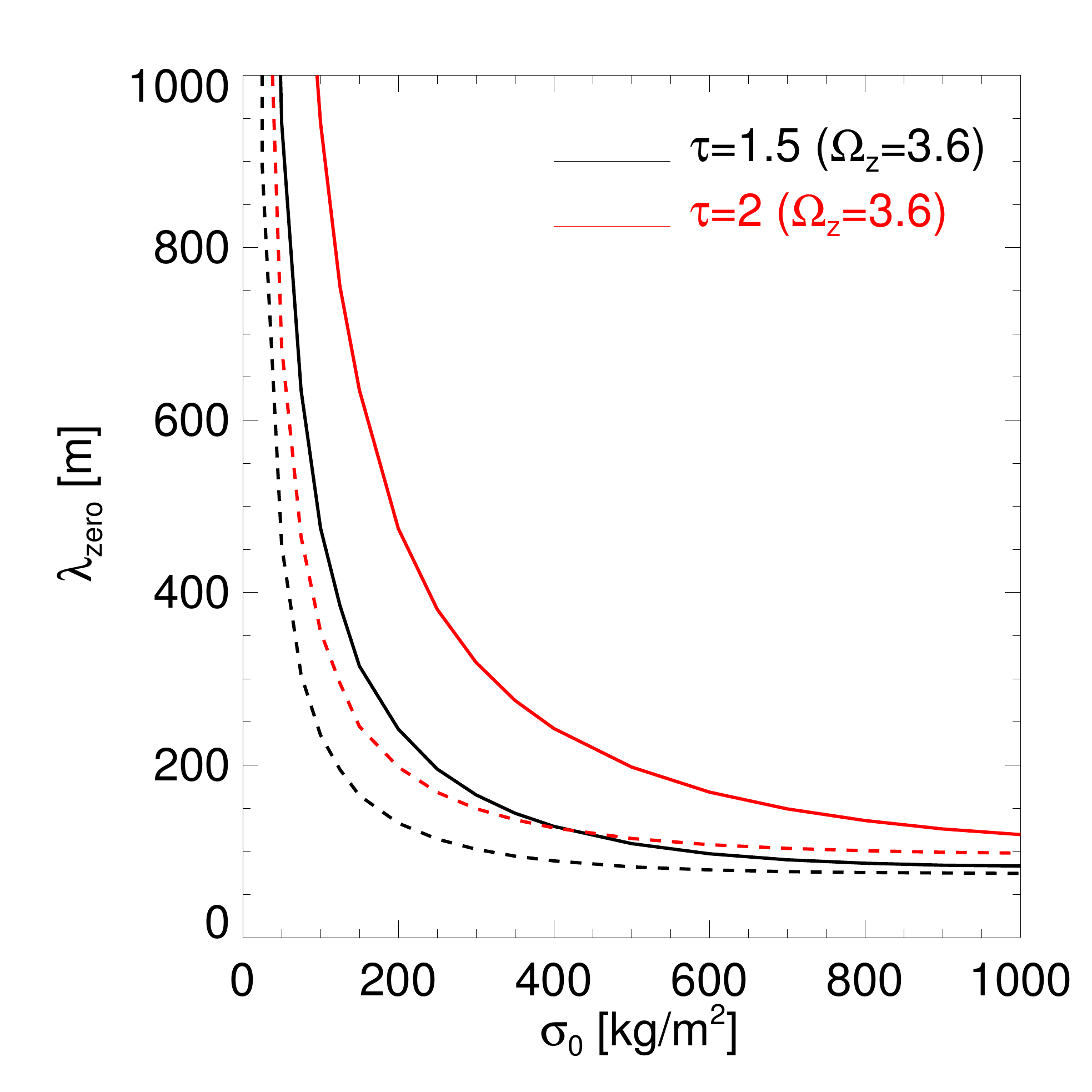}
\caption{The left panel displays linear group velocities (\ref{eq:vg}) of overstable waves for different ground state 
surface mass densities 
$\sigma_{0}$ (expressed in units $\,\text{kg}\,\text{m}^{-2}$). In the case of a non-zero self-gravity the oscillation frequency possesses a minimum 
at a certain wavelength, where the group velocity changes direction. This wavelength (right panel) decreases with increasing $\sigma_{0}$. The dashed 
curves represent the isothermal model ($G_{T}\to \infty$ and $p_{s}=1$). Parameters correspond to an optical depth of $\tau=1.5$ and $\Omega_{z}=3.6$ (see 
Section \ref{sec:hydropar}).}
\label{fig:vgroup}
\end{figure}

For the description of the pattern of sources and sinks that arise in course of the nonlinear evolution of viscously overstable modes (Section 
\ref{sec:hydronosg}) we need to define the group velocity of overstable 
waves, which measures the propagation speed of small perturbations imposed to the wave trains.
The group velocity of linear overstable waves is given by 
\begin{equation}\label{eq:vg}
 v_{g}=\frac{\mathrm{d} \omega_{I}}{\mathrm{d}k}
\end{equation}
and is displayed in Figure \ref{fig:vgroup} for the same parameters as used in Figure \ref{fig:omega}.
Only in the presence of self-gravity the group velocity changes its sign
at a certain wavelength which we name $\lambda_{zero}(\sigma_{0})$ (the subscript $zero$ indicating the vanishing of the group velocity). The wavelength 
$\lambda_{zero}(\sigma_{0})$ (Figure \ref{fig:vgroup}) is a 
decreasing function of $\sigma_{0}$.
In the isothermal model, indicated by dashed lines, the wavelengths $\lambda_{zero}(\sigma_{0})$ are shifted towards smaller values.

\FloatBarrier

\clearpage
\section{Hydrodynamic Parameters}\label{sec:hydropar}

The hydrodynamic model introduced in the previous section contains free 
parameters and transport coefficients which must be specified before 
equations (\ref{eq:nleq}) can be integrated. Sets of parameters were derived by \citet{salo2001} 
from N-body simulations which were conducted for Saturnocentric distance $10^{5}\,\text{km}$ and $\Omega=1.95\cdot 10^{-4}\,\text{s}^{-1}$, using the 
\citet{bridges1984} velocity- 
dependent coefficient of restitution and a particle radius of one meter.
Self-gravity was approximated by an enhancement of the vertical frequency of oscillations, $\Omega_z/\Omega>1$, a method introduced by \citet{wisdom1988}. 
Because we 
neglect the effects of direct particle-particle gravity, these sets of parameters 
can in principle be used directly for our hydrodynamic integrations. 

Note that the parameter $\Omega_z$ does not directly enter the hydrodynamic 
model. It affects indirectly through the altered transport coefficients. 
The enhancement of $\Omega_z$ has the effect of increasing the overall 
collision frequency, as self-consistent vertical gravity would do, which 
promotes viscous overstability. 
Throughout this paper we express values of $\Omega_{z}$ scaled with the Keplerian frequency $\Omega$.
The collective radial gravity [Equation (\ref{eq:poissoneq})] used in our model does not induce any changes in the transport 
properties of the ground state ring. This will be different if true 
gravitational encounters between individual particles are taken into 
account.

It must be kept in mind that the transport coefficients from \citet{salo2001} are determined for a pre-specified parameterization of these 
quantities and their dependence on density and temperature [Equations (\ref{eq:pres})-(\ref{eq:coolfunc})]. This 
parameterization is not unique and at this 
point it is not 
even clear if the particular form provided by equations (\ref{eq:pres})-(\ref{eq:coolfunc}) is well 
suited to follow the development into the nonlinear regime. The 
coefficients were determined in simulations from small amplitude 
perturbations of the ring ground state (\citet{salo2001}), for which they 
are assumed to be representative. In the nonlinear regime, which we 
investigate in this study, we may therefore expect deviations of the 
hydrodynamical model from the results of the N-body simulations that 
arise from this problem.

In Table \ref{tab:hydropar} we list the sets of parameters that are used for the 
hydrodynamic models in this paper. The columns specify the optical 
depths for which the parameters are valid. The parameter sets for $\tau=1.5$ and $\tau=2$ with $\Omega_{z}=3.6$ are highlighted with the labels $\tau_{15}$ 
and $\tau_{20}$, to which we will 
refer in the following. The last column, labeled $st99$, gives parameters which were 
used in the hydrodynamical model by \citet{schmit1999}.
These parameters are based on the assumption that the ground state velocity dispersion takes the value $c_{0}=2 \,\text{mm}\, \text{s}^{-1}$, which
corresponds to a vertical ring thickness of $H\approx10\,\text{m}$ by adopting the dilute estimate $H=c_{0}/\Omega$.
The viscosity in this parameter set was then estimated by assuming an optical depth $\tau\sim 1$ and using the relation $\nu\sim 0.26 \, (c_{0}^2 /\Omega)\, 
\tau^{1.26}$,
found by \citet{wisdom1988}. We 
use these parameters only to test our numerical scheme in the isothermal 
limit (Section \ref{sec:hydronumtests}). The rows of the table specify the factor of 
vertical frequency enhancement that was used by \citet{salo2001} to 
determine the parameters, followed by the ground state effective 
velocity dispersion, the kinematic shear viscosity, the kinematic heat 
conductivity, and the granular temperature.

In Table \ref{tab:scalings} we provide a list of symbols with their numerical values 
used or, respectively, the characteristic scales which were employed to 
express our model in dimensionless form.

\vspace{ -0.cm}
\begin{minipage}{\textwidth}
\centering
\captionof{table}{\textbf{List of Symbols and their Scalings}}\label{tab:scalings} 
   \renewcommand*{\arraystretch}{1.2}
    \begin{tabular}{|l|l|l|l|l|l|}
 \hline
    \multicolumn{1}{|c|}{Quantity} & \multicolumn{1}{|c|}{Scaling} & \multicolumn{4}{|c|}{Value} \\     \hline 
     $c_{0}=\sqrt{\frac{p_{0}}{\sigma_{0}}}$ (effective velocity dispersion)  &   &   \multicolumn{4}{|c|}{}\\[0.15cm]
    $Q_{0}=\frac{\Omega c_{0}}{\pi G \sigma_{0}}$ (ground state Toomre-parameter)  &   &    \multicolumn{4}{|c|}{}\\  
    $G$ (gravitational constant) &  &   \multicolumn{4}{|c|}{$6.67 \cdot 10^{-11} \, \text{m}^3 \, \text{kg}^{-1} \text{s}^{-2}$}\\ 
     $M_{S}$ (Saturn's mass) &  &   \multicolumn{4}{|c|}{$5.69 \cdot 10^{26} \, \text{kg}$}\\ 
    $\Omega$ (orbital frequency at $r=10^5 \, \text{km}$)&  & \multicolumn{4}{|c|}{$ 1.948 \cdot 10^{-4}\, \text{s}^{-1}$}\\
    $x$ (radial coordinate) & $ c_{0} \, \Omega^{-1} $ &   \multicolumn{4}{|c|}{}\\ 
    $t$ (time) & $ \Omega^{-1} $ &   \multicolumn{4}{|c|}{}\\ 
    $k$ (wavenumber) & $ c_{0}^{-1} \, \Omega  $ &  \multicolumn{4}{|c|}{}\\ 
    		$\omega$ (complex wave frequency) & $\Omega$ &  \multicolumn{4}{|c|}{}\\  
   		$\sigma$ (surface mass density) & $\sigma_{0}$ &  \multicolumn{4}{|c|}{}\\
		$u$, $v$ (planar velocity components ) & $ c_{0}$ &  \multicolumn{4}{|c|}{}\\
		$T$  (temperature) & $c_{0}^2$ &   \multicolumn{4}{|c|}{}\\
		$F_{\kappa}$ (heat flux) & $\sigma_{0} \, c_{0}^3 $ &  \multicolumn{4}{|c|}{}\\
		$\phi$  (self-gravity potential) & $c_{0}^2$ &   \multicolumn{4}{|c|}{}\\
          	$p$ (isotropic pressure) & $\sigma_{0} \, c_{0}^2 $ &  \multicolumn{4}{|c|}{}\\
     		$\eta$ (dyn. shear viscosity) & $\sigma_{0} \, c_{0}^2 \, \Omega^{-1}$ &  \multicolumn{4}{|c|}{}\\
     		$\kappa$ (dyn. heat conductivity) & $\sigma_{0} \, c_{0}^2 \, \Omega^{-1}$ &  \multicolumn{4}{|c|}{}\\
     		$\Gamma$ (collisional cooling function) & $\sigma_{0} \, c_{0}^2 \,\Omega$ &  \multicolumn{4}{|c|}{}\\
    		$\hat{P}$ (pressure tensor) & $\sigma_{0} \, c_{0}^2 $ &  \multicolumn{4}{|c|}{}\\
    		$\hat{C}$ (velocity dispersion tensor) & $ c_{0}^2 $ &  \multicolumn{4}{|c|}{}\\
    		$\hat{S} $ (rate of strain tensor) & $\Omega $ &  \multicolumn{4}{|c|}{}\\ \hline 	
    \end{tabular}\par
   \bigskip
\end{minipage}

\begin{minipage}{\textwidth}
\centering
\captionof{table}{\textbf{Hydrodynamic Parameters}}\label{tab:hydropar} 
   \renewcommand*{\arraystretch}{1.2}
    \begin{tabular}{|l|l|l|l|l|l|l|}
 \hline
    \multicolumn{1}{|c|}{} &   \multicolumn{1}{|c|}{$\tau=1$} & \multicolumn{1}{|c|}{$\tau=1.5\, (\tau_{15})$} &  
\multicolumn{1}{|c|}{$\tau=2.0\, (\tau_{20})$} & \multicolumn{1}{|c|}{$\tau=1.5$} 
& \multicolumn{1}{|c|}{$\tau=2.0$} & \multicolumn{1}{|c|}{$st99$} \\     \hline 
         $\Omega_{z} \, [\Omega]$ &    \multicolumn{1}{|c|}{3.6} &  \multicolumn{1}{|c|}{3.6} & \multicolumn{1}{|c|}{3.6} & \multicolumn{1}{|c|}{2.0} & 
\multicolumn{1}{|c|}{2.0} & 
\multicolumn{1}{|c|}{3.6}\\[0.15cm]     
     $c_{0}$ [$10^{-3} \,\text{m}\text{s}^{-1}$] &     \multicolumn{1}{|c|}{0.64} &     \multicolumn{1}{|c|}{0.86} & \multicolumn{1}{|c|}{1.06} & 
\multicolumn{1}{|c|}{0.60} & 
\multicolumn{1}{|c|}{0.71} & \multicolumn{1}{|c|}{2.0}\\[0.15cm]
 $\nu_{0}$ [$10^{-3} \,\text{m}^{2}\text{s}^{-1}$] &     \multicolumn{1}{|c|}{0.75} &     \multicolumn{1}{|c|}{1.30} & \multicolumn{1}{|c|}{1.86} & 
\multicolumn{1}{|c|}{0.65}  & 
\multicolumn{1}{|c|}{0.90} & \multicolumn{1}{|c|}{5.39}\\[0.15cm]
     $\kappa_{0}$ [$10^{-3} \,\text{m}^2 \text{s}^{-1}$] &     \multicolumn{1}{|c|}{3.14} &     \multicolumn{1}{|c|}{5.38} & \multicolumn{1}{|c|}{7.61} & 
\multicolumn{1}{|c|}{2.78} & 
\multicolumn{1}{|c|}{3.53} & \multicolumn{1}{|c|}{-}\\[0.15cm]
     $T_{0}$ [$10^{-8} \,\text{m}^{2}\text{s}^{-2}$]  &     \multicolumn{1}{|c|}{6.56} &     \multicolumn{1}{|c|}{6.72} & \multicolumn{1}{|c|}{7.22}& 
\multicolumn{1}{|c|}{6.19} & 
\multicolumn{1}{|c|}{6.18}& \multicolumn{1}{|c|}{-}\\[0.15cm] \hline
     $\gamma$  &     \multicolumn{1}{|c|}{2.14} &     \multicolumn{1}{|c|}{1.99} & \multicolumn{1}{|c|}{2.12}& \multicolumn{1}{|c|}{2.06} & 
\multicolumn{1}{|c|}{1.99} & 
\multicolumn{1}{|c|}{1}\\[0.15cm]
$\beta$ &     \multicolumn{1}{|c|}{1.15}  &     \multicolumn{1}{|c|}{1.19} & \multicolumn{1}{|c|}{1.55}& \multicolumn{1}{|c|}{1.06}  & 
\multicolumn{1}{|c|}{1.16} & 
\multicolumn{1}{|c|}{1.26}\\[0.15cm]
     $n_{T}$  &     \multicolumn{1}{|c|}{-0.13} &     \multicolumn{1}{|c|}{-0.10} & \multicolumn{1}{|c|}{0.08} & \multicolumn{1}{|c|}{-0.12} & 
\multicolumn{1}{|c|}{-0.10}& 
\multicolumn{1}{|c|}{-}\\[0.15cm]
     $p_{s}$  &     \multicolumn{1}{|c|}{2.19} &     \multicolumn{1}{|c|}{2.41} & \multicolumn{1}{|c|}{2.72} & \multicolumn{1}{|c|}{2.11} & 
\multicolumn{1}{|c|}{2.26}& 
\multicolumn{1}{|c|}{-}\\[0.15cm]
     $p_{T}$  &     \multicolumn{1}{|c|}{0.22} &     \multicolumn{1}{|c|}{0.15} & \multicolumn{1}{|c|}{0.18} & \multicolumn{1}{|c|}{0.28} & 
\multicolumn{1}{|c|}{0.28}& 
\multicolumn{1}{|c|}{-}\\[0.15cm]
     $G_{s}$ &     \multicolumn{1}{|c|}{2.17}  &     \multicolumn{1}{|c|}{2.19} & \multicolumn{1}{|c|}{2.54} & \multicolumn{1}{|c|}{2.06} & 
\multicolumn{1}{|c|}{2.16}& 
\multicolumn{1}{|c|}{-}\\[0.15cm]
     $G_{T}$   &     \multicolumn{1}{|c|}{0.62} &     \multicolumn{1}{|c|}{0.57} & \multicolumn{1}{|c|}{0.67} & \multicolumn{1}{|c|}{0.61} & 
\multicolumn{1}{|c|}{0.64}& 
\multicolumn{1}{|c|}{-}\\[0.15cm] 
     $F_{2}$   &     \multicolumn{1}{|c|}{-0.28} &     \multicolumn{1}{|c|}{-0.28} & \multicolumn{1}{|c|}{0.08} & \multicolumn{1}{|c|}{-0.26} & 
\multicolumn{1}{|c|}{-0.28}& \multicolumn{1}{|c|}{-}\\[0.15cm] 
     $F_{3}$   &     \multicolumn{1}{|c|}{-0.35} &     \multicolumn{1}{|c|}{-0.26} & \multicolumn{1}{|c|}{-0.48} & \multicolumn{1}{|c|}{-0.49} 
& 
\multicolumn{1}{|c|}{-0.48}& \multicolumn{1}{|c|}{-}\\[0.15cm] \hline
   \end{tabular}\par
\end{minipage}

\newpage

\section{Numerical Methods}\label{sec:numerics}

\subsection{Hydrodynamic Scheme}\label{sec:hydro}

\subsubsection{Discretization Technique}

For numerical solution of the full nonlinear system (\ref{eq:nleq}) we bring these equations into flux-conservative form by defining
\begin{equation}\label{eq:convar}
 \mathbf{U} = \begin{pmatrix} \sigma  \\\sigma u \\ \sigma v  \\ e \end{pmatrix}
\end{equation}
with the energy density
\begin{equation}\label{eq:energy}
 e \equiv \sigma \left(\frac{1}{2} u^2 + \frac{3}{2} T\right).
\end{equation}
The two terms describe the radial kinetic and internal energy densities, respectively.
By using these quantities the Equations (\ref{eq:nleq}) are equivalent to
\begin{equation}\label{eq:hcl}
 \partial_{t} \mathbf{U} = - \partial_{x}\mathbf{F} + \mathbf{S}.
\end{equation}
In this equation we define the flux vector
\begin{equation}\label{eq:physflux}
 \mathbf{F} = \begin{pmatrix} \sigma u   \\ \sigma u^2 + p \\ \sigma u v  \\ u\left(e+p\right) \end{pmatrix}
\end{equation}
and the source term 
\begin{equation*}
 \mathbf{S} = \begin{pmatrix} 0  \\ 2\Omega  \sigma v - \sigma \partial_{x} \phi + \partial_{x} \Pi_{xx} \\ -\frac{1}{2}\Omega \sigma u + \partial_{x} 
\Pi_{xy}  \\  2 \Omega \sigma u v + \hat{\Pi}:\mathbf{\nabla} \mathbf{u} -\Gamma -\partial_{x}F_{\kappa} - \sigma u \, \partial_{x} \phi  + u \,\partial_{x} 
\Pi_{xx}  \end{pmatrix}
\end{equation*}
where 
\begin{equation*}
\Pi_{kl} = p\delta_{kl}-P_{kl}  
\end{equation*}
denotes the viscous stress tensor.

We solve the system (\ref{eq:hcl}) with a conservative finite difference method on a uniform mesh with nodes $x_{j}$, where $j=1,2,\ldots,n$, with grid 
spacing $h=x_{j+1}-x_{j}$.
The spatial derivatives of the stress tensor components and the heat flux in $\mathbf{S}$ are evaluated with simple central discretizations of at least 6th 
order accuracy. The treatment of the radial self-gravity force $\partial_{x}\phi$ is 
described in the following section.

Since the solutions of Equations (\ref{eq:nleq}) are typically smooth structures, possibly interspersed with 
discontinuities (LO2010), a computation of the flux term is required which is highly accurate in the smooth regions while being 
able to resolve discontinuities without generating spurious oscillations. The popular Total Variation Diminishing (TVD) schemes are not suitable to distinguish 
an extremum from a discontinuity, which, in our case, would result in a loss of accuracy for large parts of the numerical solution, as we typically expect 
traveling 
periodic wave structures. 

In order to discretize the flux term we follow \citet{shu1988} and write 
\begin{equation}\label{eq:defnumflux}
 \partial_{x}\mathbf{F}\left(x_{j}\right)= \frac{\mathbf{f}_{j+1/2} -\mathbf{f}_{j-1/2}}{h},
\end{equation}
where we introduce the ``numerical flux function'' $\mathbf{f}$ and  where the subscripts $j\pm 1/2$ denote 
evaluation at the half nodes $x_{j}\pm h/2$.
Equation (\ref{eq:defnumflux}) holds exactly if $\mathbf{f}$ is defined through
\begin{equation}\label{eq:numflux}
 \mathbf{F}(x_{j})=\frac{1}{h} \int_{x_{j-1/2}}^{x_{j+1/2}} \mathbf{f}(\xi)\mathrm{d}\xi.
\end{equation}
Our conservative scheme is then formulated by approximating the interface values between two neighboring cells of the mesh, $\mathbf{f}_{j+1/2}$ and 
$\mathbf{f}_{j-1/2}$, by using relation 
(\ref{eq:numflux}) and the fact that the cell averages $\mathbf{F}(x_{j})$ are known for all $j$, since these are 
the values of the physical flux (\ref{eq:physflux}) at the nodes $x_{j}$. 

The basic procedure is as follows (e.g.\ \citet{shu2009}). One defines the primitive function of $\mathbf{f}(x)$ by
\begin{equation}
\mathbf{p_{f}}(x) = \int_{x_{0}}^{x} \mathbf{f}(\xi)\mathrm{d}\xi,
\end{equation}
with arbitrary lower limit $x_{0}$.
The function $\mathbf{p_{f}}(x)$ is then approximated by a Lagrange-polynomial of order $2r-1$ which interpolates through the $2r$ (with 
integer $r\geq 2$) data points  
$[x_{i+1/2},\mathbf{p_{f}}(x_{i+1/2})]$ with $i=j-r,j-r+1,\ldots,j+r-1$. Here
\begin{equation}
\begin{split}\label{eq:halfnodes}
 \mathbf{p_{f}}(x_{i+1/2}) & = \int_{x_{0}}^{x_{i}+h/2} f(\xi) \mathrm{d}\xi \\
 \quad & = \hat{C} +\sum_{l=j-r}^{i} \int_{x_{l}-h/2}^{x_{l}+h/2} f(\xi) \mathrm{d}\xi \\
 \quad & = \hat{C} +\sum_{l=j-r}^{i} h\,  \mathbf{F}(x_{l})
 \end{split}
\end{equation}
are the interface values of the primitive function. The constant $\hat{C}$ depends on the choice of $x_{0}$.
Differentiation of the interpolating polynomial for $\mathbf{p_{f}}(x)$ with respect to $x$ then yields a polynomial of order $2r-2$  
which can be used to obtain approximations $\mathbf{\hat{f}}_{j-1/2}$, $\mathbf{\hat{f}}_{j+1/2}$ of the exact numerical flux values in 
Equation (\ref{eq:defnumflux}). The hats indicate that these values fulfill (\ref{eq:defnumflux}) up to an error, which depends on the degree of the 
interpolating polynomial. 

In order to handle discontinuities, we utilize the MP5 algorithm (\citet{suresh1997}) which applies monotonicity preserving bounds on the interface values 
$\mathbf{\hat{f}}_{j-1/2}$, $\mathbf{\hat{f}}_{j+1/2}$ which are obtained with the method described above from a 6-point stencil ($r=3)$. This scheme is 
uniformly 5th order accurate, such that
\begin{equation*}
 \partial_{x}\mathbf{F}\left(x_{j}\right)= \frac{\mathbf{\hat{f}}_{j+1/2} -\mathbf{\hat{f}}_{j-1/2}}{h} + \mathcal{O}(h^5),
\end{equation*}
anywhere, except for discontinuities. 

Time integration is performed with a 5-stage 4th order accurate TVD Runge-Kutta method (SSPRK(5,4)) developed in \citet{ruuth2006}.
The reason for using a strong stability preserving (SSP) time discretization instead of a regular variant is that 
the computational costs remain the same while these methods can improve stability when solving hyperbolic conservation laws (\citet{gottlieb2001}).

Attempting to conduct a stability analysis of only the linearized version of above equations leads to an eigenvalue problem which cannot be solved 
analytically.
As a simple criterion for the time steps we take as a guide the time step restriction which arises for a simple one dimensional advection-diffusion problem
\begin{equation*}
 \partial_{t} u + \lambda \partial_{x} u = \nu \partial_{x}^2 u,
\end{equation*}
integrated with a first order Euler forward method (the building block of any multi-stage SSP Runge-Kutta scheme) and second order central differences 
for the discretization of the spatial derivatives.
This restriction reads
\begin{equation}\label{eq:timestep}
 \Delta t \leq   \text{min}\left(\frac{\Delta x^2}{2 \nu},\frac{2 \nu}{\lambda^2}\right) 
\end{equation}
where $\lambda$ is identified with the maximal eigenvalue of the Jacobian 
\begin{equation}\label{eq:jacobian}
\hat{A}= \frac{\partial \mathbf{F}(\mathbf{U})}{\partial  \mathbf{U}}
\end{equation}
of the actual system of equations (\ref{eq:hcl}) for the whole grid.
Since we are using higher order spatial discretizations and since we are solving a \emph{system} of equations we multiply this
time step restriction in practice by a factor $0<f_{t}<1$ , as was done in LO2010. We use $f_{t}=0.1$ in most cases.
The resulting typical time 
steps lie in 
the range of $10^{-2}-10^{-4}$ orbital periods, depending on the grid resolution, the used parameter set and the evolutionary stage of the integration. 
The scaled (Table \ref{tab:scalings}) eigenvalues of the non-isothermal Jacobian (\ref{eq:jacobian}) read
 \begin{align}
\begin{split}\label{eq:evalni}
\lambda_{(1/2)}  & = u ,\\
\lambda_{(3/4)} & = u \pm \frac{3^{(-p_{T}) }}{\sigma (-2 e + \sigma u^{2})} \left[   T_{0}\, \sigma^{(2+p_{s})} \left(\frac{2 e 
-\sigma u^{2}}{\sigma T_{0}} \right)^{(1+p_{T})} \right. \\
\quad & \left. \left(3^{(p_{T})}\, p_{s} \, (2 e -\sigma u^2 ) +2  p_{T}\, \sigma^{(p_{s})} \left(\frac{2 e 
-\sigma u^2}{\sigma T_{0}} \right)^{(p_{T})}\right) \right]^{1/2}  .
\end{split}
\end{align}
In the isothermal limit with the ideal gas relation of state, these reduce to the three eigenvalues
\begin{align}
\begin{split}\label{eq:evali}
\lambda_{1}  & = u, \\
\lambda_{(2/3)} & = u \pm 1.
\end{split}
\end{align}

The homogeneous version of Equation (\ref{eq:hcl}), i.e. the case $\mathbf{S}=0$, 
is a hyperbolic system of partial differential equations such that the Jacobian possesses a complete set of independent eigenvectors with 
only real eigenvalues [(\ref{eq:evalni}), (\ref{eq:evali})]. Its (eigen)solutions follow characteristics.
This is accounted for by a correct upwinding of the numerical solution through a splitting of the physical flux (\ref{eq:physflux}) prior to 
the reconstruction of the numerical flux $\mathbf{f}$. The splitting is performed such that
\begin{equation}\label{eq:fvs1}
 \mathbf{F}(\mathbf{U})=\mathbf{F}^{+}(\mathbf{U})+\mathbf{F}^{-}(\mathbf{U})
\end{equation}
with
\begin{equation}\label{eq:fvs2}
 \hat{A}^{+}\equiv \frac{\partial \mathbf{F}^{+}(\mathbf{U})}{\partial \mathbf{U}}\geq 0 \hspace{0.3cm} \text{and}  \hspace{0.3cm} \hat{A}^{-}\equiv 
\frac{\partial \mathbf{F}^{-}(\mathbf{U})}{\partial \mathbf{U}}\leq 0.
\end{equation}
The notation means that $\hat{A}^{+}$ has only non-negative eigenvalues whereas $\hat{A}^{-}$ has only non-positive eigenvalues. 
In order to obtain correct upwinding for a general splitting (\ref{eq:fvs1}), (\ref{eq:fvs2}), $\mathbf{f}^{+}_{j+1/2}$ and $\mathbf{f}^{-}_{j+1/2}$ are 
reconstructed from  $2r$ data points 
$[x_{i+1/2},\mathbf{F}(x_{i+1/2})]$ with $i=j-r,j-r+1,\ldots,j+r-1$ and $i=j-r+1,j-r+2,\ldots,j+r$, respectively [cf.\ Equation \ (\ref{eq:halfnodes})].
In this paper we apply the Liou-Steffen splitting (\citet{liou1993}).

All hydrodynamic integrations are performed by assuming periodic boundary conditions in a radial domain whose size we denote by $L_{x}$.
This means that each component $U_{i}$ (i=1,2,3,4) of the numerical solution vector (\ref{eq:convar}) at any time possesses the Fourier representation
\begin{equation}\label{eq:fourier}
 U_{i} =  \sum_{m=0}^{n/2-1} \hat{U}_{i}^{m} \cos \left( k_{m} \,  x + \varphi_{m} \right) 
\end{equation}
with real-valued Fourier amplitude $\hat{U}_{i}^{m}$, wavenumber $k_{m}=m \, 2\pi /L_{x}$ and phase $\varphi_{m}$ of each mode $m$.

For later use we also define the mean kinetic energy density within the computational domain as
\begin{equation}\label{eq:ekin}
 e_{kin} = \frac{1}{L_{x}} \int\limits_{[L_{x}]} \mathrm{d}x \, \frac{1}{2} \sigma \left( u^2 + v^2 \right). 
\end{equation}

\subsubsection{Implementing Radial Self-Gravity}\label{sec:hydronumsg}
The implementation of self-gravity forces in a hydrodynamic simulation is in general a difficulty of its own.
Fortunately, the wave structures we study here can be treated as purely radial to a good approximation, so that the computation of the self-gravity is greatly 
simplified.

We neglect curvature and describe the axisymmetric density pattern as a collection of straight wires of infinite azimuthal extent.
A wire at radial location $x$ has a surface mass density
\begin{equation}
 \sigma(x)=\frac{1}{h}\frac{\mathrm{d}M}{\mathrm{d}y}
\end{equation}
where $\mathrm{d}M/\mathrm{d}y=const.$ denotes the mass of the wire per unit length in $y$-direction and $h$ is its radial size.
From this follows that a cell which has a radial distance $x$ from a reference location generates a radially directed gravitional force:
\begin{equation}\label{eq:sgint}
 \Delta f^{disk}=Gh\sigma_{0} \int\limits_{-\infty}^{\infty} \mathrm{d} y \,s(x)\, \frac{x}{\left(x^2 + y^2\right)^{3/2}}=2 G h \sigma_{0} \frac{ x}{|x|^2} 
s\left(x\right)
\end{equation}
where we defined $s(x)=\sigma(x)/\sigma_{0}$.
This model applied to our scheme then yields a self-gravity force at grid point $j$:
\begin{equation}\label{eq:sgsum}
f^{disk}_{j}=-2 G h \sigma_{0} \sum_{l=1, l\neq j}^{n}  s(x_{l})\, \frac{x_{j} - x_{l}}{|x_{j} - x_{l}|^2}.
\end{equation}
As it stands, relation (\ref{eq:sgsum}) neglects the force 
generated by mass contained in the bin $j$ itself, which is given by
\begin{equation}\label{eq:cbin}
 \Delta f^{disk}(0) = 2 G \sigma_{0} \left[\partial_{x}s(0) \, h +  \mathcal{O}\left(h^3\right)\right].
\end{equation}

Evaluating Equation (\ref{eq:sgsum}) for the whole grid by direct summation would involve of the order of $n^2$ operations, where the number of grid points can 
get as large as 
$n\sim 10^4$. However, using the periodicity of $s(x_{l})$, the sum (\ref{eq:sgsum}) can be written as a convolution 
\begin{equation}\label{eq:conv}
f^{disk}_{j}= \sum_{l=-n}^{n-1}  s_{l}\, f^{kern}_{j-l}
\end{equation}
of $s_{l}=s(x_{l})$. The force kernel reads
\begin{equation}\label{eq:kernel}
f^{kern}_{j-l} = -2 G h \sigma_{0} \frac{x_{j} - x_{l}}{|x_{j} - x_{l}|^2}.
\end{equation}
Equation (\ref{eq:conv}) can be solved efficiently with a FFT method, which needs only of the order of $n \log n$ steps.
The discrete Fourier convolution theorem states that (see for example \cite{binney1987}) the Fourier transforms are related as
\begin{equation}\label{eq:convol}
\mathcal{F}_{k}\left( f^{disk}\right) = \mathcal{F}_{k}\left(s \right) \cdot \mathcal{F}_{k}\left( f^{kern}\right)
\end{equation}
and therefore the self-gravity force is obtained as a back transformation
\begin{equation}\label{eq:backconvol}
 f^{disk}_{j}=   \mathcal{F}_{j}^{-1}\left[ \mathcal{F}_{k}\left(s \right) \cdot \mathcal{F}_{k}\left( f^{kern}\right) \right].
\end{equation}
Equations (\ref{eq:convol}) and (\ref{eq:backconvol}) can be solved efficiently by using a FFT.
In our simulations we will exclusively make use of periodic boundaries, so that the FFT is directly applicable without the need to pad extended
arrays with zero's (\cite{binney1987}). 
The self-gravity force (\ref{eq:sgsum}) neglects far field contributions (with $|x_{j}-x_{l}|>L_{x}$), since the kernel (\ref{eq:kernel}) ranges over a 
limited 
region. This restriction can be overcome by adding contributions from additional neighboring replicas of the original density field to the kernel.

\subsubsection{Tests}\label{sec:hydronumtests}

Before we apply our hydrodynamical integration scheme to viscous overstability in planetary rings we perform several tests.
Here we resort to established results from linear theory, checking the accuracy of our scheme by measuring growth rates and oscillation frequencies of linear 
overstable modes (\ref{eq:linper}).
These should agree with values obtained with numerical solution of (\ref{eq:evp}).
In these test integrations the calculation of self-gravity is performed with the corrective term (\ref{eq:cbin}), as well as an extension of the self-gravity 
kernel 
(\ref{eq:kernel}) with 5 adjacent replica's on each side of the computational domain. 
For both the isothermal and the non-isothermal schemes we use the $\tau_{15}$-parameters (Table \ref{tab:hydropar}).
Surface mass densities $\sigma_{0} = 0$,  $\sigma_{0} = 350\,\text{kg}\,\text{m}^{-2}$ and $\sigma_{0} = 
600\,\text{kg}\,\text{m}^{-2}$ are adopted.

To measure linear growth rates, we use a computational domain with $L_{x}=1\,\text{km}$ and seed all modes down to about $\lambda=30\,\text{m}$ at once, with
small amplitudes. This initial state is then integrated for about 15 orbital periods.
To obtain the growth rate of a mode $m$, we perform a linear 
fit to the corresponding Fourier-amplitude $\log(\hat{U}_{i}^{m})$ [cf.\ (\ref{eq:fourier})].

Linear frequencies are obtained from short integrations (up to about 30 orbital periods), where we follow the evolution of a single seeded small amplitude 
mode in a domain of radial size $L_{x}=\lambda$. 
The oscillation frequencies are measured by analyzing the time evolution of the radial velocity field of a linear overstable mode at a fixed radial location 
with the Lomb-periodogram (\citet{press1992}) to obtain the dominant frequency.

During these integrations, we make sure that amplitudes of all other (non-seeded) modes 
remain less than about 0.1 $\%$ of the amplitude of the seeded mode. 
For convenience we use as a seed the eigensolutions of the 
isothermal hydrodynamic equations, which are available in analytical form (\citet{schmidt2001b}). 

\begin{figure}[h!]
\centering
\includegraphics[width = 0.4 \textwidth]{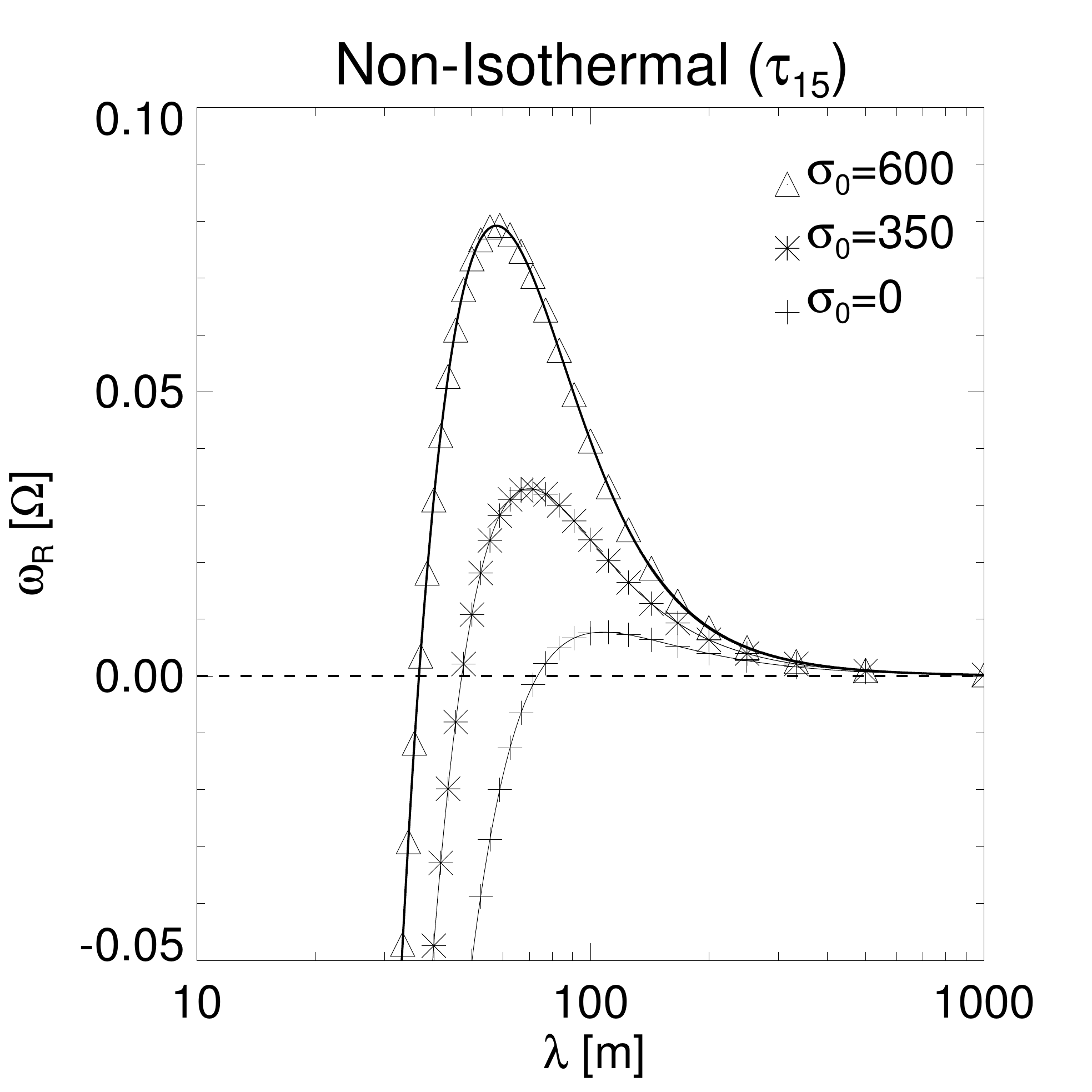}
\includegraphics[width =0.4 \textwidth]{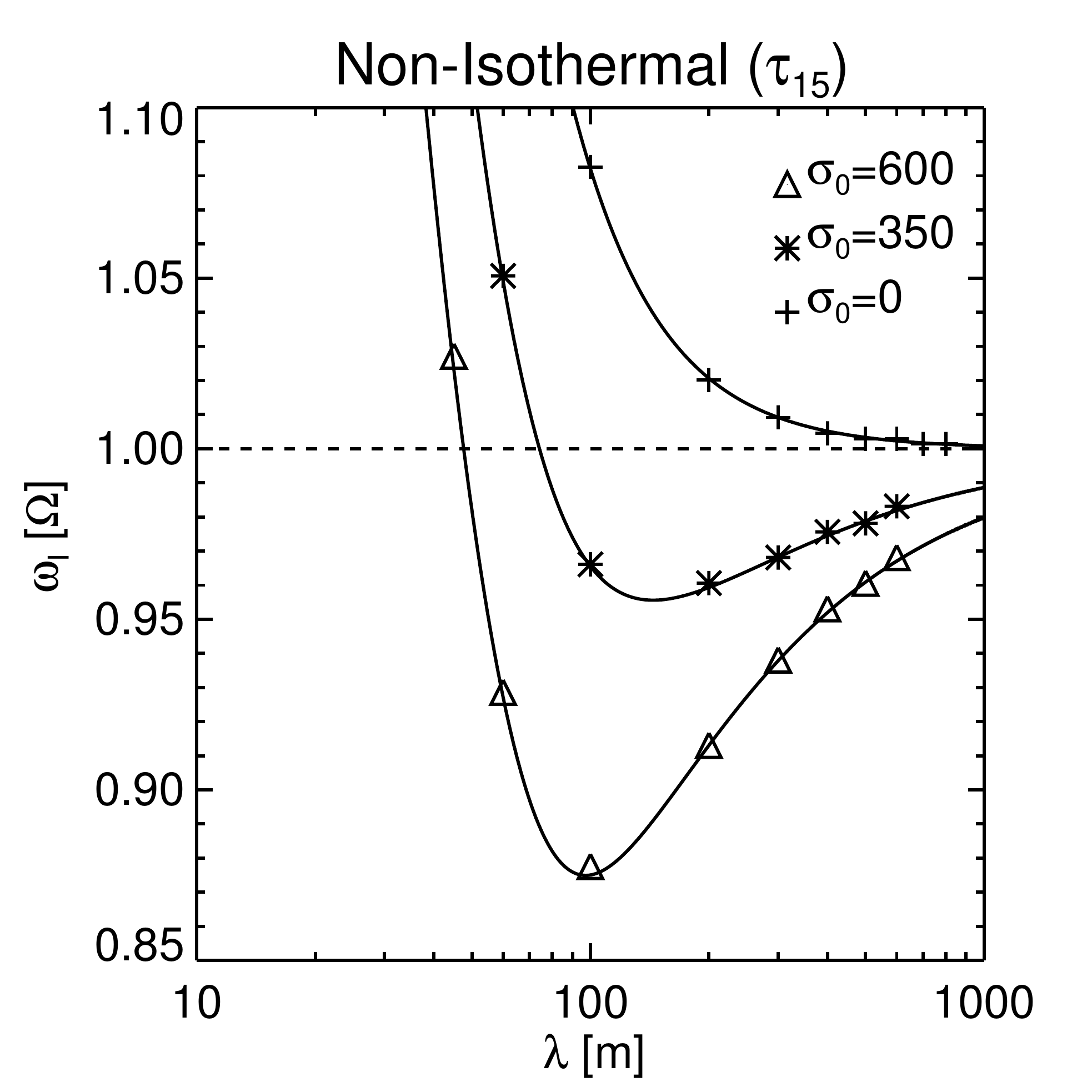}
\caption{Non-isothermal hydrodynamic linear growth rates (left panel) and oscillation frequencies (right panel) of overstable modes with the 
$\tau_{15}$-parameters. Symbols 
denote the values obtained with the flux-conservative integration scheme and the lines represent the theoretical curves obtained with numerical solution of 
(\ref{eq:evp}).}
\label{fig:grates}
\end{figure}

For brevity, we present here only the results for the non-isothermal model, the isothermal case being very similar.
Figure \ref{fig:grates} displays measured growth rates and oscillation frequencies.
With a grid resolution $h=1\,\text{m}$ the error of the computed growth rates and oscillation frequencies is less than one percent on all relevant length 
scales.
Neglecting the above mentioned corrections to the computation of self-gravity leads to mild drops of the growth rates, as well as mild enhancements of the 
oscillation 
frequencies, however, being less than 5 percent on all relevant length scales.
We do not include these corrections in our integrations of the nonlinear evolution of overstability, because they do not affect the outcome in a 
significant manner.
The neglect of the finite bin size correction can be interpreted as a smoothing of structures on very short length scales.
Especially the kernel extension is not necessary since the box sizes are in all integrations presented in the following sections chosen large enough to 
comprise many 
wavelengths 
of each relevant mode.

\FloatBarrier

\clearpage

\subsection{N-Body Simulations}\label{sec:nbsim}

We adopt the local simulation method (\citet{wisdom1988}) which was used by \citet{salo2001} (see also \citet{salo1992a} and \citet{salo1995}).
Thus, we simulate $N_{p}$ particles contained in a small rectangular region, co-moving with the mean Keplerian angular frequency $\Omega=\sqrt{G M_{S}/r^3}$ at 
distance $r=10^5 \, \text{km}$ from the planet. The simulation region has dimensions $L_{x}\times L_{y}$ in a cartesian coordinate system $[x,y,z]$ where the 
$x$-axis points radially outward 
and the $y$-axis is directed along the orbital motion.
In radial and azimuthal direction we apply periodic boundary conditions. Particles crossing the radial box boundary re-enter with appropriately modified
velocities to 
account for the shear.

Particles are identical, smooth, spin-less spheres, with radius $R_{p}=1\, \text{m}$. 
Furthermore, to describe collisional energy loss in simulations,  we use either a constant normal coefficient of restitution $\epsilon=0.5$, or the 
\citet{bridges1984} collision law
\begin{equation}\label{eq:bridges}
 \epsilon_{b}(v_{n}) = \left(\frac{v_{n}}{v_{c}}\right)^{-0.234}
\end{equation}
with $v_{n}$ being the normal component of the relative velocity vector of two impacting particles and the scale parameter $v_{c}=0.077 
\,\text{mm}\,\text{s}^{-1}$. This relation was used in the simulations described in \citet{salo2001}.
The constant $\epsilon=0.5$ leads to a system which mimicks very well the cool, flattened ring state resulting from $\epsilon_{b}(v_{n})$ and for our purposes 
the two cases yield practically the same results.
Particle collisions are modeled with the same visco-elastic impact model that was originally devised by  \citet{dilley1993} to parameterize measurements
of elasticity (see \citet{salo1995}).

The motion of each particle is described with the Hill-equations
\begin{align}\label{eq:hill}
\begin{split}
 \ddot x - 2\Omega \dot y + \left(\kappa^2 - 4 \Omega^2\right) x & = F_{x} + F^{g}(x) \\
 \ddot y + 2\Omega \dot x & = F_{y}\\
 \ddot z + \Omega_{z}^2 z & = F_{z}
 \end{split}
\end{align}
where $\kappa$ is the epicyclic frequency which equals $\Omega$ in this study, because we neglect effects from the oblateness of the planet. Furthermore, 
$F_{x}$, $F_{y}$, $F_{z}$ describe forces per unit mass due to 
particle impacts and $F^{g}(x)$ denotes the radial collective self-gravity force per unit mass, which we discuss in the next section. Note that the 
vertical frequency $\Omega_{z}$ is enhanced compared to $\Omega$, mimicking the effect of vertical self-gravity (\citet{wisdom1988}). Parameters used in the 
N-body simulations are given in Table \ref{tab:nbstat}.

\vspace{ 0.5 cm}
\begin{minipage}{\textwidth}
\centering
\captionof{table}{\textbf{N-Body Simulation Parameters of Large-Scale Runs}}\label{tab:nbstat} 

   \renewcommand*{\arraystretch}{1.2}
    \begin{tabular}{| p{4cm}| p{2.5cm}| p{2.5cm}|  p{2.5cm}| p{2.5cm}| }
 \hline
    $\tau$  & \multicolumn{1}{|c|}{$1.5$}  & \multicolumn{1}{|c|}{$1.5$} & \multicolumn{1}{|c|}{$1.5$} & \multicolumn{1}{|c|}{$2$} \\     \hline 
    $\Omega_{z} \, [\Omega]$ & \multicolumn{1}{|c|}{3.6} & \multicolumn{1}{|c|}{3.6}  & \multicolumn{1}{|c|}{2} & \multicolumn{1}{|c|}{3.6} \\     \hline  
    $\epsilon$ & \multicolumn{1}{|c|}{$\epsilon_{b}$}  & \multicolumn{1}{|c|}{0.5} & \multicolumn{1}{|c|}{$\epsilon_{b}$} & \multicolumn{1}{|c|}{0.5} \\     
\hline  
  $L_{x}$ & \multicolumn{1}{|c|}{$2\,\text{km}$}  &  \multicolumn{1}{|c|}{$5\,\text{km}$} & \multicolumn{1}{|c|}{$2\,\text{km}$} & 
\multicolumn{1}{|c|}{$5\,\text{km}$} \\     \hline 
    $L_{y}$ & \multicolumn{1}{|c|}{$10\,\text{m}$} & \multicolumn{1}{|c|}{$31.41\,\text{m}$} & \multicolumn{1}{|c|}{$10\,\text{m}$} & 
\multicolumn{1}{|c|}{$25.14\,\text{m}$}\\     \hline  
  $N_{p}$ & \multicolumn{1}{|c|}{9,550} & \multicolumn{1}{|c|}{75,000} & \multicolumn{1}{|c|}{9,550} & \multicolumn{1}{|c|}{80,000}     \\     \hline   
 $R_{p}$ & \multicolumn{4}{|c|}{1m} \\     \hline  
           $\sigma_{0}$ & \multicolumn{4}{|c|}{\textit{free parameter}} \\     \hline  

    \end{tabular}\par
\end{minipage}

\subsubsection{Treatment of Self-Gravity in N-Body Simulations}\label{sec:nbsg}
It is well known from theoretical treatments and from simulations that the vertical component of self-gravity leads to a flattening of the ring, thereby 
increasing the collision frequency of the ring particles.
The high collision frequency in principle promotes viscous overstability, as it increases the relative contribution of nonlocal momentum transport,
resulting in an effective shear viscosity which increases steeply with increasing optical depth. 

The effects of the planar components of self-gravity on
instability mechanisms such as the viscous overstability are less well understood.
According to kinetic treatments (e.g.\ \citet{shu1985c}) and simulations (e.g.\ \citet{salo1995}), self-gravitational encounters contribute to 
the local 
viscosity of the system, transferring energy from systematic motion to random motions. This is efficient if the velocity dispersion is smaller than the 
mutual two-body escape 
speed $v_{esc}=\sqrt{2 G m_{0}/R_{p}}$ of ring particles with mass $m_{0}$. Therefore, $v_{esc}$ becomes a lower limit for the velocity dispersion in 
self-gravitating 
particulate systems.
Theoretically, in a dilute inviscid disk, the planar self-gravity can lead to local instability of axisymmetric modes if the radial component of the velocity 
dispersion fulfills $c_{r}<c_{crit}=3.36 G \sigma/ \kappa$ (\citet{Toomre1964}). 
In terms of the Toomre-parameter this threshold reads $Q<1$.
Nevertheless, in realistic self-gravitating simulations of Saturn's dense rings it is found (\citet{salo1992a,salo1995}) that $Q$ often adjusts to values 
around 2. The system 
is found to be no 
longer uniform and self-gravity wakes form and dissolve on orbital timescales. These non-axisymmetric structures contribute to the angular momentum 
transport through gravitational torques as well as through their systematic motion. It turns out that in the presence of wakes the 
nonlocal viscosity becomes unimportant, compared to the roughly equal contributions from the (strongly enhanced) local and gravitational viscosity 
(\citet{daisaka2001}). The wakes 
heat up the particle system, establishing a steady state Toomre-parameter above 1, 
depending on the precise particle properties, like $\epsilon$, internal density and particle radius.

In this study we neglect the direct gravitational interactions during particle encounters. 
Consequently, the simulated ring states lack the related heating processes, which results in significantly lower (ground state) velocity dispersions.
But most importantly, since wake structures do not appear, the ground state is homogeneous and we can use the transport coefficients determined 
by \citet{salo2001} when comparing to our hydrodynamic scheme (section \ref{sec:hydro}).

In the N-body simulations presented in this paper two aspects of self-gravity are taken into account.
First, vertical self-gravity is approximated by an artificially increased frequency of vertical oscillations in the Hill-equations of motion 
(\ref{eq:hill}), resulting in the effects described above. 
We adopt for most simulations the factor $\Omega_{z}=3.6$, which was originally introduced by 
\citet{wisdom1988}, and which was later also used by \citet{latter2013}. Furthermore, the transport coefficients used in the hydrodynamical model in LO2009 and 
LO2010 were obtained from N-Body simulations with the same $\Omega_{z}=3.6$ (\citet{salo2001}).
In this approximation the vertical self-gravity is assumed to be generated by a homogeneous slab of material with vertical thickness $H$ and with 
($z$-independent) volume density $\rho=\sigma_{0}/H$ so that vertical integration of the Poisson equation results in
\begin{equation}
 \partial_{z}\phi(z) = \frac{4 \pi G \sigma_{0} }{H}z,  
\end{equation}
where $|z|\leq H/2$, i.e. within the homogeneous slab.
Combined with the planet's vertical force this results in a total vertical force
\begin{equation}\label{eq:wtfac}
 F_{z}(z)=-\left(1 + \frac{4 \pi G \sigma_{0} }{\Omega^2\, H}\right)\Omega^2 \,z\equiv -\Omega_{z}^2  \, z,
\end{equation}
defining thereby the (scaled) effective vertical frequency $\Omega_{z}$. Note that the chosen value $\Omega_{z}/\Omega=3.6$ is larger than 
the vertical enhancement in rings. It is chosen mainly to enable a direct comparison with the aforementioned studies.

Moreover, following \citet{salo2009}, we model the radial component of self-gravity in a manner which is similar to the method used in our 
hydrodynamic scheme (section \ref{sec:hydronumsg}).
The force calculation is based on a radial Fourier-decomposition of the tangentially averaged surface density
\begin{equation}\label{eq:fdecomp}
 \sigma(x)=\sigma_{0}\, \left[1 + \sum_{m=1}^{m_{max}} A_{m} \cos \left( k_{m} \,  x + \varphi_{m} \right) \right]
\end{equation}
with wavenumbers $k_{m}=m \, 2 \pi /L_{x}$ and where $A_{m}$ and $\varphi_{m}$ denote the amplitude and phase of the corresponding Fourier mode.
The cutoff $m_{max}$ is to be chosen sufficiently high (typically a few hundreds) in order to avoid aliasing effects.
Each of the $m_{max}$ plane waves in (\ref{eq:fdecomp}) contributes to the radial self-gravity potential through relation (\ref{eq:wkbsg}), i.e.
\begin{equation}
 \phi_{m}\left(x\right) =  \frac{2 \pi G \sigma_{0}}{k_{m}}\,   A_{m} \cos \left( k_{m} \, x + \varphi_{m} \right) .
\end{equation}
The total radial self-gravity force per unit mass then reads
\begin{equation}\label{eq:sgmodes}
 F^{g}(x) = - 2 \pi G \sigma_{0} \, \sum_{m=1}^{m_{max}}  A_{m} \sin \left( k_{m} \, x + \varphi_{m} \right) .
\end{equation}

One notes that the ground state surface density $\sigma_{0}$ is now a free model parameter, as the simulated particles are otherwise massless. 
The tangential component of self-gravity is not considered since we assume that the ring retains azimuthal symmetry.
Both, this method and the self-gravity implementation applied in the hydrodynamic scheme (Section \ref{sec:hydronumsg}) neglect curvature, consistent with the 
hydrodynamic model presented in Section \ref{sec:theo}. From a theoretical point of view the mode calculation (\ref{eq:sgmodes}) is more accurate than the
straight wire model (\ref{eq:sgsum}) since it automatically assumes infinite extent of waves, whereas the kernel (\ref{eq:kernel}) ranges 
over a limited region, thus neglecting far distance contributions. As stated in Section \ref{sec:hydronumsg} this restriction can in principle be overcome 
by adding 
contributions from additional neighboring replicas of the original density field to the kernel.

\subsubsection{Growth Rates and Oscillation Frequencies of Overstable Modes in the Linear Regime}\label{sec:nbtest}

In the determination of linear growth rates from N-body simulations we seed one single mode with a small initial amplitude and a wavelength $L_x/m$ ($m$ is the 
mode number). Only this mode is taken into 
account in the calculation of radial self-gravity. As in the hydrodynamic measurements (Section \ref{sec:hydronumtests}) the growth rate is computed from a 
linear fit to the time evolution of the corresponding 
Fourier amplitude, while
the oscillation frequency is obtained with the Lomb normalized periodogram.
Care is taken to use a time interval during which the oscillation amplitude remains small, typically about 20 
orbital periods. The box size used in these simulations is $L_x=1\,\text{km}$, and the measured modes cover $m=1 -25$ (down to 40 meters).

\begin{figure}[h!]
\centering
\includegraphics[width = 0.75 \textwidth]{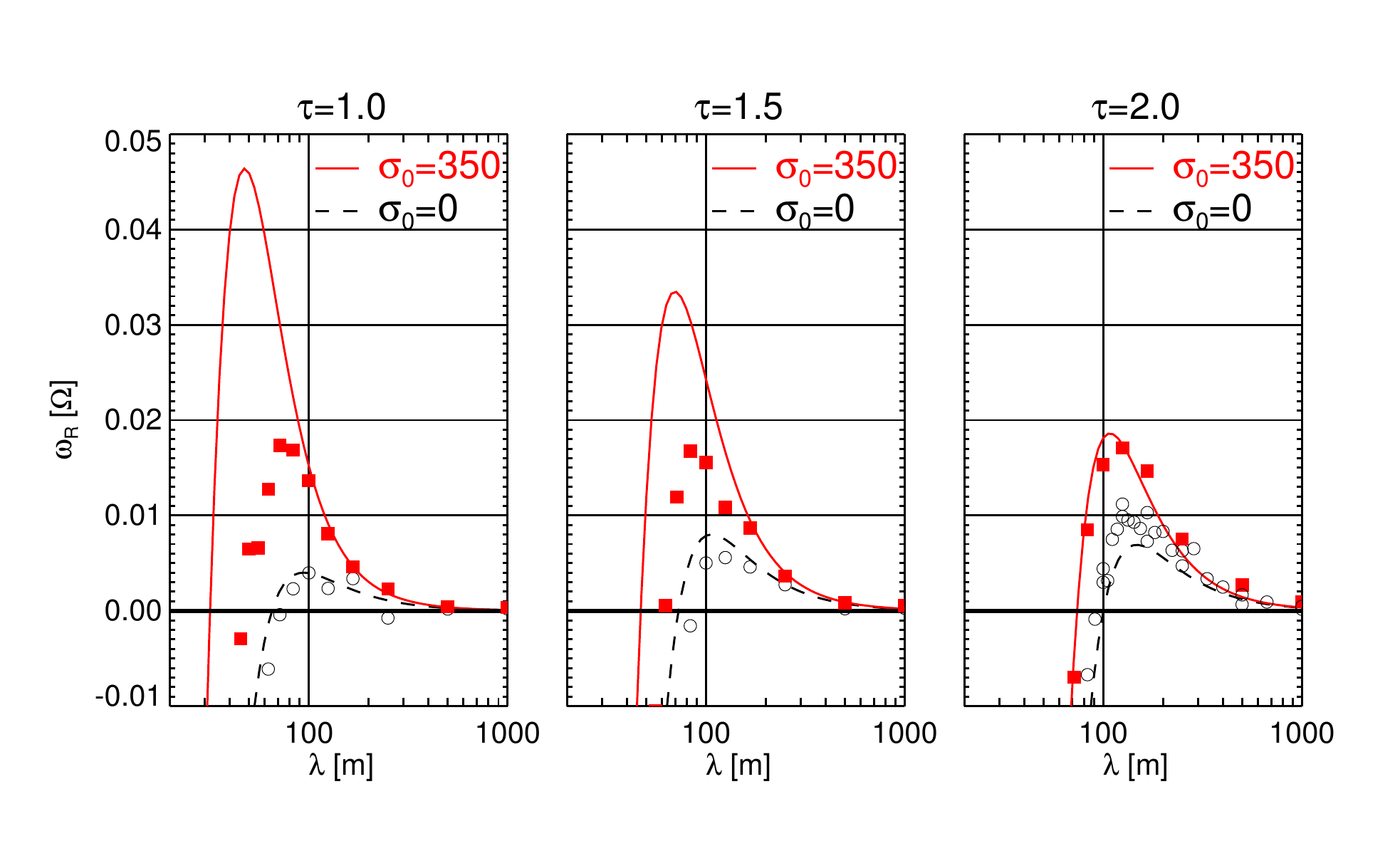}
\caption{Growth rates of linear overstable modes from N-body simulations with different optical depths for $\Omega_{z}=3.6$. Filled squares 
correspond to $\sigma_{0}=350\,\text{kg}\, \text{m}^{-2}$ and open circles to $\sigma_{0}=0$.
The curves represent the corresponding hydrodynamic predictions for which we used the parameter sets listed in Table \ref{tab:hydropar}.}
\label{fig:nblin1}
\end{figure}

Figure \ref{fig:nblin1} shows the linear growth rates in simulations with different optical depths $\tau$, along with theoretical curves resulting from the 
non-isothermal model.
We find that for $\sigma_{0}=0$ the growth rates of the hydrodynamic model and the N-Body simulations match reasonably well for all optical 
depths (cf.\ 
\citet{schmidt2001b}). With the higher optical depth $\tau=2$ we obtain good agreement also for the moderate surface density $\sigma_{0}=350\, 
\text{kg}\text{m}^{-2}$.
For smaller optical depths though, there develops discrepancy with increasing $\sigma_{0}$.
From Figure \ref{fig:nblin1} follows that, contrary to the hydrodynamic prediction, decreasing the optical depth from $\tau=2$ towards $\tau=1$ does not 
lead to an overall increase of the growth rates, but merely produces a shift toward shorter wavelengths.

\begin{figure}[h!]
\centering
\includegraphics[width = 0.75 \textwidth]{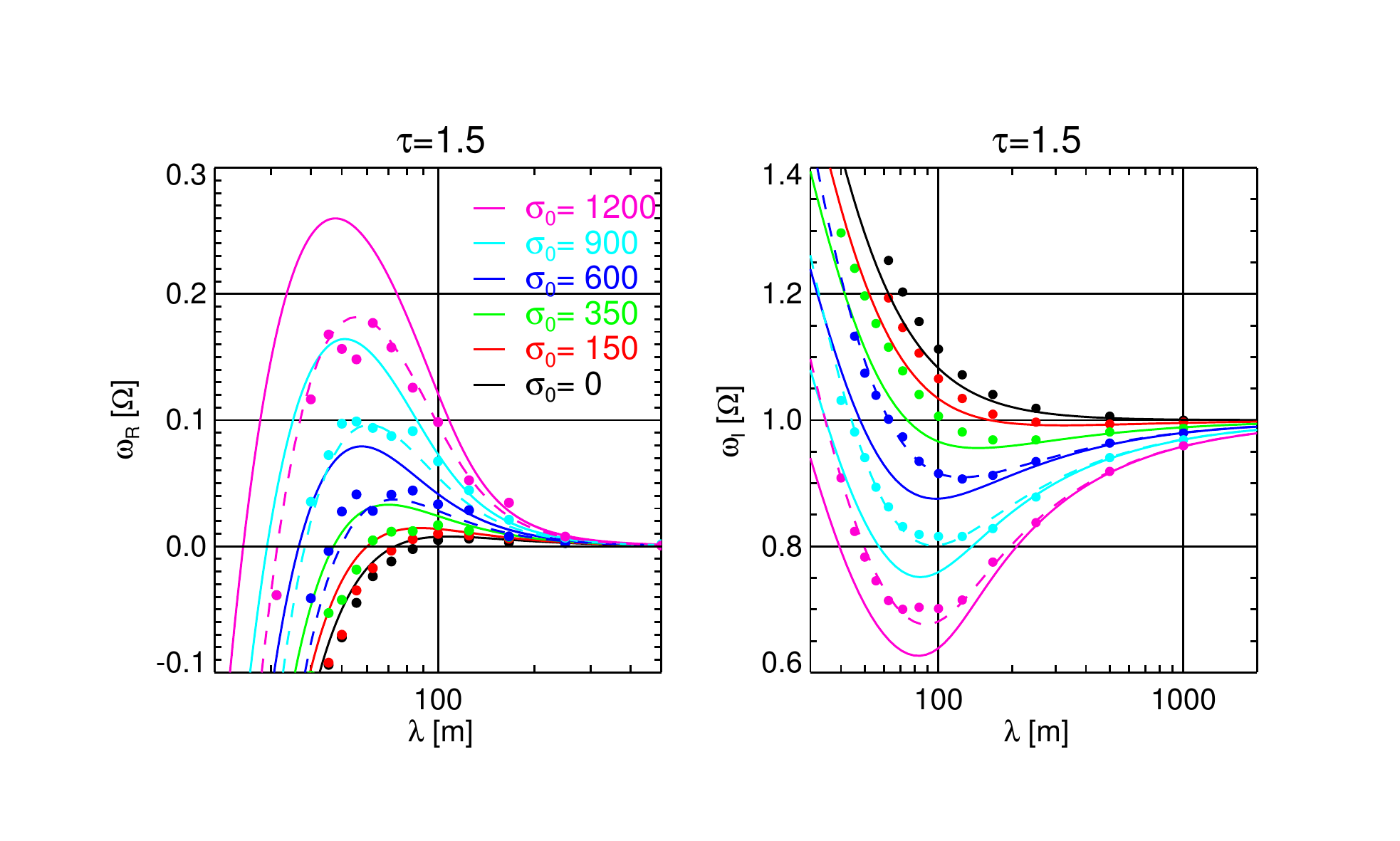}
\caption{Linear growth rates (left) and oscillation frequencies (right) from N-body simulations with optical depth $\tau=1.5$ for various surface densities. 
Non-isothermal hydrodynamic model curves (solid) for the same surface densities $\sigma_{0}$ (units $\text{kg}\,\text{m}^{-2}$) are shown in the 
same color. The dashed curves result from the hydrodynamic model with an increased value of $p_{s}$ by $40\%$.}
\label{fig:nblin2}
\end{figure}

In Figure \ref{fig:nblin2} we present results for growth rates and oscillation frequencies for fixed optical depth $\tau=1.5$ and $\Omega_{z}=3.6$ but varying 
surface density $\sigma_{0}$.
The solid curves represent the non-isothermal model, computed from Equation (\ref{eq:evp}). 
While the hydrodynamic model overestimates the growth rates, it tends to underestimate the oscillation frequency. 
Overall, it provides a good match for modes of larger wavelength.

The deviations might to some extent arise from the vertical dynamics 
of the simulated particulate disk. Namely, vertical expansions of the disk will affect the isotropic 
pressure and the transport coefficients on the orbital timescale and this interferes with the growth of 
overstable modes. For the self-gravitating runs we observe a related vertical splashing of ring particles in the compressed phases of the oscillations (cf.\ 
Figure \ref{fig:fluctnb} and Figure 1 in \citet{salo2001}) already during the linear growth phase. Splashing occurs since the ring flow is nearly 
incompressible (\citet{borderies1985}).

The dashed curves in Figure \ref{fig:nblin2} are non-isothermal model curves computed with a pressure coefficient $p_{s}$ [cf.\ Equation 
(\ref{eq:pres})] that was increased by 40 percent from its nominal value (Table \ref{tab:hydropar}). 
For clarity, we plot modified curves in both panels only for the three largest values of $\sigma_{0}$. This modification of a single parameter 
leads overall to a considerably better agreement with the results from N-body simulations in the linear regime. 
However, significant deviations remain in the nonlinear regime.
This will be further discussed in Section \ref{sec:nldisp} and Appendix \ref{sec:pmod}.

\FloatBarrier

\clearpage

\section{Results}\label{sec:results}

We begin by presenting results of the hydrodynamic model in the limit of vanishing self-gravity in Section \ref{sec:hydronosg}. 
Here we distinguish between the isothermal and non-isothermal cases. We perform a qualitative comparison to the hydrodynamic results of LO2010 
and to 
the non-gravitating N-Body simulations of RL2013. In section \ref{sec:hydrosg} we present the hydrodynamic model with radial self-gravity. 
We restrict the description mainly to integrations with the vertical frequency $\Omega_{z}=3.6$ to facilitate the comparison with
the aforementioned work. Our integrations with $\Omega_{z}=2$ behave qualitatively similar. The Section \ref{sec:nbres} is devoted to the results of 
our gravitating N-Body simulations and comparison to the hydrodynamic model.

\subsection{Hydrodynamical Integrations Without Self-Gravity}\label{sec:hydronosg}

\subsubsection{Isothermal Model}\label{sec:isonosg}
Our isothermal model without self-gravity produces results very similar to those presented in LO2010.

\begin{figure}[h!]
\centering
\includegraphics[width = 0.4 \textwidth]{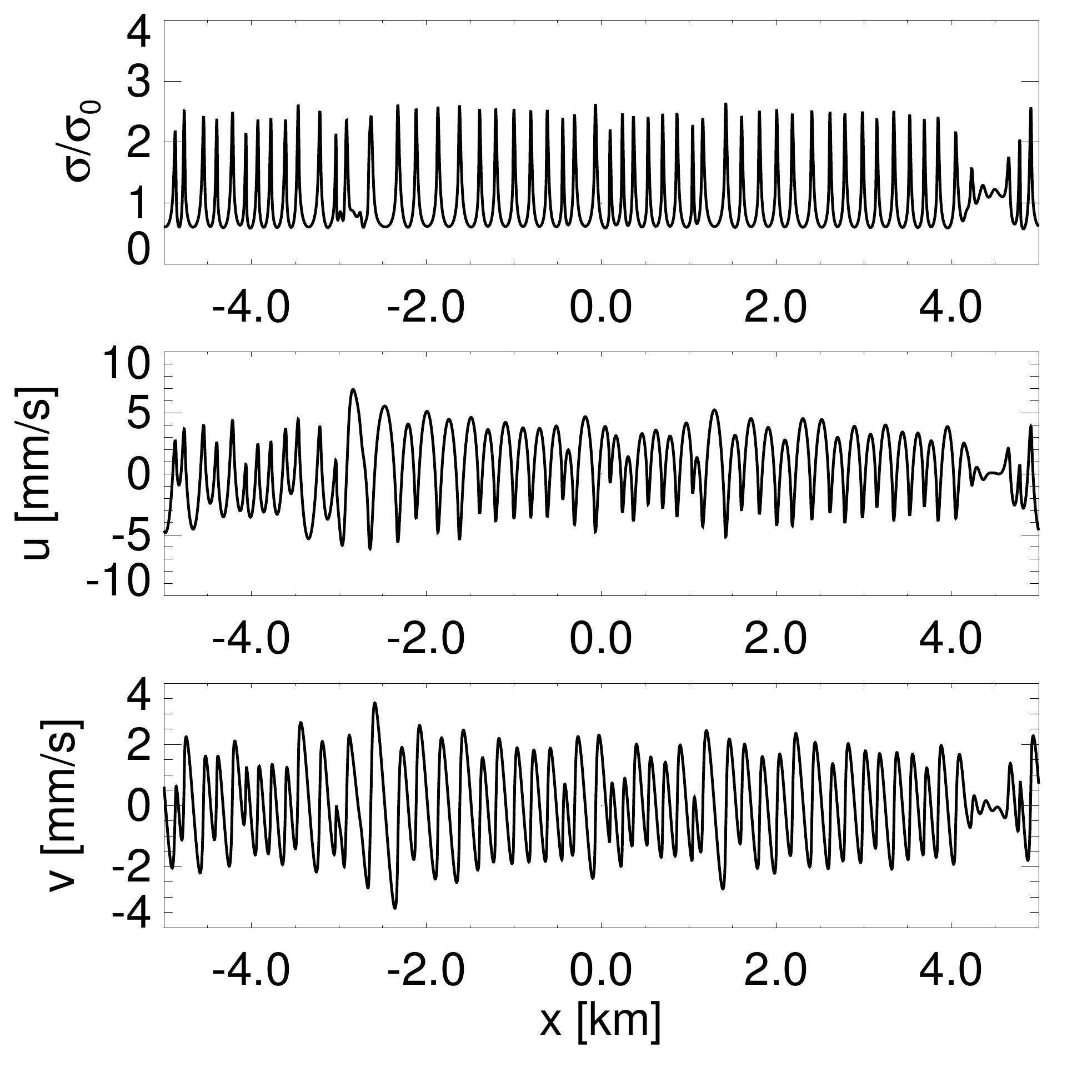}
\includegraphics[width = 0.4 \textwidth]{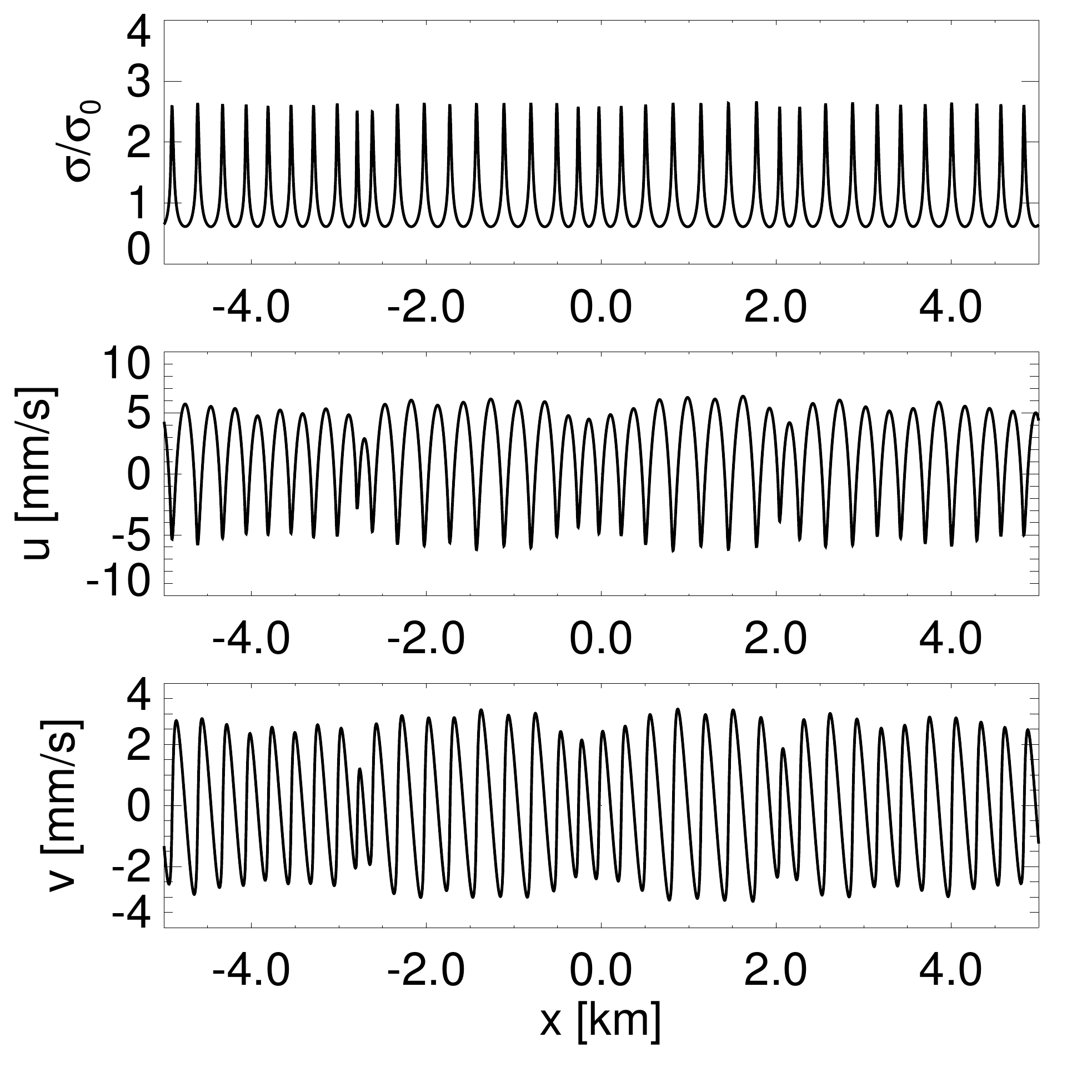}
\caption{Hydrodynamic fields characterizing intermediate (t=4,000 ORB, left panel) and advanced states (t=20,000 ORB, right panel) of an isothermal integration 
with 
the $\tau_{15}$-parameters. Radial self-gravity is not included ($\sigma_{0}=0$).}
\label{fig:tau15ngb}
\end{figure}

Figure \ref{fig:tau15ngb} shows snapshots of the hydrodynamic field quantities during two different stages of nonlinear evolution
with the $\tau_{15}$-parameters. The seed for this integration is spectral white noise consisting of both left and right traveling, small amplitude waves.
The left panel represents an intermediate state of the evolution in which the initial perturbations have already attained substantial amplitudes and different 
modes begin to interact
with each other. The 
plots reveal the presence of a source/sink pair for traveling waves in the system. The source coincides with the density depletion near 
$x=4.5\,\text{km}$. At this location also the velocity amplitudes are small.
The corresponding sink is less easy to locate. It reveals itself through a reversal in the shape of the radial velocity profile $u$ (across 
$x=-3\,\text{km}$), indicating that the two nonlinear traveling waves collide at this point.
In the advanced wave state, displayed in the right panel, these structures have disappeared, leaving a unidirectional wave train which fills out the entire 
domain and which undergoes small amplitude and phase fluctuations.

\begin{figure}[h!]
\centering
\includegraphics[width = 0.4 \textwidth]{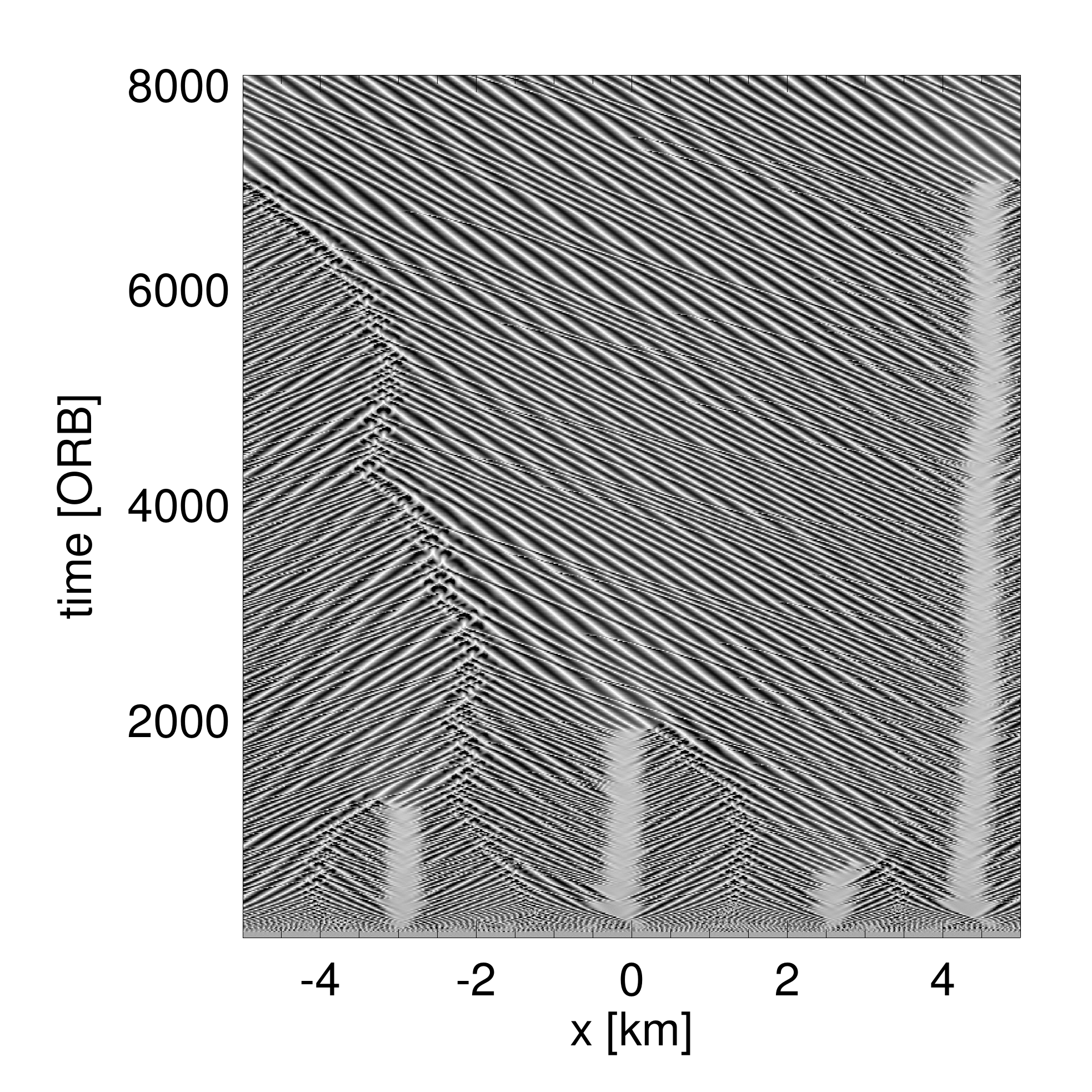}
\includegraphics[width = 0.4 \textwidth]{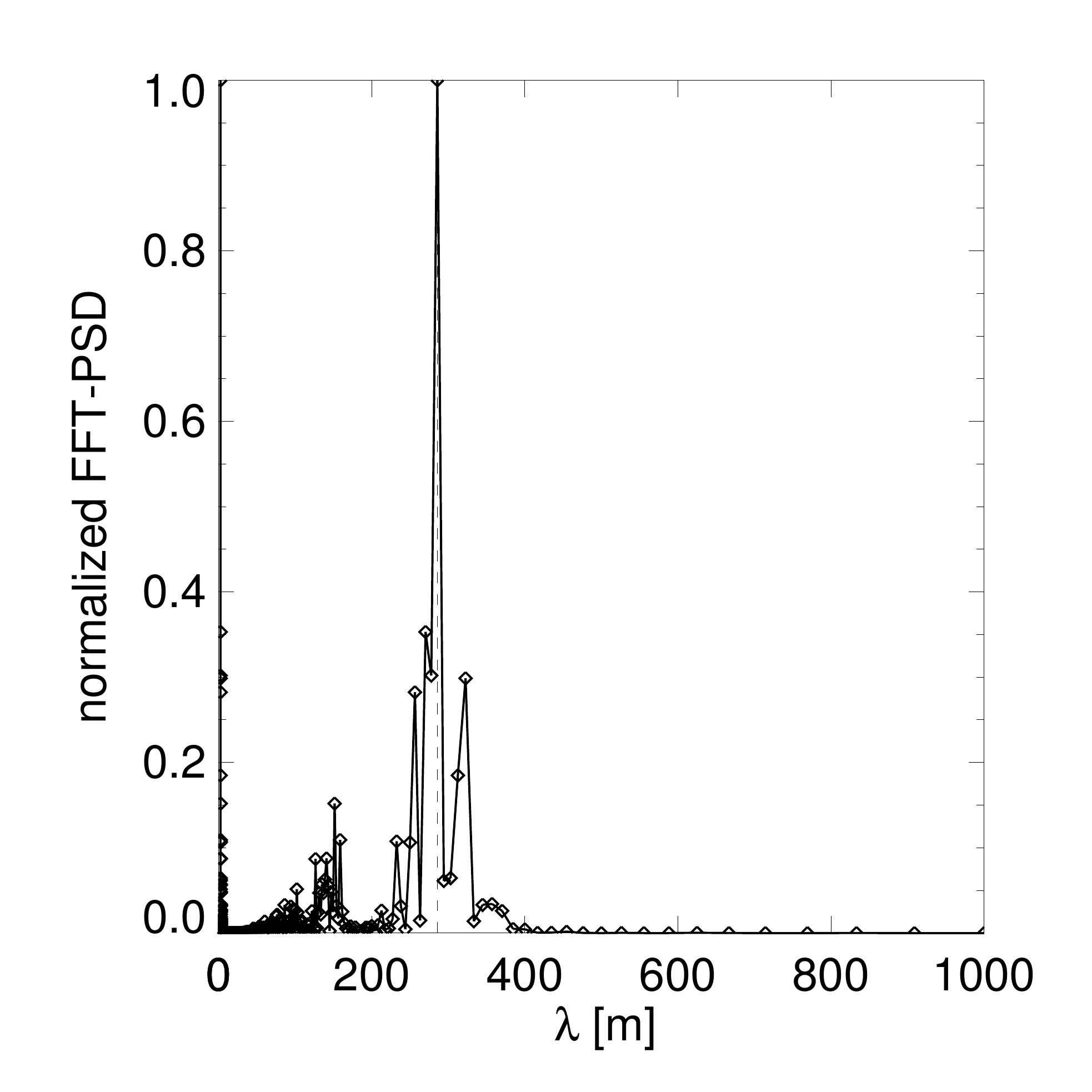}
\caption{Stroboscopic space-time diagram (left panel) and final normalized power spectral density (PSD) of the surface mass density field (right panel) for the 
same 
integration as in Figure \ref{fig:tau15ngb} with the $\tau_{15}$-parameters.}
\label{fig:tau15nga}
\end{figure}

Figure \ref{fig:tau15nga} shows for the same integration a stroboscopic space-time 
diagram, as well as the final power spectrum of the surface mass density field.
In the space-time diagram lines of constant gray-shading indicate lines of constant phase of the wave structures.
The term stroboscopic means that the diagram is plotted with a sampling rate of $1/\text{orbit}$ (see Appendix \ref{sec:strob}). 
Source and sink structures are clearly visible in this diagram. The sources are the gray stripes which remain at nearly fixed locations, showing only small 
radial excursions. These are interconnected by the (less pronounced) sinks. We observe initially four source/sink pairs. The sources emit a complicated 
sequence of phase modulations, which are expected to travel 
with the corresponding group velocity for these wavelengths. It is also seen that the sinks wander in a stochastic manner towards the sources, resulting 
eventually in an annihilation of the two. One pair survives for about 7,000 orbits.

Following \citet{vanhecke1999}, sources are active structures which send out waves, while sinks are locations where the waves meet and disappear.
The distinction between sources and sinks in a space-time diagram is to be made according to the sign of the group velocity of the adjacent wave 
patches. The group velocity points away from sources and towards sinks.
In the usual definition, sources and sinks are coherent structures which can appear in solutions of the complex coupled Ginzburg-Landau (CGL) equations.
As outlined for instance in \citet{vanhecke1999}, this applies to systems which undergo a supercritical Hopf-bifurcation from a homogeneous ground 
state into a 
traveling wave state (such as the viscously overstable fluid disk investigated here), where the interaction between counter-propagating waves is 
large enough so that these can suppress each other. Then the system can develop unidirectional wave patches, separated by (stable) sinks and sources. 
In some parameter regimes of the CGL equations, these structures can exhibit highly complex dynamics.

\begin{figure}[h!]
\centering
\includegraphics[width = 0.4 \textwidth]{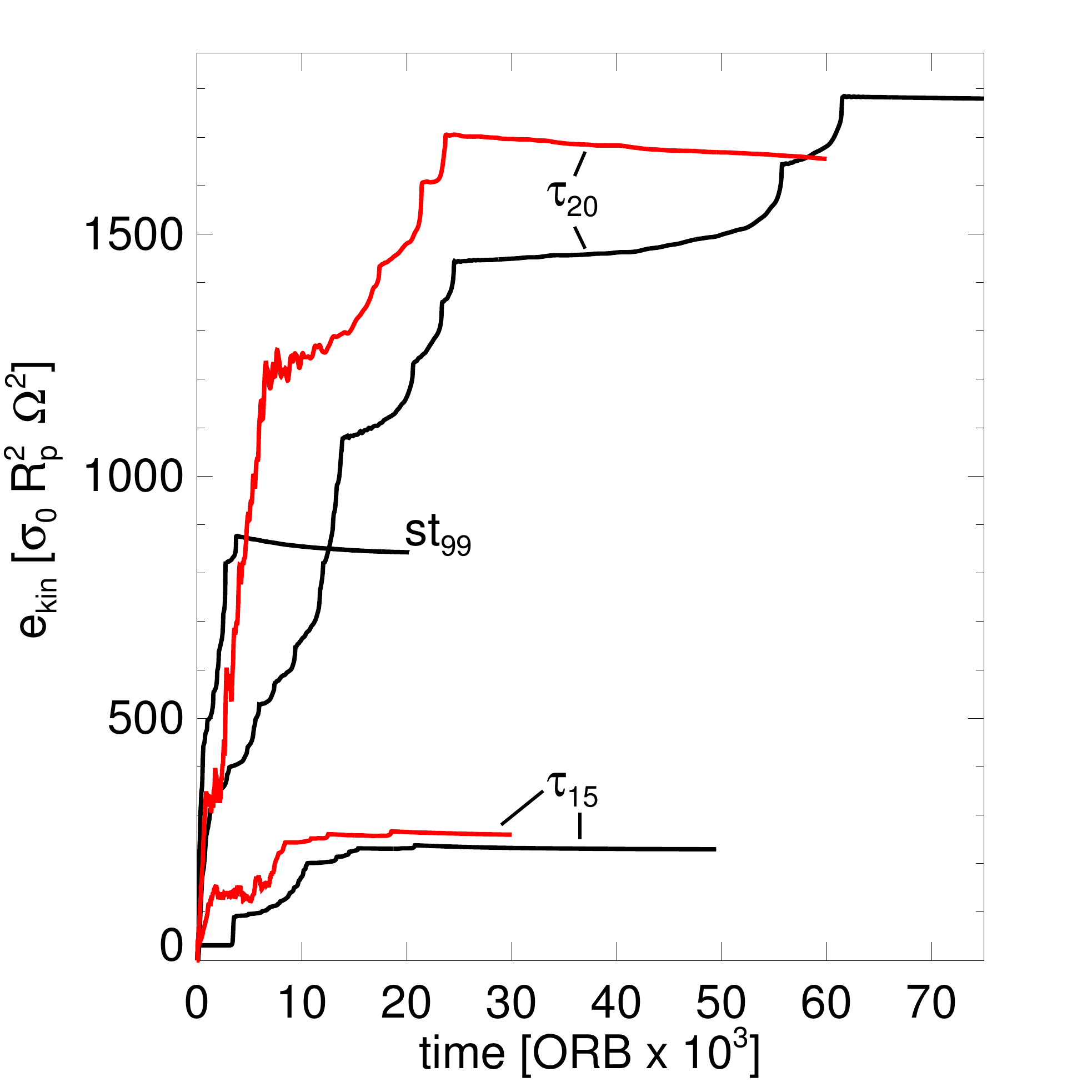}
\includegraphics[width = 0.4 \textwidth]{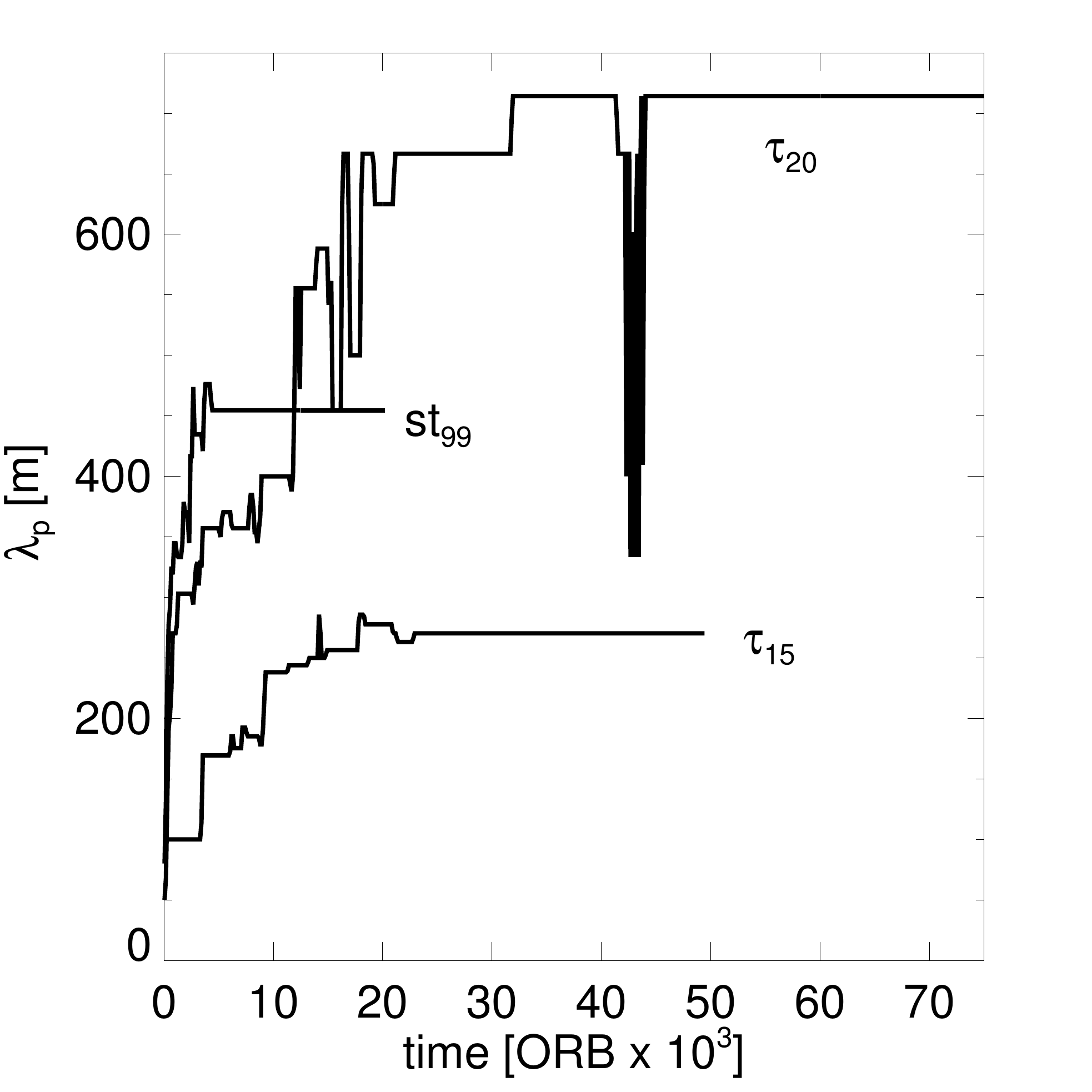}
\caption{\emph{Left}: Evolution of kinetic energy densities of hydrodynamic integrations in the isothermal model without self-gravity with the $\tau_{15}$, the 
$\tau_{20}$, and the $st99$-parameters (see Table \ref{tab:hydropar}). The initial states corresponding to the red curves are small amplitude spectral white 
noise, those corresponding to the black curves consist of a small amplitude wave with $\lambda=100\,\text{m}$. The red curve labeled $\tau_{15}$ is from the 
same 
integration as described in Figures \ref{fig:tau15ngb} and \ref{fig:tau15nga}.
\emph{Right}: Evolution of the prevalent wavelength corresponding to the three black curves in the left panel. The final values of $\lambda_{p}$ are in good 
agreement with the values $\lambda_{st}=233\,\text{m},455\,\text{m},\,659\,\text{m}$, computed for the same parameters by LO2009. In all integrations we used 
$L_{x}=10\, \text{km}$ with $h=2.5\, \text{m}$. Note that for the units of $e_{kin}$ we adopt $R_{p}=1\,\text{m}$ to compare with our N-Body simulations in 
Section \ref{sec:nbres}.}
\label{fig:ekinisonosg}
\end{figure}

Once all sources and sinks have disappeared, the perturbations eventually develop into a single traveling wave mode which 
subsequently increases its wavelength through a so called staircase process (LO2010). 
The different stages of development are visible in the evolution of the kinetic 
energy density $e_{kin}$ (\ref{eq:ekin}), presented in Figure \ref{fig:ekinisonosg} (left panel: lower red curve marked $\tau_{15}$).
This plot also shows the evolution of $e_{kin}$ of an initial white noise state with the $\tau_{20}$-parameters (upper red curve marked $\tau_{20}$).
The remaining three black curves correspond to integrations of initial states consisting of a single mode ($\lambda=100\,\text{m}$) with the $\tau_{15}$, the 
$\tau_{20}$, as well as the $st99$-parameters.

The kinetic energy densities describing the integrations from white noise (the red curves) exhibit fluctuations during the intermediate stage of the evolution.
These are caused by the presence of the source/sink structures (Figure \ref{fig:tau15nga}, left panel), since the nonlinear waves
connecting these structures undergo wavelength and amplitude fluctuation.
In contrast, the systems which evolve from a $\lambda=100\,\text{m}$ mode do not exhibit source/sink pairs. This explains the lack of fluctuations 
in their kinetic energy curves (the three black curves).

Figure \ref{fig:ekinisonosg} (right panel) shows for these three integrations starting from the $\lambda=100\,\text{m}$ mode the evolution of the prevalent 
wavelength $\lambda_{p}$. This wavelength corresponds to the maximum of the power spectral density (cf.\ Figure \ref{fig:tau15nga}, right panel), as a function 
of time. Nonlinear self-interactions of the wave train on the average result in a growth of $\lambda_{p}$, accompanied by strong fluctuations, until it 
eventually settles on a constant value.
The final values of $\lambda_{p}$ are in good agreement with the wavelengths $\lambda_{st}$ found by LO2009, who have shown that all nonlinear wave trains with 
a wavelength larger than $\lambda_{st}$, are stable with respect to 
perturbations, while those with $\lambda<\lambda_{st}$ are not. 
This picture explains the observed growth of 
$\lambda_{p}$ towards these critical values, induced by small perturbations of the wave trains.
It is, however, in principle possible that the system settles on a considerably larger wavelength or even on a set of multiple wavelengths, depending on
the precise initial conditions. Values for $\lambda_{st}$, determined by LO2009 are given in the caption of Figure 
\ref{fig:ekinisonosg}.  

\clearpage

\begin{figure}[h!]
\centering
\includegraphics[width = 0.4 \textwidth]{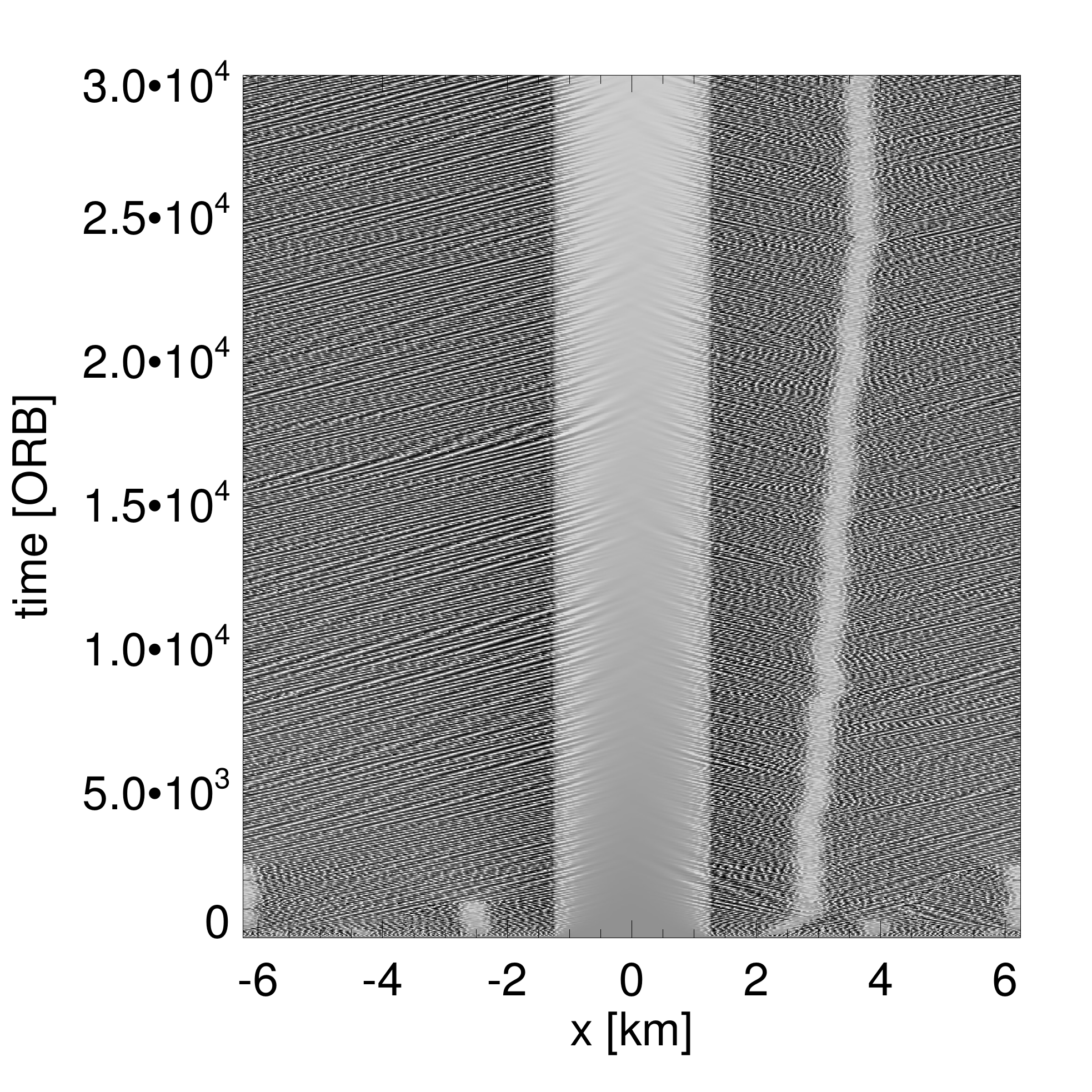}
\includegraphics[width = 0.4 \textwidth]{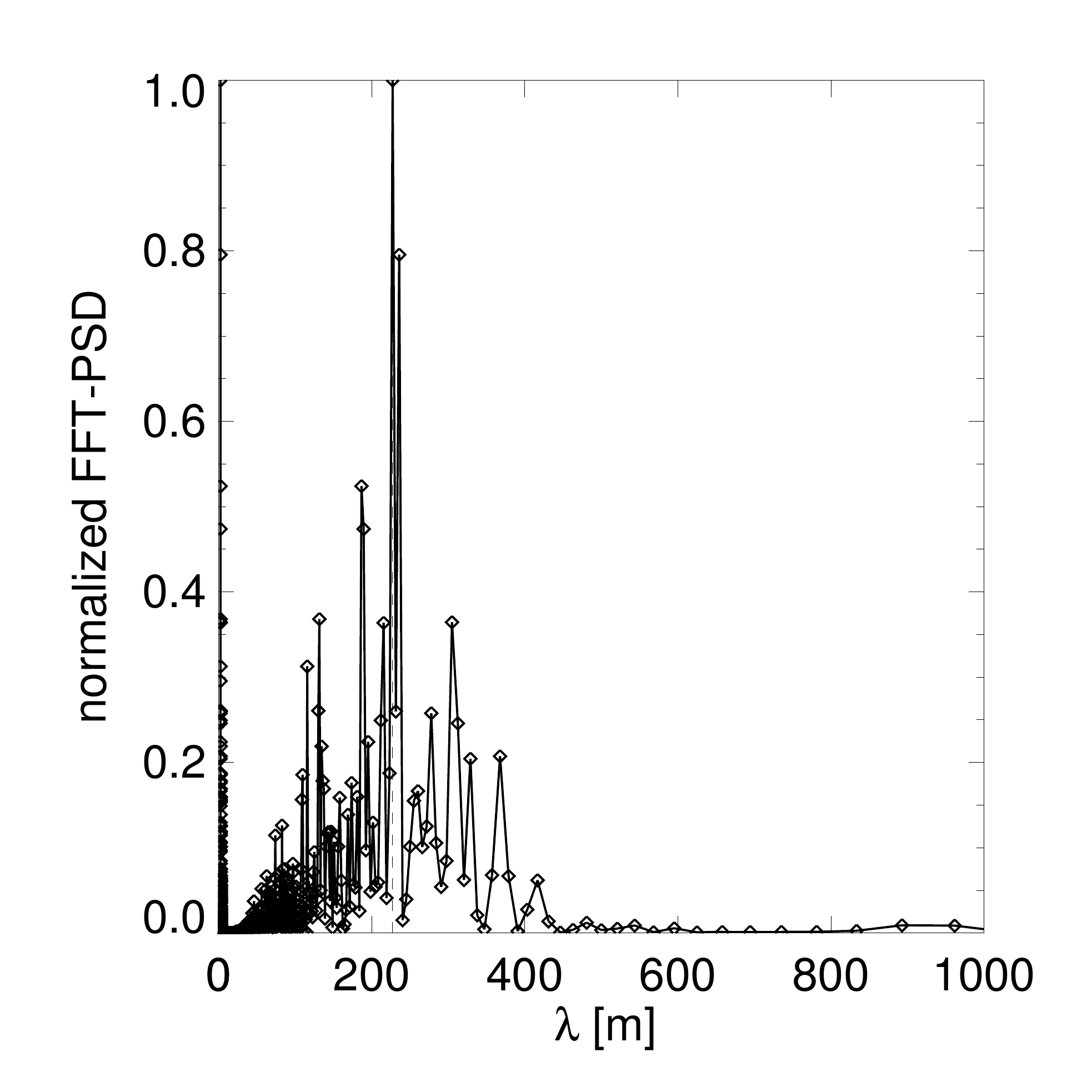}
\caption{Isothermal integration with the $\tau_{15}$-parameters and $\sigma_{0}=0$ (no radial self-gravity) where a buffer-zone is included such that 
$\beta=-0.5$ in the 
radial domain 
$[-1.25\,\text{km};+1.25\,\text{km}]$. In this zone the hydrodynamic state is modified, so that overstable waves will damp (see text). Left and right panels 
show a stroboscopic space-time diagram and the final power spectrum, respectively.}
\label{fig:tau15bufnosg}
\end{figure}

We also perform a few integrations where we include a buffer-zone in the calculation region, i.e.\ a small radial sub-region where the density exponent of the 
viscosity in Equation (\ref{eq:shearvis}) takes values $-1<\beta<\beta_{c}$. In such a region the condition for viscous overstability is not 
fulfilled so that the linear growth rate of 
overstable modes (\ref{eq:osgrate}) is negative. Waves which travel into this region are consequently damped. This modification 
introduces an obstacle for traveling waves, a situation that might typically occur in Saturn's rings when the background properties change. A buffer-zone leads 
in all considered cases to a state with one source and one sink 
structure, where the buffer-zone serves as the latter.
The long term prevalent wavelengths in these integrations are concentrated around those wavelengths which we also find for the final traveling waves in the 
homogeneous 
boxes. Figure \ref{fig:tau15bufnosg} displays the outcome of an isothermal integration with a buffer-zone.

\FloatBarrier

\subsubsection{Non-isothermal Model}\label{sec:nonisonosg}

\begin{figure}[h!]
\centering
\includegraphics[width =0.4 \textwidth]{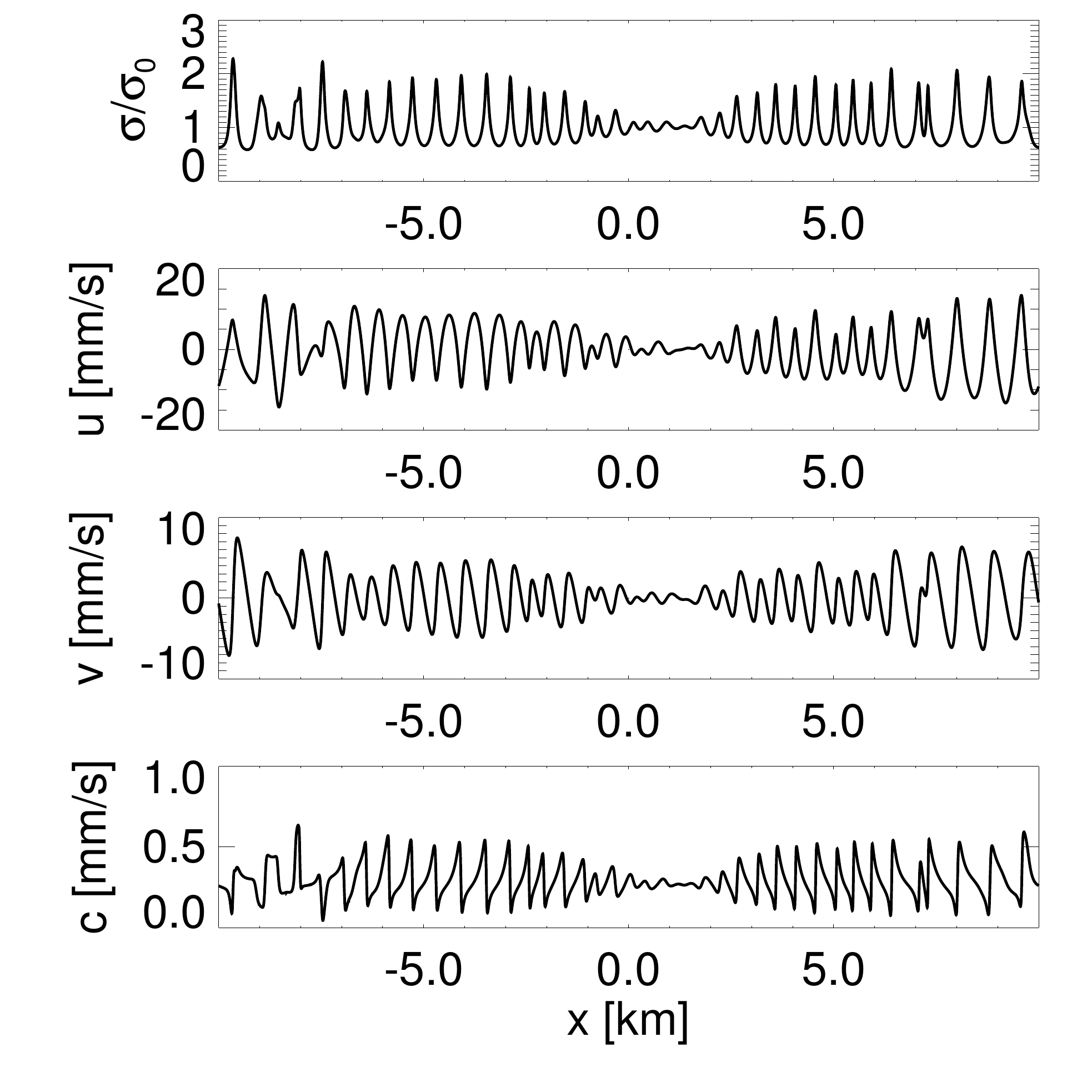}
\includegraphics[width =0.4 \textwidth]{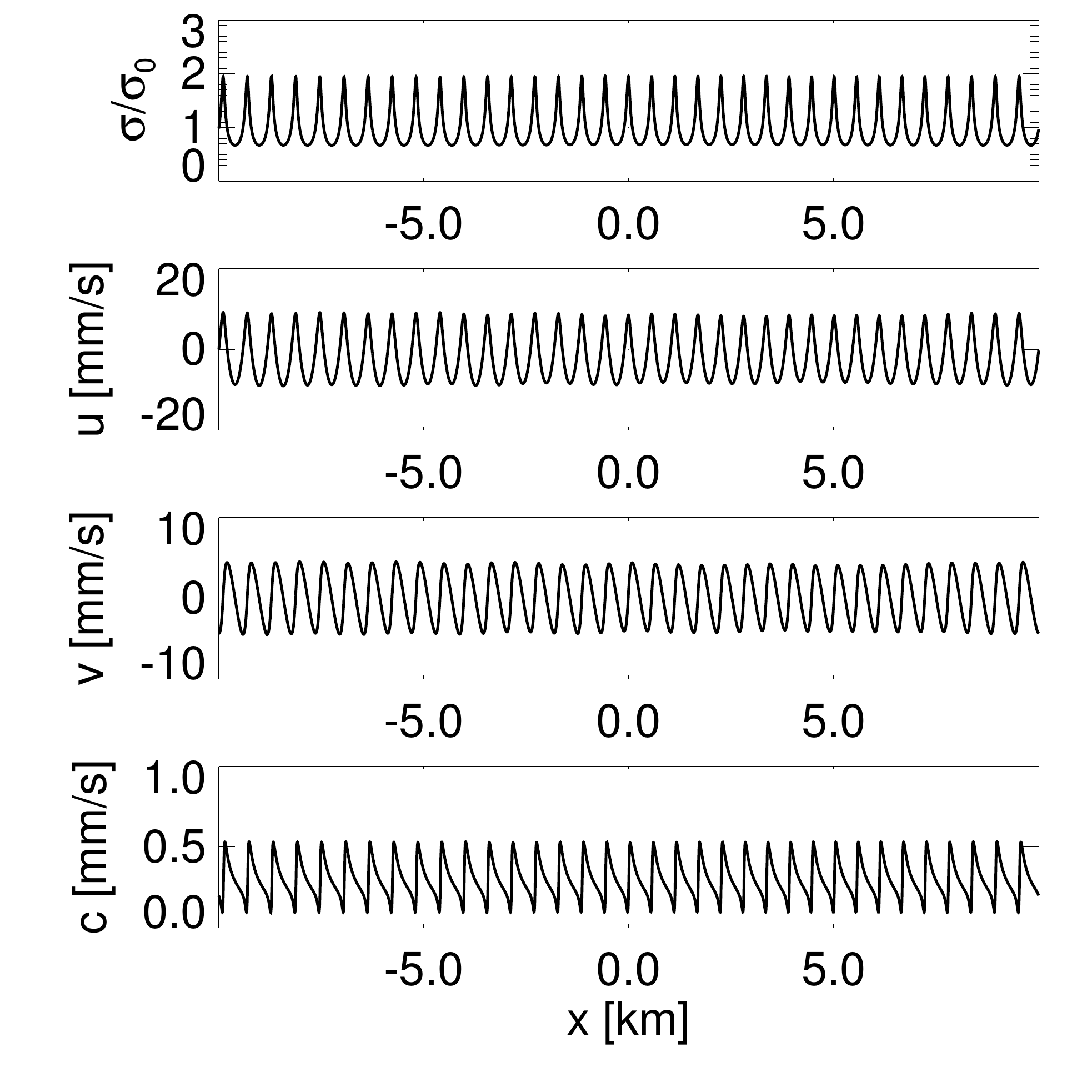}
\caption{Hydrodynamic fields characterizing intermediate (t=5,000 ORB, left panel) and final states (t=14,000 ORB, right panel), respectively, of a 
non-isothermal integration 
without self-gravity with the 
$\tau_{20}$-parameters.}
\label{fig:tau20nonisonosgb}
\end{figure}

\begin{figure}[h!]
\centering
\includegraphics[width = 0.4 \textwidth]{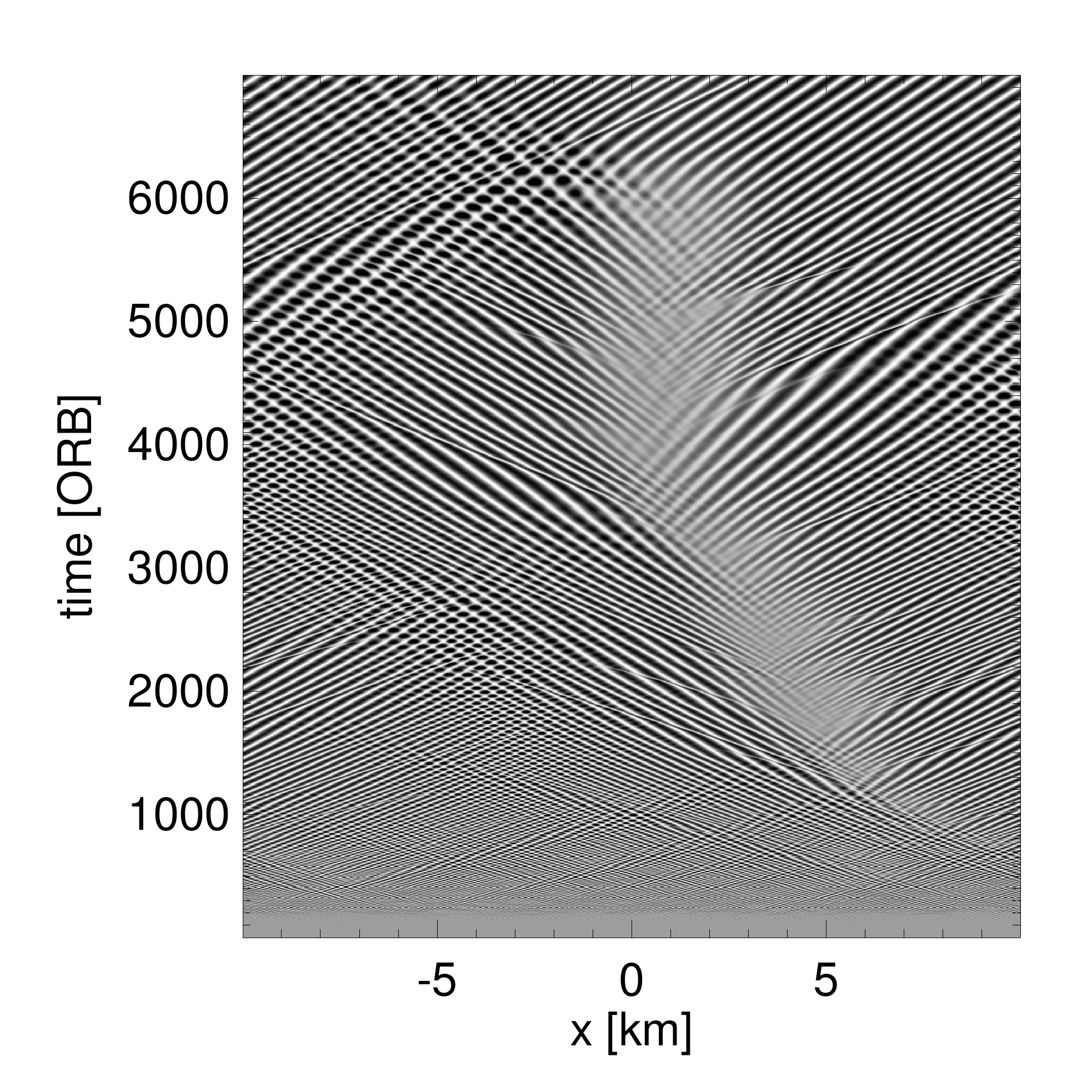}
\includegraphics[width = 0.4 \textwidth]{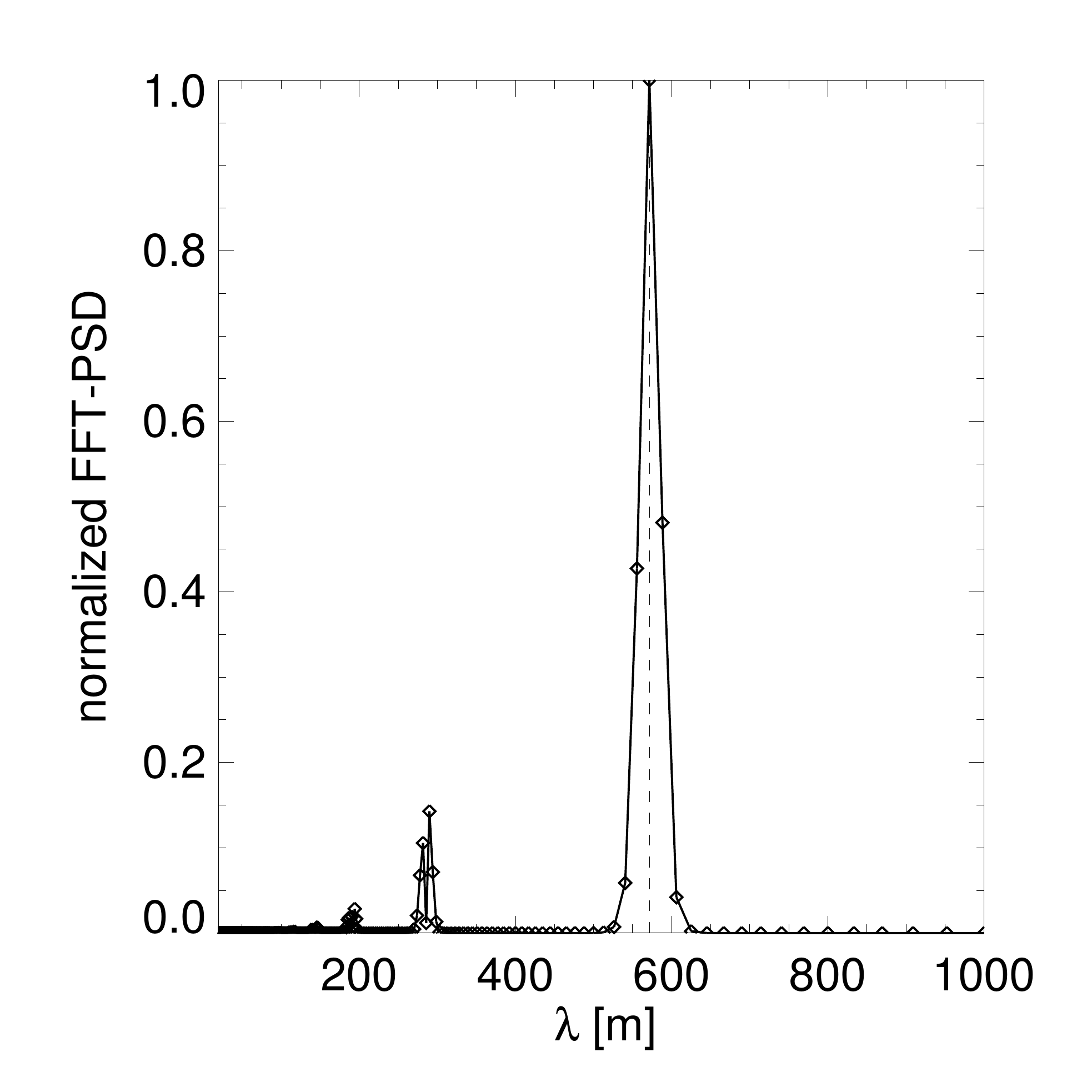}
\caption{Stroboscopic space-time diagram (left panel) and final power spectrum of the surface mass density field (right panel) for the same non-isothermal 
integration as in 
Figure 
\ref{fig:tau20nonisonosgb} with the $\tau_{20}$-parameters.}
\label{fig:tau20nonisonosga}
\end{figure}

The non-isothermal scheme in the limit of vanishing radial self-gravity produces results that agree reasonably well with the non self-gravitating N-body 
simulations presented in RL2013.
Figures \ref{fig:tau20nonisonosgb} and \ref{fig:tau20nonisonosga} describe an integration with the $\tau_{20}$-parameters. The initial state of the 
integration is spectral white noise with wavelengths down to $\lambda\sim 50\,\text{m}$. After complicated, disordered transient states, 
with strongly asymmetric 
sink and source 
structures, the system settles on a single traveling wave state.

For comparison, the final state wavelengths we find with the $\tau_{15}$ and $\tau_{20}$-parameters are about $\lambda=360\, \text{m}$  and $\lambda=570\, 
\text{m}$ (Figure \ref{fig:tau20nonisonosga}, right panel), respectively. This is in good agreement with the final wavelength of the ``fiducial run'' from 
RL2013 with $\tau=1.64$ which is close to $\lambda=450\, \text{m}$. Overall, these results indicate a trend of increasing final state wavelength with 
increasing ground state optical 
depth, which was expected also by LO2009 on theoretical grounds.

The colliding waves penetrate each other over many wavelengths in the sink structures before they damp. In contrast, we find fairly narrow 
sink and source structures in the isothermal model (cf.\ Figure \ref{fig:tau15nga}). RL2013 discussed the same discrepancy between the appearance of 
sources and sinks in their N-body simulations, compared to those in the isothermal hydrodynamic model of LO2010. They attributed it to the particulate nature 
of the nonlinear wave-wave interactions of their N-Body simulations. 
Because we find the large zones of co-existence of left and right traveling wave modes also around the sinks in our non-isothermal hydrodynamic model,
we conclude that this is not an effect tied to the particulate nature of the system.
We believe that it is a consequence of the shape of the equation of state, mediating the action of pressure, as well as thermal 
effects, modeled by the temperature equation. Generally, the nonlinear interaction of left and right traveling modes, and thus their competition, seems to be 
much stronger in the isothermal model.

 \clearpage

  \FloatBarrier

\subsection{Hydrodynamical Integrations Including Radial Self-Gravity}\label{sec:hydrosg}

\subsubsection{Isothermal Model}\label{sec:isosg}
In this section we describe our hydrodynamic model results with radial self-gravity, starting with the isothermal model. The initial state for all integrations 
is spectral white noise down to length scales $\lambda \sim 
50\,\text{m}$. The computational regions for most integrations have radial dimension $L_{x}=5\,\text{km}$ with a grid resolution $h=2.5 \,\text{m}$.
We find that our isothermal integrations show three qualitatively distinct types of behavior with increasing strength of
self-gravity, i.e.\ with increasing ground state surface mass density $\sigma_{0}$.

For $\sigma_{0} \lesssim 200 \, \text{kg}\, \text{m}^{-2}$ and the $\tau_{15}$-parameters ($Q_{0}\gtrsim 4$), the influence of self-gravity is weak, and the 
system constantly generates modes corresponding to the largest linear growth rates. These waves experience nonlinear 
interactions, resulting in modes with longer wavelengths. The spectrum accordingly shows a concentration of 
power on wavelengths $\lambda\lesssim 100 \,\text{m}$, and energy scattered over a wide range of larger wavelengths which exceed the wavelengths characterizing 
the 
final state of the non self-gravitating integrations by large amounts. 
It is unclear whether this wavelength growth would halt at some finite value. The numerical time step becomes very small in this state, resulting in
an impracticably slow integration.
Thus, we find that in the isothermal model the regime of small self-gravity forces is difficult to 
probe with our numerical method. In the next section we will see that this difficulty does not occur in the non-isothermal model.

\begin{figure}[h!]
\centering
\includegraphics[width =0.35 \textwidth]{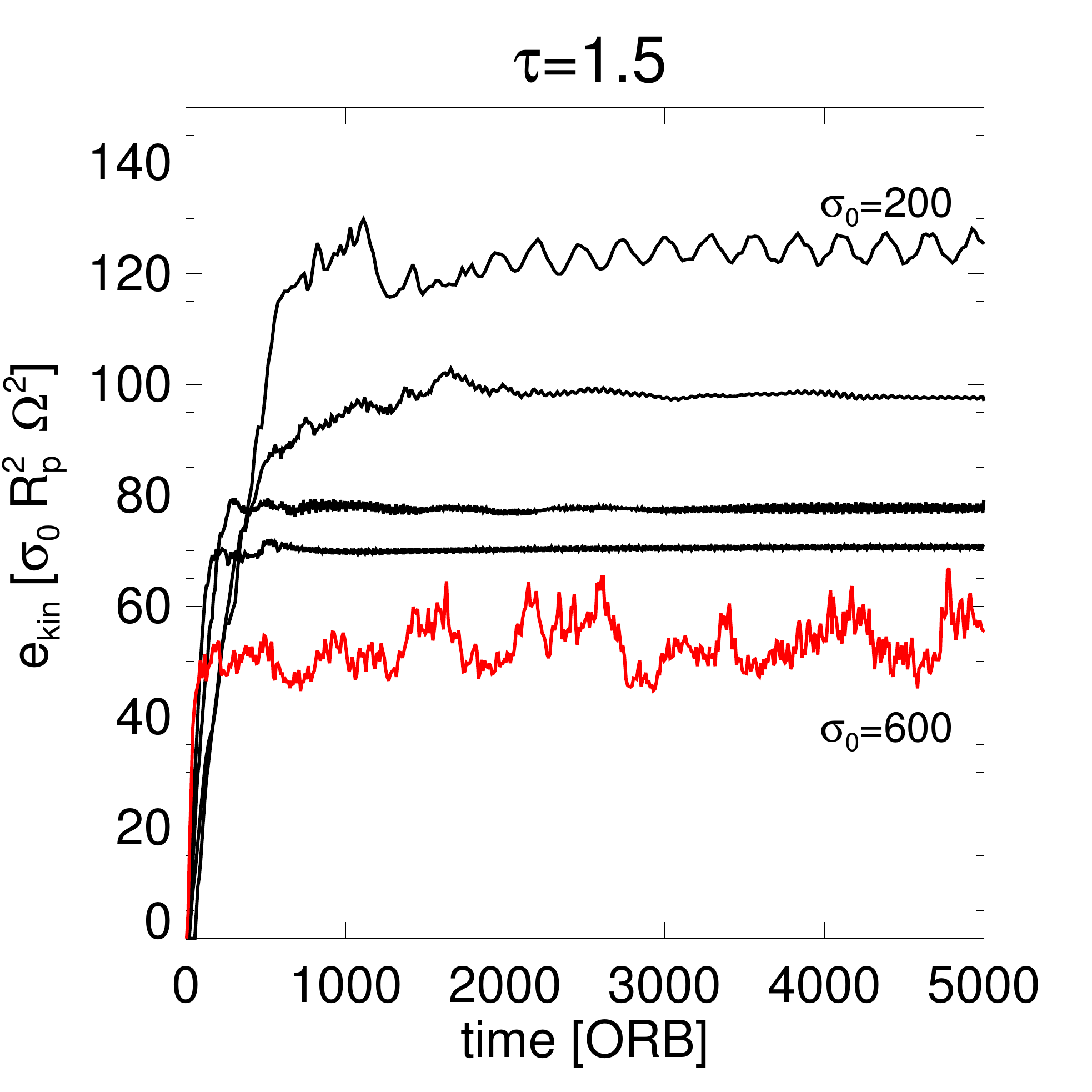}
\includegraphics[width =0.35 \textwidth]{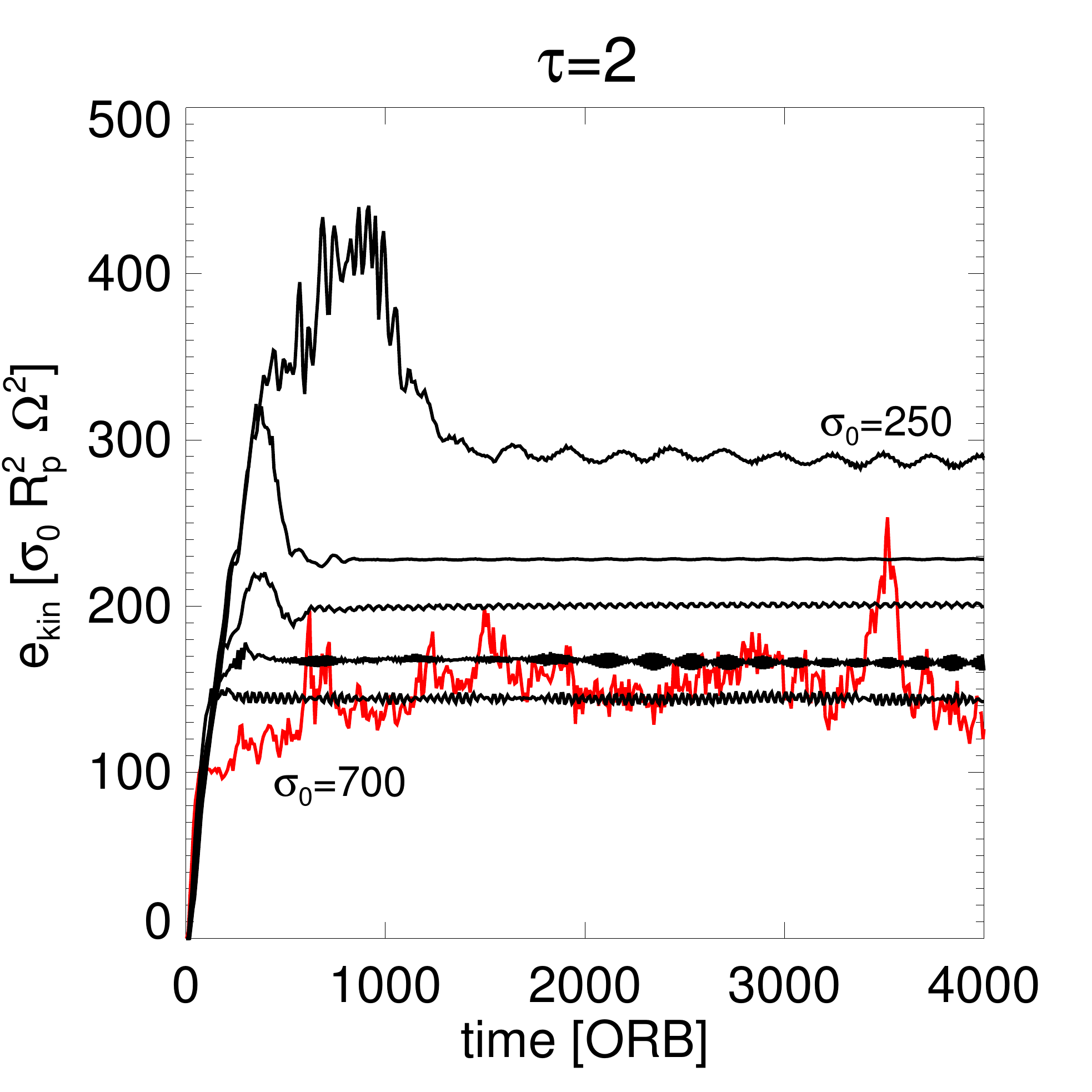}
\caption{Kinetic energy densities 
of isothermal integrations with different values $\sigma_{0}$. Curves with higher saturation energy 
correspond to smaller values of $\sigma_{0}$.
\emph{Left}: $\tau_{15}$-parameters with $\sigma_{0}=200, 250, 300, 400 \, \text{and} \,600\, \text{kg}\, \text{m}^{-2}$.
\emph{Right}: $\tau_{20}$-parameters with $\sigma_{0}=250, 300, 350, 400, 500\,  \text{and} \,700\, \text{kg}\, \text{m}^{-2}$. The red curves in both panels 
correspond to the highest value of $\sigma_{0}$, respectively.}
\label{fig:ekincompare}
\end{figure}

\begin{figure}[h!]
\vspace{-0.2cm}
\centering
\includegraphics[width = 0.4 \textwidth]{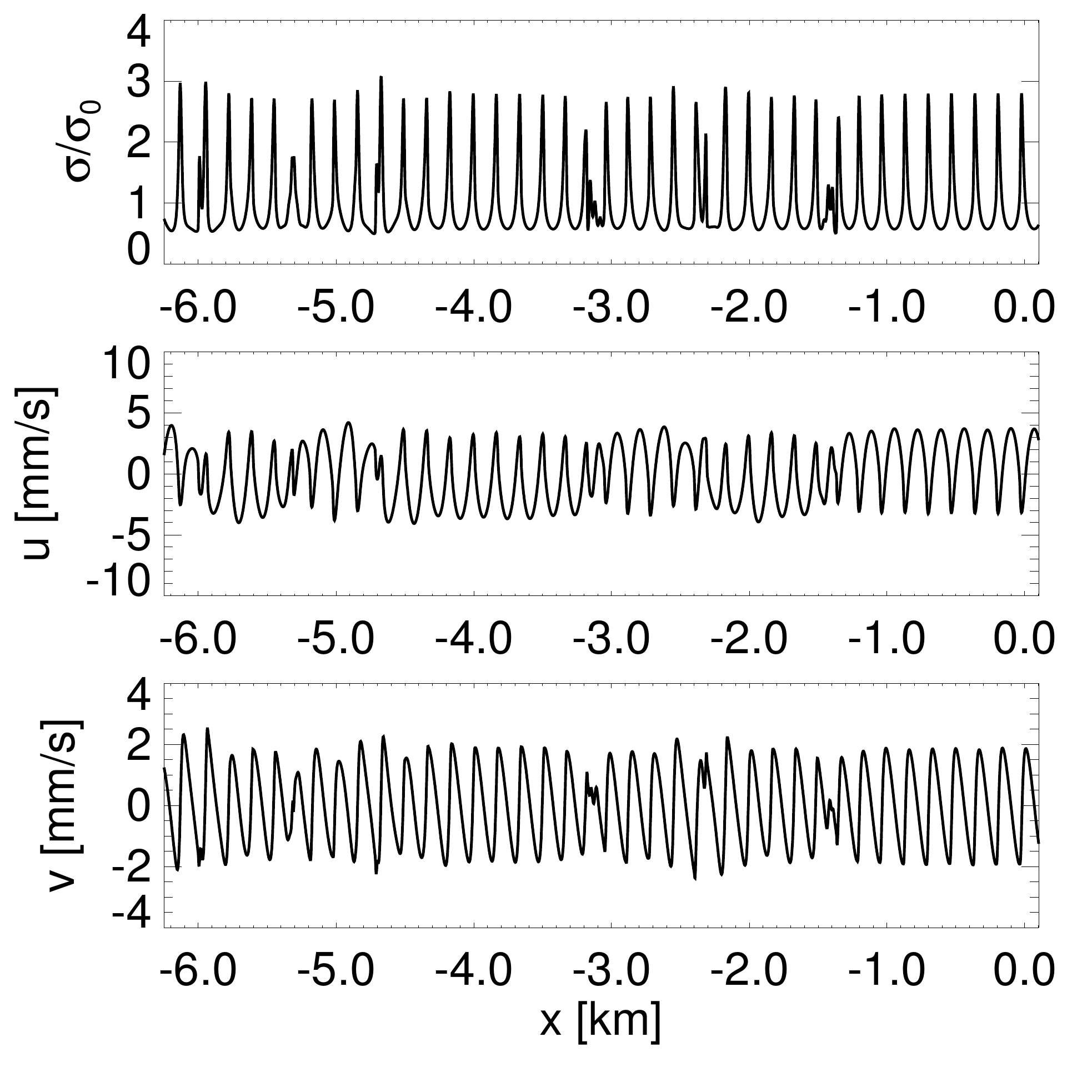}
\includegraphics[width = 0.4\textwidth]{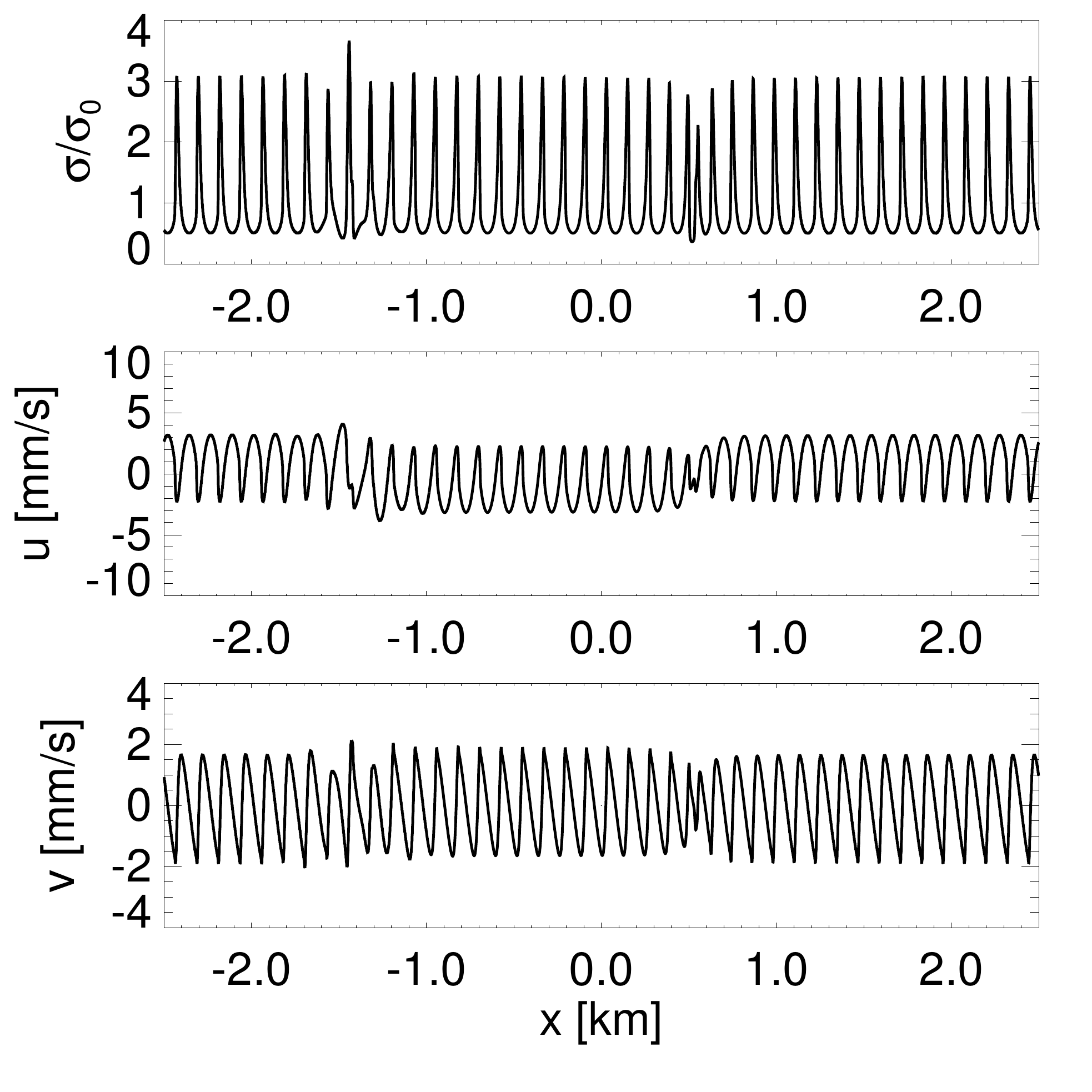}\\
\hspace{-0.6cm}\includegraphics[width = 0.42 \textwidth]{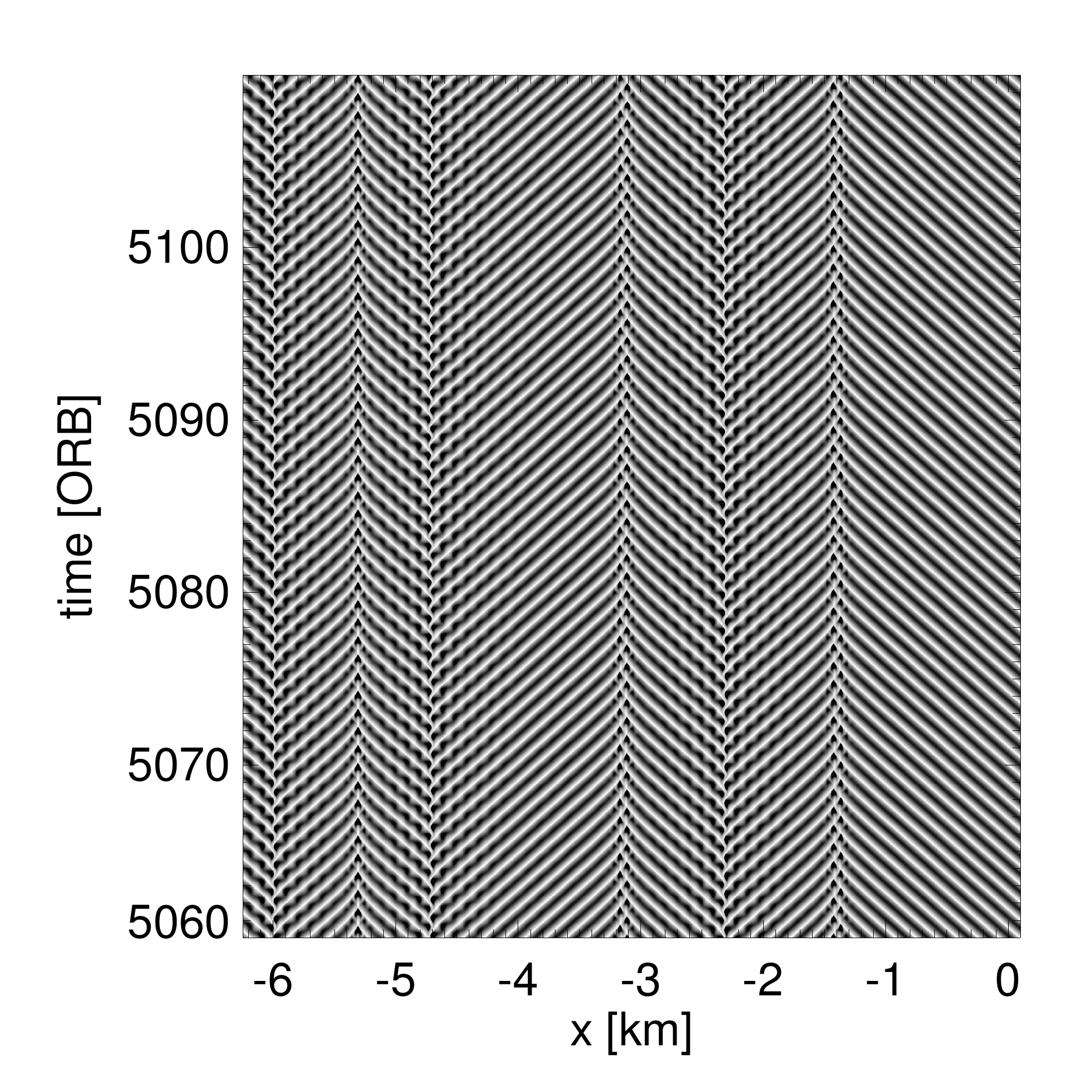}
\hspace{-0.2cm}\includegraphics[width = 0.42\textwidth]{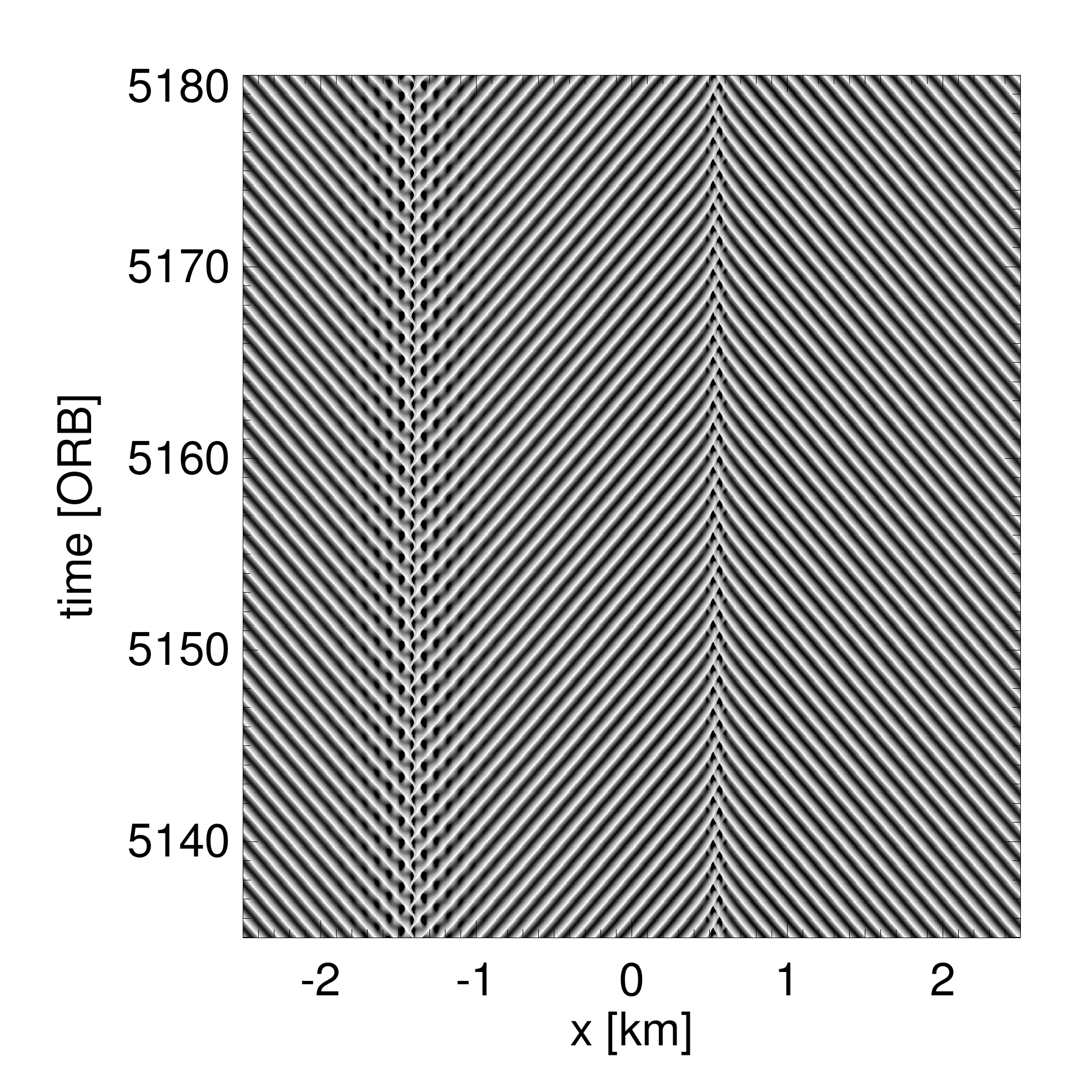}
\caption{Snapshots of hydrodynamic quantities (upper panels) and orbit-resolved (sampling interval of $0.02\,\text{ORB}$) space-time diagrams (lower panels) of 
stable source/sink states resulting from 
isothermal self-gravitating integrations with the $\tau_{15}$-parameters. \textit{Left}: 
$\sigma_{0}=250\, \text{kg}\, 
\text{m}^{-2}$ 
 ($Q_{0}=3.2$, $\lambda \approx 170\, \text{m}$). \textit{Right}: $\sigma_{0}=400\, \text{kg}\, \text{m}^{-2}$ 
 ($Q_{0}=2.0$, $\lambda \approx 120 \, \text{m}$). 
}
\label{fig:sgstable}
\end{figure}

When increasing the value of $\sigma_{0}$, such that $200 \, \text{kg}\, \text{m}^{-2}\lesssim \sigma_{0} \lesssim 600 \, \text{kg}\, \text{m}^{-2}$ 
($4\gtrsim Q_{0} \gtrsim 1.5 $ for 
$\tau=1.5$), the system behaves very differently.
Initially, we observe a fast development of multiple source/sink structures. Moreover, the wavelengths of the interacting waves grow fast.
However, this growth slows down and halts at a certain wavelength. We 
observe that upon 
gradually increasing $\sigma_{0}$, this prevalent wavelength reduces in a monotonic manner. Since the wavelength of a nonlinear saturated overstable wave 
is proportional to its amplitude (SS2003, LO2009), the kinetic energy density is also a good proxy for the dominant wavelength of the overstable waves. 
Figure \ref{fig:ekincompare} shows the development of the kinetic energy densities of integrations with intermediate and high values of $\sigma_{0}$, 
confirming that higher values of the surface mass density result in states with smaller overall kinetic energy density, indicating a smaller 
dominant wavelength. 
We find that the outcomes of these computations exclusively consist of (quasi-)stable source/sink states.
Examples are presented in Figure \ref{fig:sgstable}.
By comparing Figures \ref{fig:tau15nga} and \ref{fig:sgstable} one can see that the sources and sinks in the integrations with self-gravity are more narrow 
than in the case of vanishing self-gravity, indicating a stronger interaction between the counter-propagating wave trains.
The stable source and sink structures connect patches of counter-propagating traveling waves with spatially constant wavelength. Integrations for more than 
20,000 orbits have been 
performed throughout which 
these configurations persisted, without any signs of numerical instability or merging of sinks and sources.
\FloatBarrier
Further increasing $\sigma_{0}$ leads to numerical instability of our scheme, unless we reduce the time steps by a large factor. 
We performed two integrations in this regime ($\tau=1.5$ with $\sigma_{0}= 600\, \text{kg}\, \text{m}^{-2}$ and $\tau=2$ with $\sigma_{0}= 
700\, \text{kg}\, \text{m}^{-2}$) with time steps $f_{t}\sim 0.01$. In these we find source/sink structures which become chaotic such that these disappear and 
reappear continuously, showing a stochastic peculiar motion. 
The kinetic energy for these integrations (red curves in Figure \ref{fig:ekincompare}) undergoes strong fluctuations, caused by fluctuations in the dominant 
wavelength.

Notable is that none of our self-gravitating isothermal integrations presented here produces final states consisting of a single traveling wave mode.
We additionally performed integrations where the seed consisted of a single overstable mode with $50\,\text{m} \lesssim \lambda \lesssim 100\,\text{m}$.
These integrations either develop source/sink structures, leading to the states described above, or they exhibit a unidirectional wave train with ever 
growing 
wavelength
until the numerical time step becomes so small that the further evolution cannot be followed anymore.

\FloatBarrier
\clearpage
\subsubsection{Non-isothermal model}\label{sec:nonisosg}
We now turn to the results of the non-isothermal integrations including the radial component of self-gravity.
Here the calculation box for most integrations is $L_{x}=20\,\text{km}$ with $h=2.5 \,\text{m}$. 
Integrations with high surface densities $\sigma_{0}\geq 700\, \text{kg}\, \text{m}^{-2}$ are conducted in smaller boxes ($L_{x}=5\,\text{km}$).
For these we find it necessary to employ a finer grid ($h=1 \,\text{m}$), as the hydrodynamic quantities exhibit very sharp transitions.

\begin{figure}[h!]
\centering
\includegraphics[width = 0.4 \textwidth]{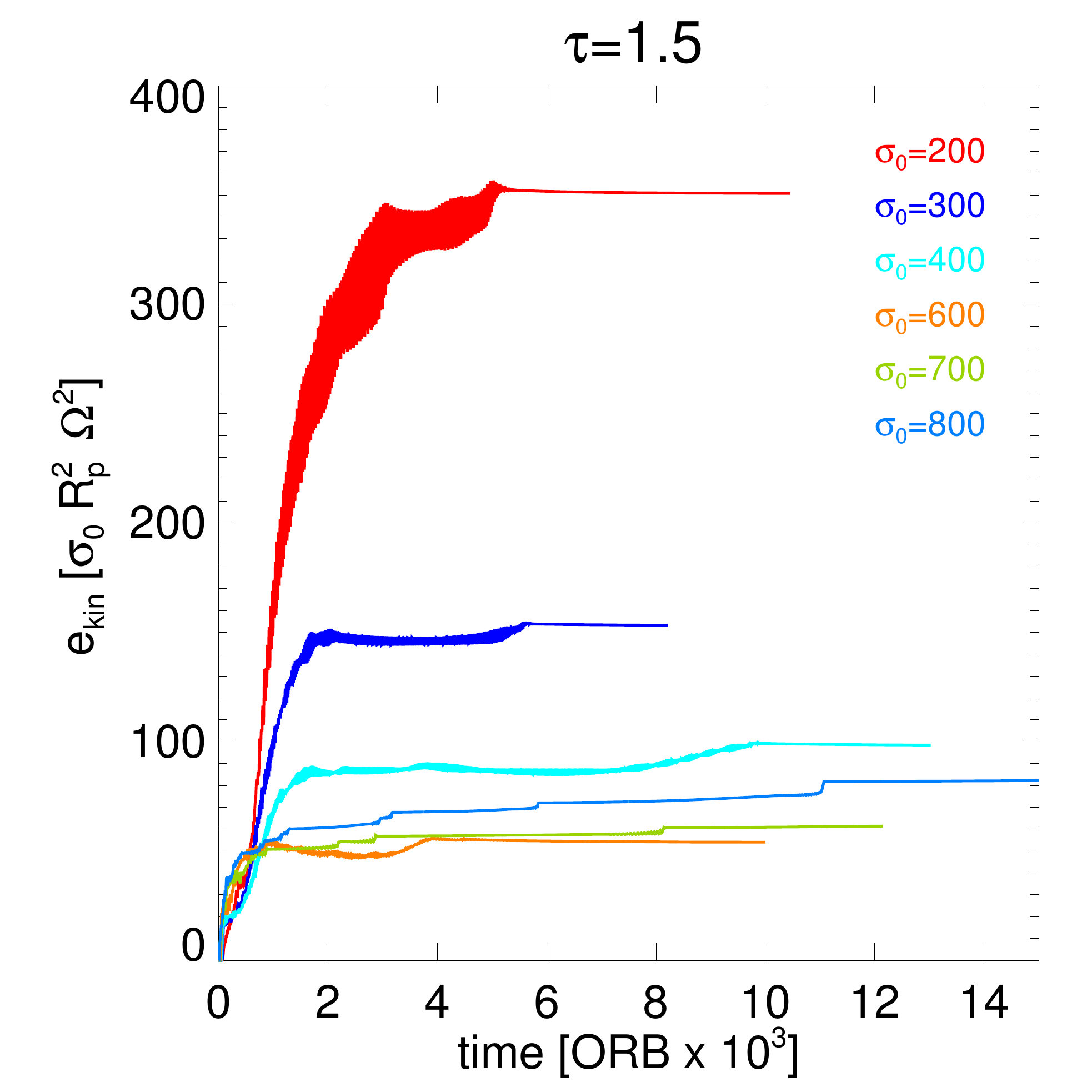}
\includegraphics[width = 0.4 \textwidth]{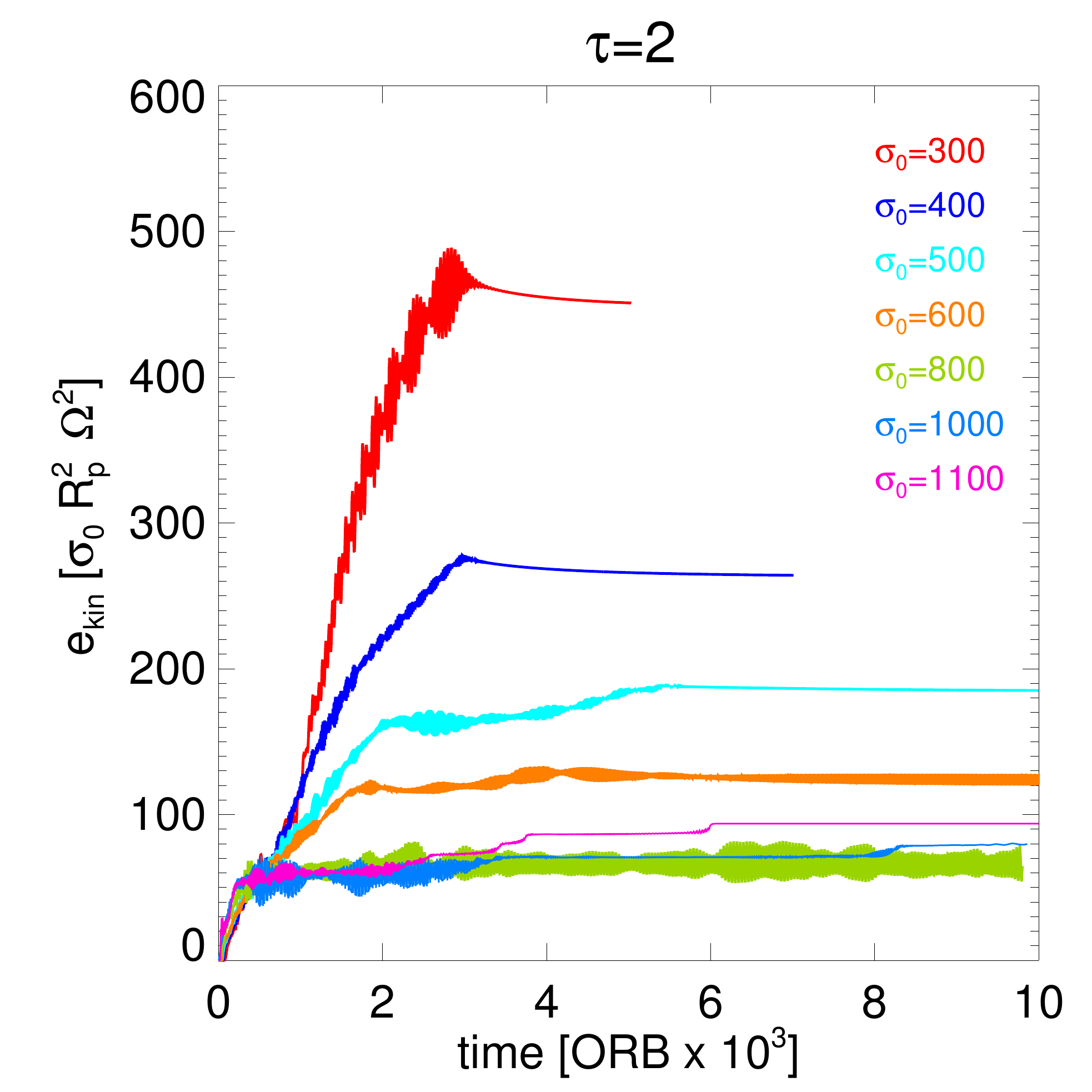}
\caption{Evolution of the kinetic energy for non-isothermal hydrodynamic integrations with different $\sigma_{0}$-values. The final wave train energies show a 
monotonic 
decrease as a function of $\sigma_{0}$ which reverses around $\sigma_{0}=600\, \text{kg}\, \text{m}^{-2}$ for $\tau=1.5$ ($\sigma_{0}=700\, \text{kg}\, 
\text{m}^{-2}$ for $\tau=2$).}
\label{fig:etauhydro}
\end{figure}

\begin{figure}[h!]
\centering
\includegraphics[width = 0.4 \textwidth]{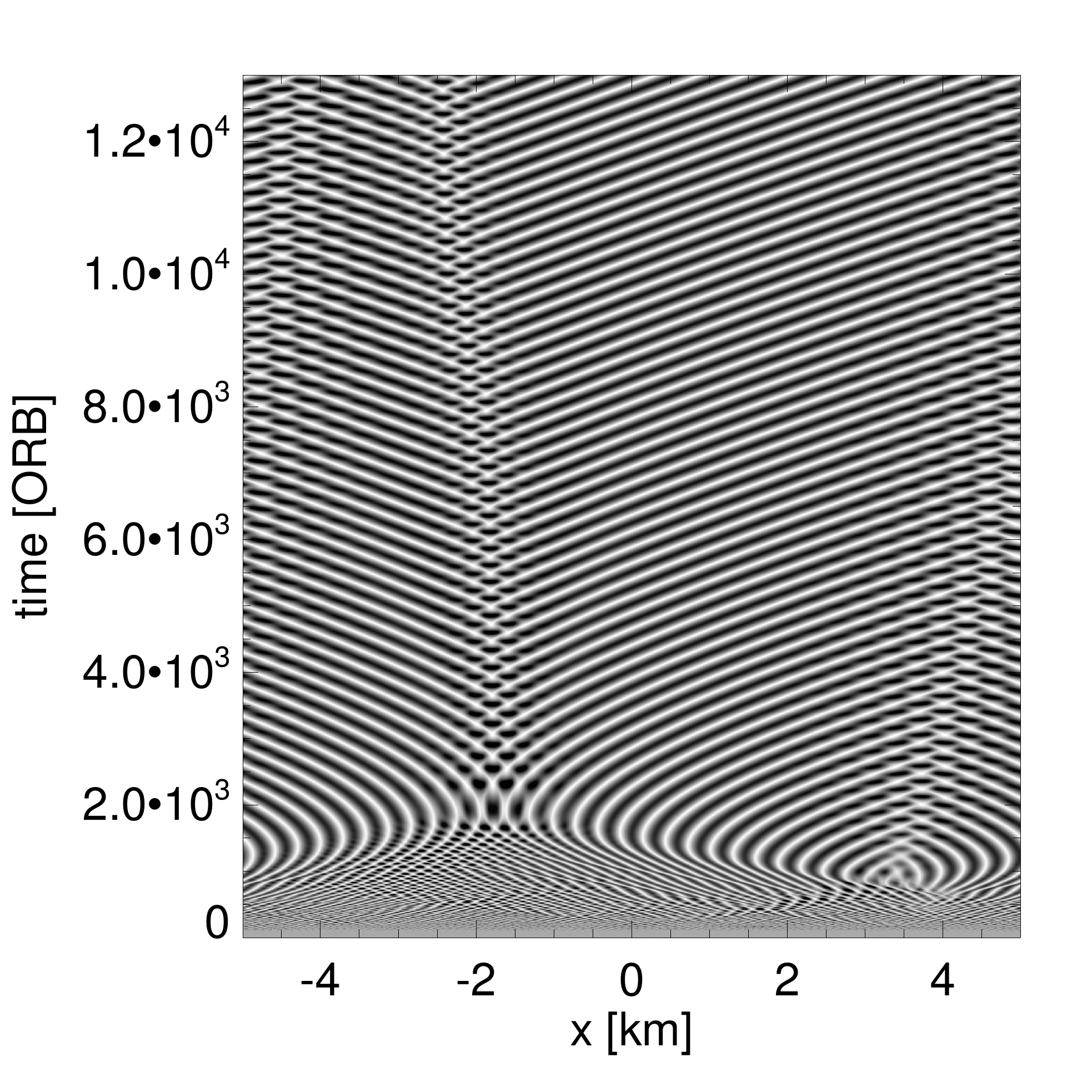}
\includegraphics[width = 0.4 \textwidth]{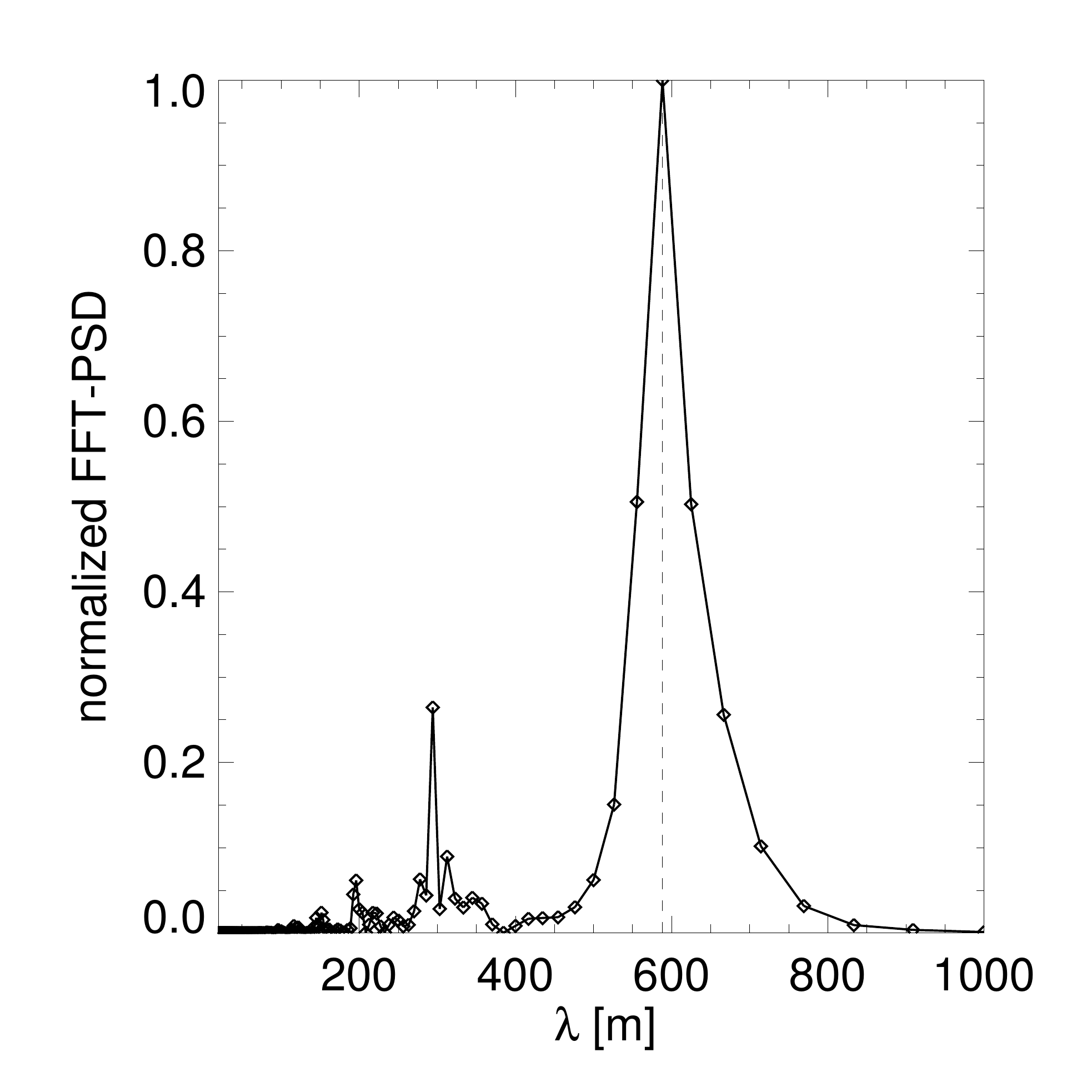}
\caption{Non-isothermal integration with the $\tau_{15}$-parameters and $\sigma_{0}=125\, \text{kg}\, \text{m}^{-2}$. The right panel shows the final power 
spectrum. 
The 
space-time diagram (left panel) is stroboscopic, giving the impression of a reversal of the phase velocity of the waves (see also Figure 
\ref{fig:vgph}). 
The system 
evolves into an ordered 
source/sink state with a prevalent wavelength exceeding that of the case $\sigma_{0}=0$. However, for larger $\sigma_{0}$ the wavelengths in gravitating 
integrations are shorter than those with $\sigma_{0}=0$.}
\label{fig:tau15nonisoweaksg}
\end{figure}

Figure \ref{fig:etauhydro} shows the evolution of the kinetic energy density of integrations with the $\tau_{15}$ and the $\tau_{20}$-parameters with different 
surface densities $\sigma_{0}$.
The results show similarities to the isothermal results (\ref{fig:ekincompare}).
For sufficiently large $\sigma_{0}$ the main effect of self-gravity is a reduction of the final state wavelengths, and 
thus the energy of ordered motions $e_{kin}$. 
For small self-gravity forces, the final state is dominated by modes with wavelengths larger than those found for the non self-gravitating case 
(Section \ref{sec:nonisonosg}).
This result was also found within the isothermal model. Nevertheless, the non-isothermal system evolves into an ordered final state which is not 
polluted 
by modes with smaller wavelengths, as can be seen for example in Figure \ref{fig:tau15nonisoweaksg}.

Similar to the non self-gravitating integrations (Section \ref{sec:hydronosg}), the initial stage for all $\sigma_{0}$-values is chaotic, with strong 
spatio-temporal 
fluctuations 
of all hydrodynamic quantities. During this stage $e_{kin}$ is increasing more or less strongly with time.
The intermediate source/sink phase, though, is different if a moderate radial self-gravity is included. The sources and sinks are more numerous and narrower 
than for the non self-gravitating case. The source/sink pattern resembles the one found for the isothermal model with 
intermediate values of  $\sigma_{0}$ (cf.\ Figure \ref{fig:sgstable}), i.e. the structures appear more stable, showing less fluctuations. 
Quasi-stable source/sink states, persisting for more than 10,000 orbits, are found for small, intermediate and large values of 
$\sigma_{0}$. For instance the integration shown in Figure \ref{fig:tau15nonisoweaksg} represents such a case with small 
$\sigma_{0}=125\,\text{kg}\,\text{m}^{-2}$. Another example 
is the case $\sigma_{0}=600\, \text{kg}\, \text{m}^{-2}$ for $\tau=2$, where a source/sink pair reveals itself through small, persistent fluctuations in the 
(leveled) kinetic energy curve (Figure \ref{fig:etauhydro}), as the waves connecting these structures undergo small fluctuations in phase and amplitude.

For larger $\sigma_{0}$ with the $\tau_{20}$-parameters the hydrodynamic field quantities become increasingly distorted during the initial and 
intermediate stages where the different wave patches exhibit in many cases standing wave-like amplitude fluctuations. This phase can persist for a long time, 
as for the case $\sigma_{0}=800\, \text{kg}\, \text{m}^{-2}$ with $\tau=2$ (right panel in Figure \ref{fig:etauhydro}, see also Figure \ref{fig:fluctnb} in 
Section \ref{sec:nbres}). This behavior and the fact that the 
overstable oscillation frequency for large $\sigma_{0}$ is 
in general significantly different from the orbital frequency makes it harder to identify source and sink structures in 
(stroboscopic) space-time plots.
For integrations with $Q_{0}\lesssim 1$ with $\tau=1.5$ we find, after a short initial chaotic stage, an elongated 
stair-case process in which wavelength and kinetic energy undergo a slow stepwise increase. This is seen in the curves for $\sigma_{0}=700\, 
\text{kg}\, \text{m}^{-2}$ and $\sigma_{0}=800\, \text{kg}\, \text{m}^{-2}$ with $\tau=1.5$ (left panel in Figure \ref{fig:etauhydro}), and also in the curve 
for the case $\sigma_{0}=1,100\, \text{kg}\, \text{m}^{-2}$ with $\tau=2$ (right panel in Figure \ref{fig:etauhydro}), subsequent to a highly distorted 
standing 
wave phase. The final state traveling waves of these 
integrations possess small phase and amplitude perturbations.

Although not verified for all our integrations, it is likely that sink and source structures eventually merge and vanish, thus resulting in single 
mode traveling waves. This is a notable difference to the results of our isothermal integrations with radial self-gravity, where we did not find stable (on 
timescales of at least some 10,000$\, \text{ORB}$) final 
states consisting of a single unidirectional traveling wave.

\FloatBarrier

\FloatBarrier

Our isothermal and non-isothermal models utilize a significantly different equation of state, given by Equations (\ref{eq:pres}) and 
(\ref{eq:ideal}), 
respectively.
In Section \ref{sec:theo} we have shown that, on a linear level, the effects of the temperature equation on overstable waves are mildly stabilizing (see Figure 
\ref{fig:omega}).
For an assessment of thermal effects in the nonlinear regime we perform hydrodynamic integrations with the 
\emph{isothermal} $\tau_{15}$ and $\tau_{20}$-parameters, but adopting the density dependence of pressure 
($p_{s}$) of the non-isothermal model (Table \ref{tab:hydropar}), instead of 
$p_{s}=1$ for the ideal gas relation. 
In this case we find with both parameter sets a saturation of overstability 
similar to the one obtained in the non-isothermal system, but with considerably larger saturation 
wavelengths. Thus, the inclusion of temperature variations leads to a saturated state of the viscous overstability with considerably less kinetic 
energy contained in the nonlinear wave trains, which amounts to a smaller saturation wavelength.

\FloatBarrier

\subsection{N-Body Simulations}\label{sec:nbres}

In the following we turn to the results of our N-body simulations (cf.\ Section \ref{sec:nbsim}) with varying magnitude of the radial self-gravity 
force.
In all conducted simulations the waves undergo a chaotic initial stage with standing wave 
like patterns, similar to those encountered in hydrodynamic integrations with large surface densities $\sigma_{0}$. The duration of this stage is found 
to increase with increasing $\sigma_{0}$. Systems with small and intermediate $\sigma_{0}$ evolve into uniform 
traveling wave states within a few thousand orbits.
Source and sink structures are not found in any of the runs. 
This absence might be explained by the relatively small size of the simulation box ($L_{x}\leq 5 \, \text{km}$) used here, when compared to our hydrodynamic 
integrations (Figures \ref{fig:tau20nonisonosga} and \ref{fig:tau15nonisoweaksg}) as well as the non-selfgravitating N-body simulation presented in Figure 6 in 
RL2013. 

\begin{figure}[h!]
\vspace{-0.2cm}
\centering
\includegraphics[width = 0.35\textwidth]{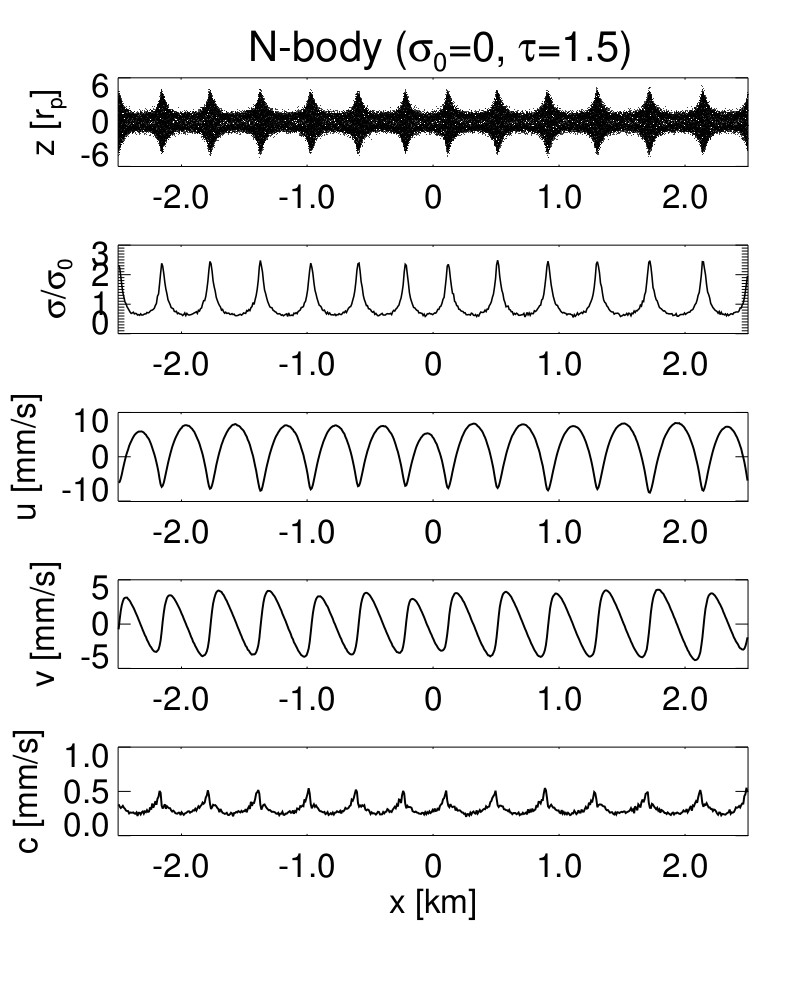}
\includegraphics[width = 0.35\textwidth]{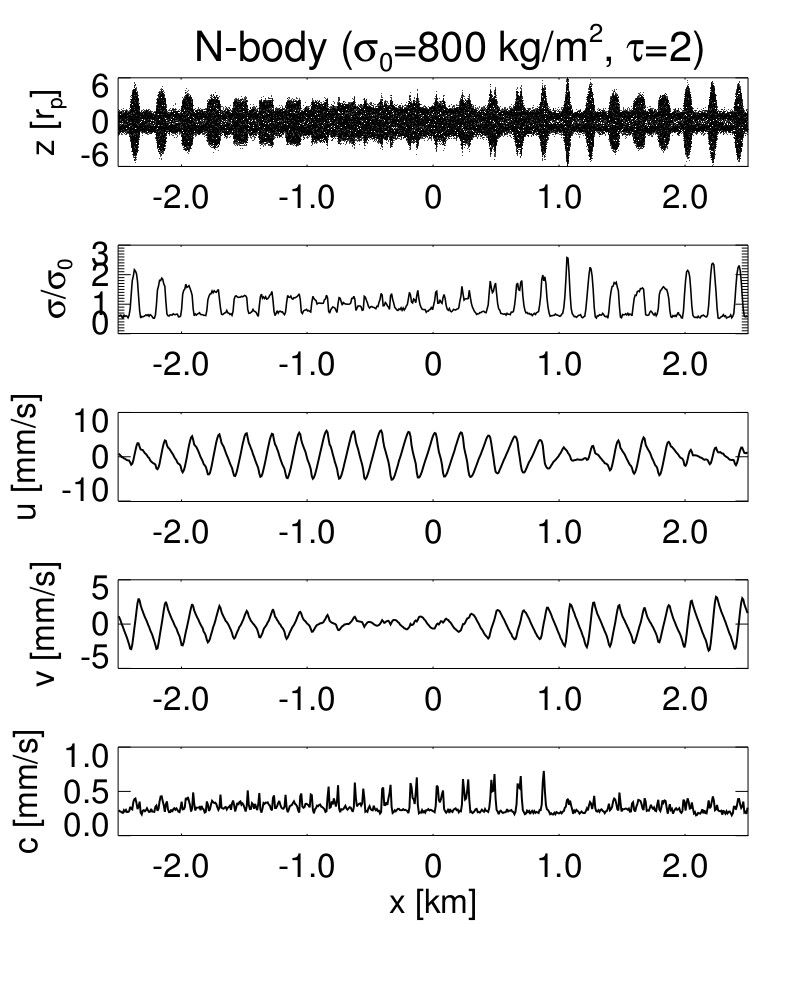}\\
\hspace{-0.4cm}\includegraphics[width = 0.375\textwidth]{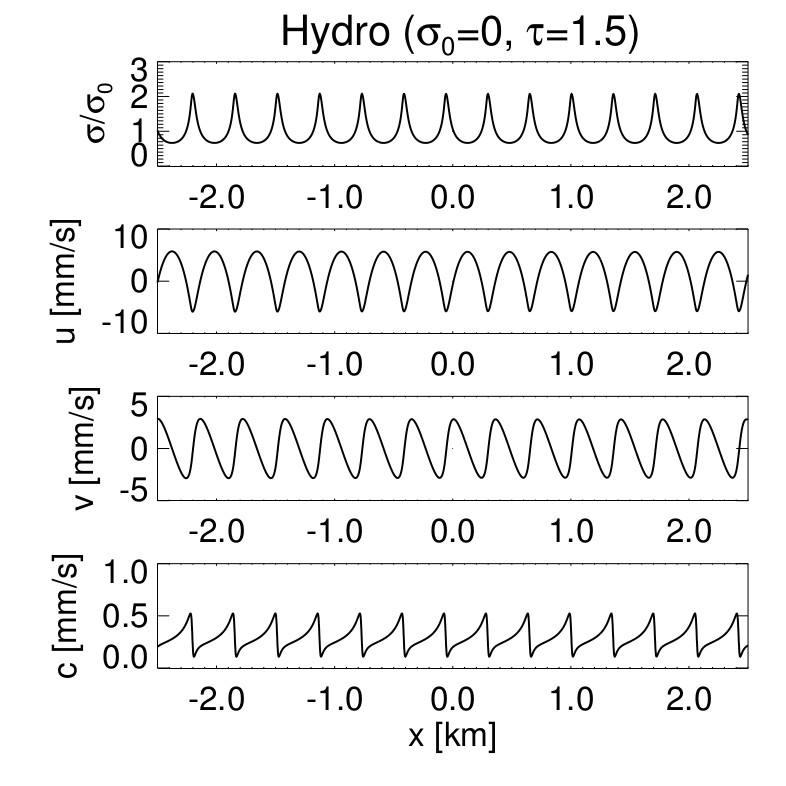}
\hspace{-0.4cm}\includegraphics[width = 0.375\textwidth]{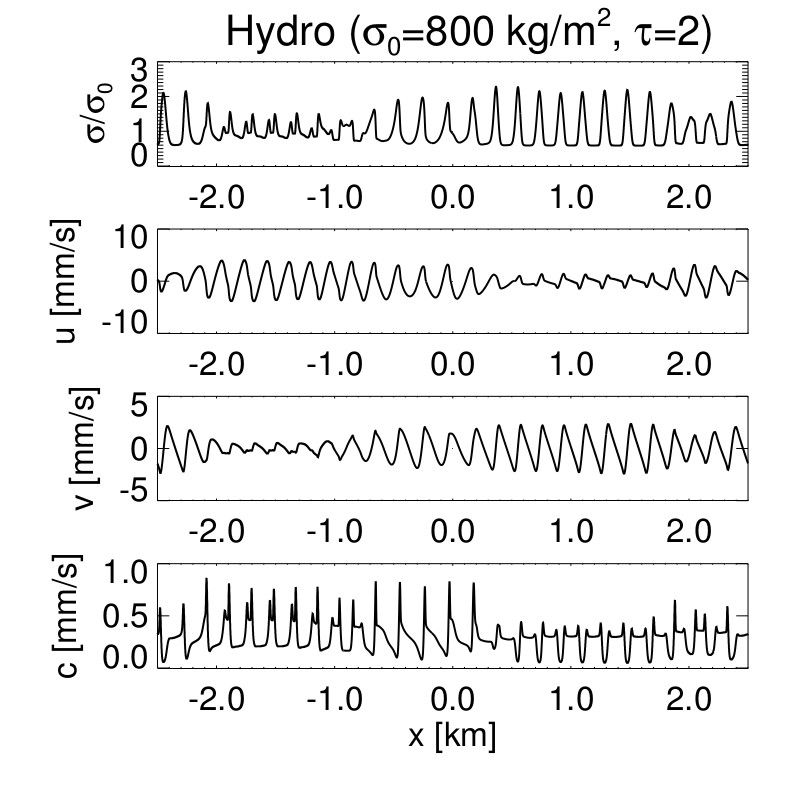}
\caption{Comparison of saturated states of the viscous overstability in simulations (upper panels) and the non-isothermal hydrodynamic model (lower panels). 
The left column displays non-selfgravitating systems with $\tau=1.5$. The final states in both panels are left traveling waves.
The right column shows systems with strong self-gravity. For this case both the N-body simulation and the hydrodynamic integration result in persistent 
standing wave patterns, exhibiting strong amplitude fluctuations.
All displayed cases correspond to $\Omega_{z}=3.6$.
For the hydrodynamic integration with $\sigma=0$ we used $L_{x}=10\, \text{km}$ with $h=2.5 \,\text{m}$. For the case $\sigma_{0}=800\,\text{kg}\,\text{m}^{-2}$ 
we utilized $L_{x}=5\, \text{km}$ with a finer grid ($h=1 \,\text{m}$), 
required to 
accurately capture the sharp spatial transitions of numerical quantities. The N-body simulations were conducted in boxes with $L_{x}=5\, \text{km}$.}
\label{fig:fluctnb}
\end{figure}

As an illustration (Figure \ref{fig:fluctnb}) we plot snapshots of various quantities across the simulation box for two runs ($\sigma_{0}=0$ with $\tau=1.5$ 
and $\sigma_{0}=800\,\text{kg}\,\text{m}^{-2}$ with $\tau=2$) and compare with results from non-isothermal hydrodynamical integrations.
Overall the hydrodynamic description is able to capture quite well most of the salient features of the wave trains, such as the dominant wavelength and the 
shapes of the velocity fields and the surface density.
In the case $\sigma_{0}=800\,\text{kg}\,\text{m}^{-2}$ N-body and hydrodynamic systems both exhibit complicated standing wave like patterns. 
The most notable differences are in the profiles of the velocity dispersion.
In the simulations the velocity 
dispersion does not attain values smaller than $R_{p} \Omega \sim 0.2 \, 
\text{mm/s}$, due to nonlocal viscous heating. The hydrodynamic description does not capture this lower bound and, on the other hand, overestimates the 
temperature peaks in systems with large $\sigma_{0}$.

For the N-body simulations shown in Figure \ref{fig:fluctnb}, and also for the computation of the kinetic energy (see below), a tabulation is performed of 
different quantities across the simulation box into $n$ radial zones of width $\Delta x=L_{x}/n$, covering 
the whole azimuthal and vertical extent of the simulation box.
For the computation of velocity fields we tabulate the particle's individual radial, vertical and azimuthal velocities relative to the Keplerian motion. 
The mean values of the radial and azimuthal velocities, taken over all particles in the zone at radial location $x$, are then identified with the 
hydrodynamic velocity fields $u(x)$ and $v(x)$, respectively [cf.\ Equation (\ref{eq:linper})]. These describe the collective particle motion in radial and 
azimuthal direction, respectively, which in our simulations is due to viscous overstability. The resulting vertical velocity field takes negligible 
values, since the collective vertical particle motion in (overstable) wave trains is anti-symmetric with respect to the plane $z=0$, so
that contributions from particles above and below the plane cancel. 
This is clearly seen in the particle's vertical coordinates $Z$ (Figure \ref{fig:fluctnb}) and is a consequence of the near 
incompressibility of the simulated ring state.
Furthermore, the standard deviations of the velocity components in a given zone define the diagonal components of 
the velocity dispersion tensor (Section \ref{sec:hydro}), $\hat{C}_{xx}$, $\hat{C}_{yy}$ and $\hat{C}_{zz}$.
These determine the velocity dispersion 
\begin{equation*}
 c(x)=\left[\frac{1}{3}\left(\hat{C}_{xx}^2 +\hat{C}_{yy}^2+\hat{C}_{zz}^2\right)\right]^{1/2},
\end{equation*}
which relates to the hydrodynamic temperature via $c=T^{1/2}$.
The scaled surface density $\sigma(x)/\sigma_{0}$ is obtained by scaling the number of particles in the zone at radial location $x$ with the average 
number of particles per bin in the simulation box. 

\begin{figure}[h!]
\vspace{-0.2cm}
\centering
\includegraphics[width = 0.4 \textwidth]{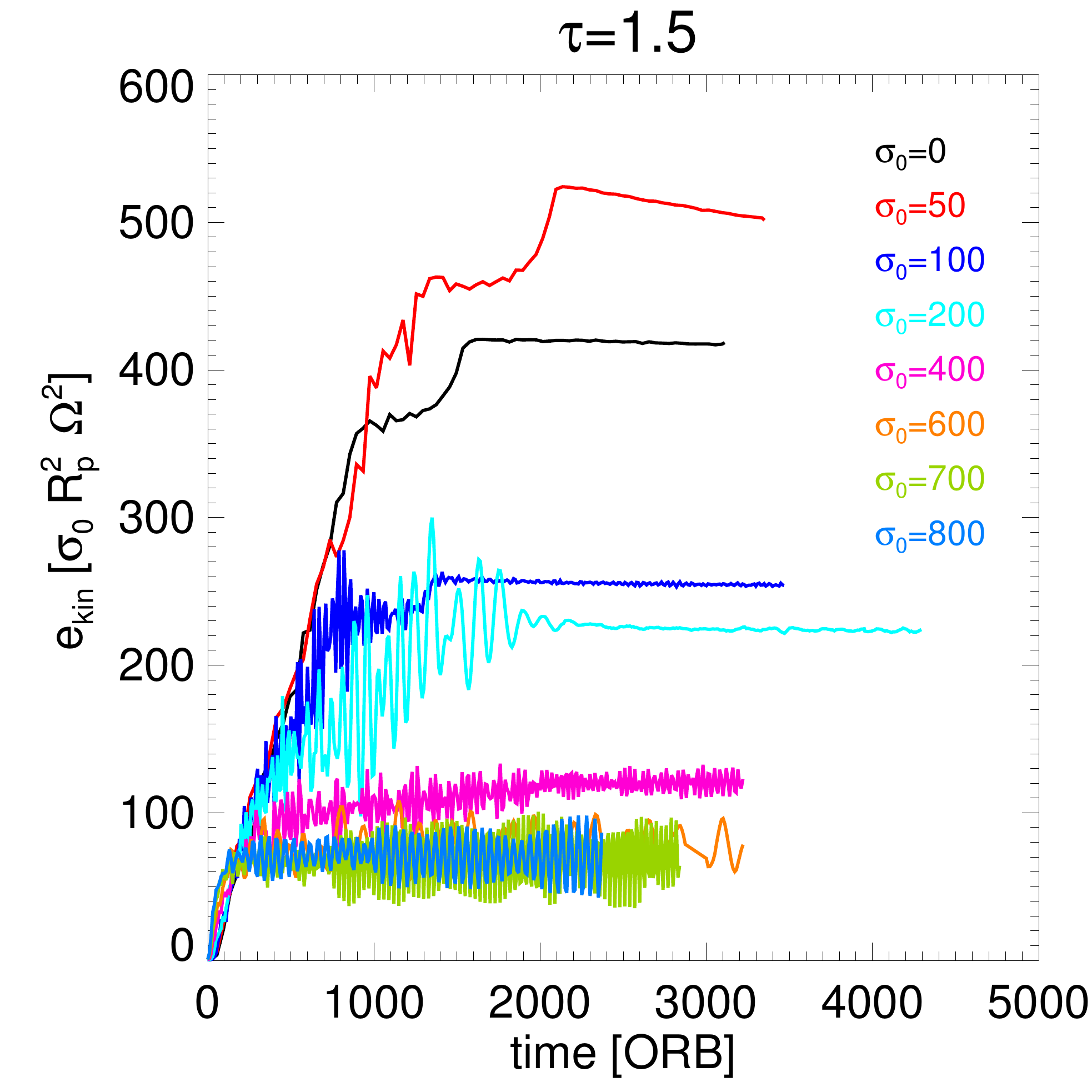}
\includegraphics[width = 0.4 \textwidth]{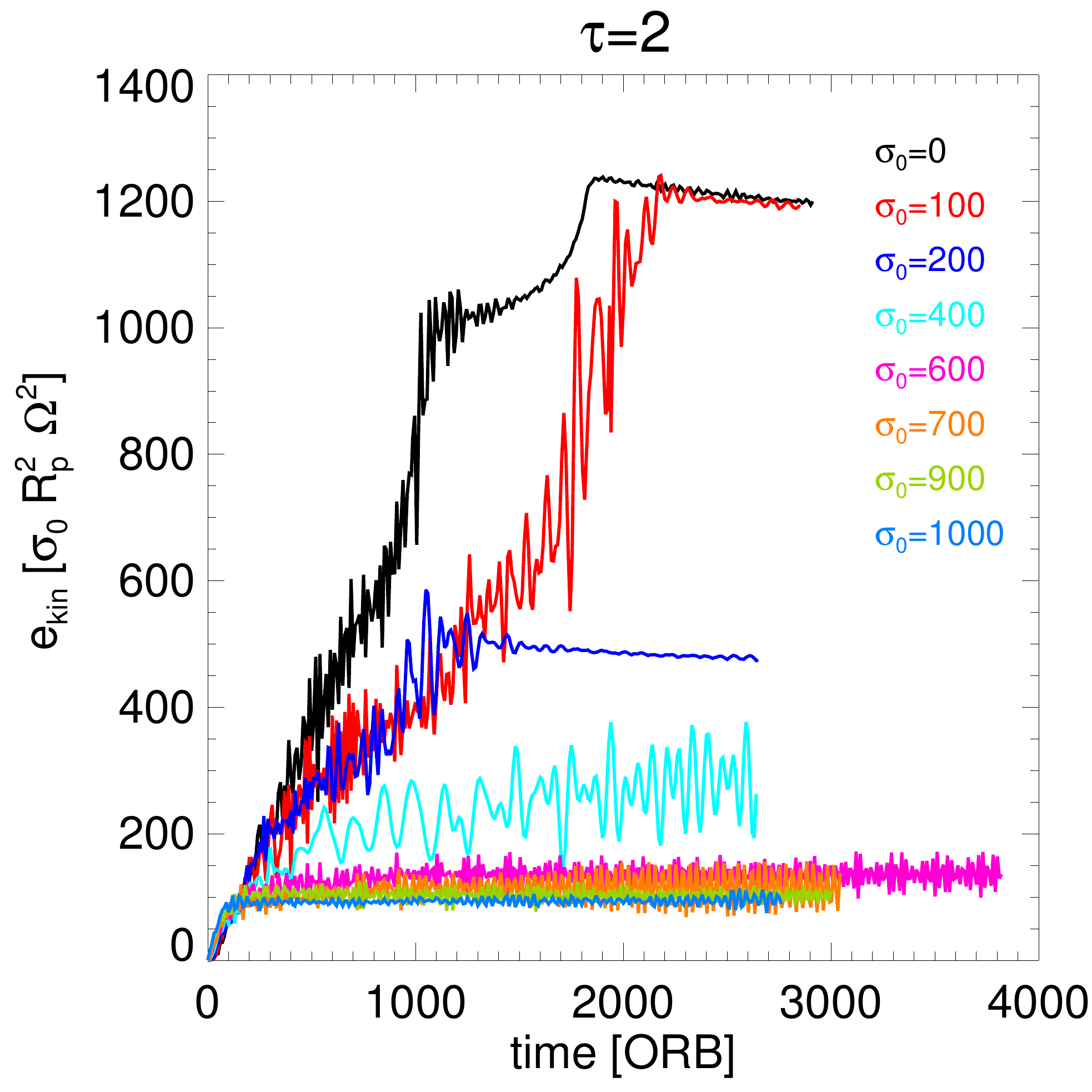}
\caption{Evolution of the kinetic energy densities in simulations with optical depths $\tau=1.5$ and $\tau=2$ and $\Omega_{z}=3.6$ for different 
$\sigma_{0}$ (in $\text{kg}\,\text{m}^{-2}$) (see Table \ref{tab:nbstat} for parameters). The behavior of $e_{kin}$ with changing 
$\sigma_{0}$ is qualitatively similar as in the non-isothermal hydrodynamic model (Figure \ref{fig:etauhydro}), but the overstable 
waves are temporarily and spatially less uniform, leading to the large fluctuations in the kinetic energy curves.}
\label{fig:enbody}
\end{figure}

Figure \ref{fig:enbody} displays the evolution of the mean kinetic energy (\ref{eq:ekin}) for simulations with $\Omega_{z}=3.6$ and with optical depths 
$\tau=1.5$ and 
$\tau=2$ for different values of $\sigma_{0}$. These simulations were conducted in boxes of radial size $L_{x}=5\,\text{km}$ (cf.\ Table \ref{tab:nbstat}).
Similar to the results of the hydrodynamic computations we find that the kinetic energy $e_{kin}$ in the overstable oscillations drops with increasing 
$\sigma_{0}$. However, 
in contrast to hydrodynamics, this trend holds within a wider range of surface mass densities $\sigma_{0}$. 
Deviations 
from this monotonic behavior occur only for very small and very large values of $\sigma_{0}$. 
For very small nonzero $\sigma_{0}$ the spectral range of the developing nonlinear overstable modes is relatively wide. It extends to larger 
wavelengths than for the non-selfgravitating case, leading to an increased kinetic energy density. For high 
$\sigma_{0}$ the trend of a decreasing 
kinetic energy seems to level off. 
Similar to the hydrodynamic results (Figure \ref{fig:etauhydro}) this occurs at about $\sigma_{0}=700\,\text{kg}\,\text{m}^{-2}$ 
for $\tau=1.5$ and at a slightly larger value for $\tau=2$. 

\begin{figure}[h!]
\vspace{-0.2cm}
\centering
\includegraphics[width = 0.4 \textwidth]{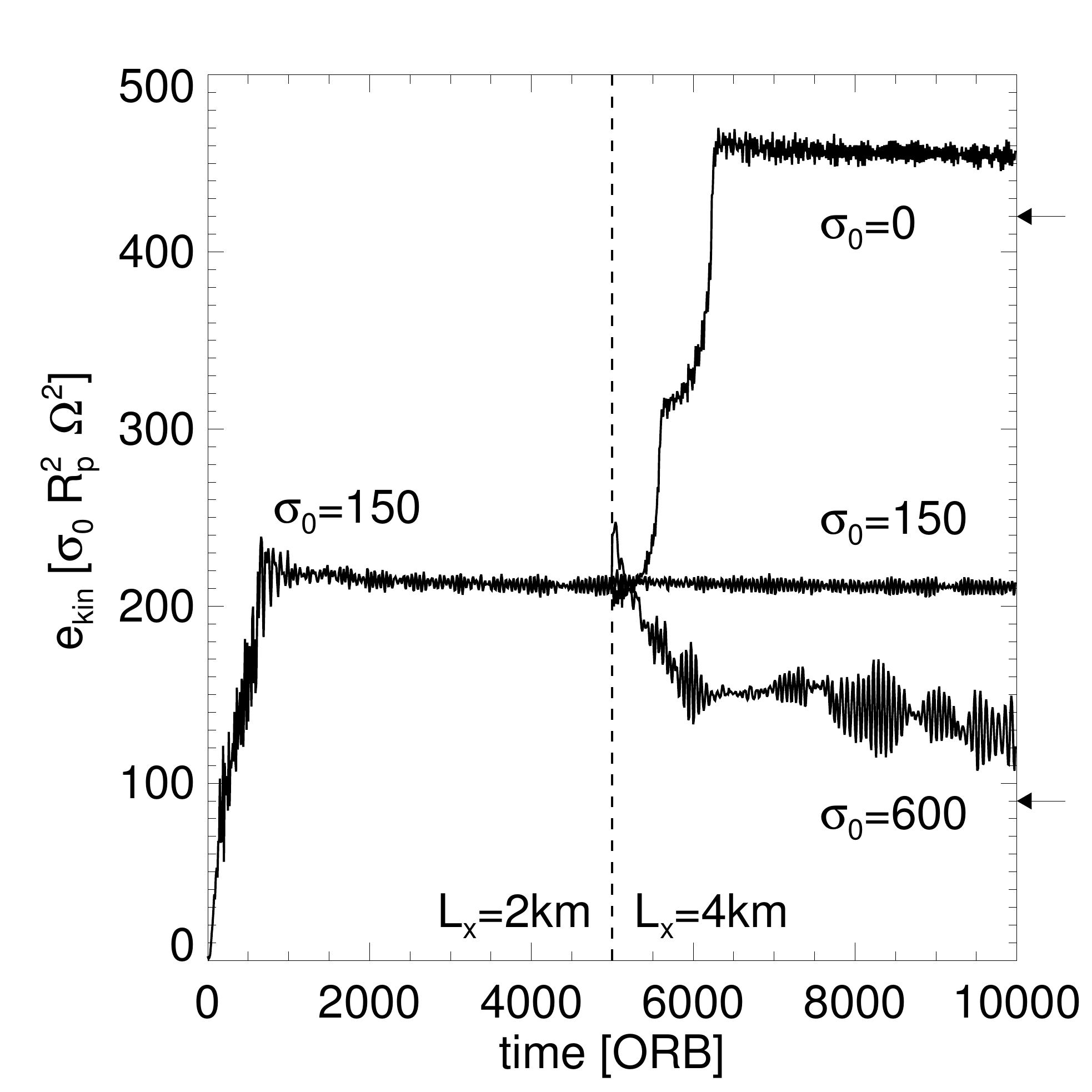}
\includegraphics[width = 0.4 \textwidth]{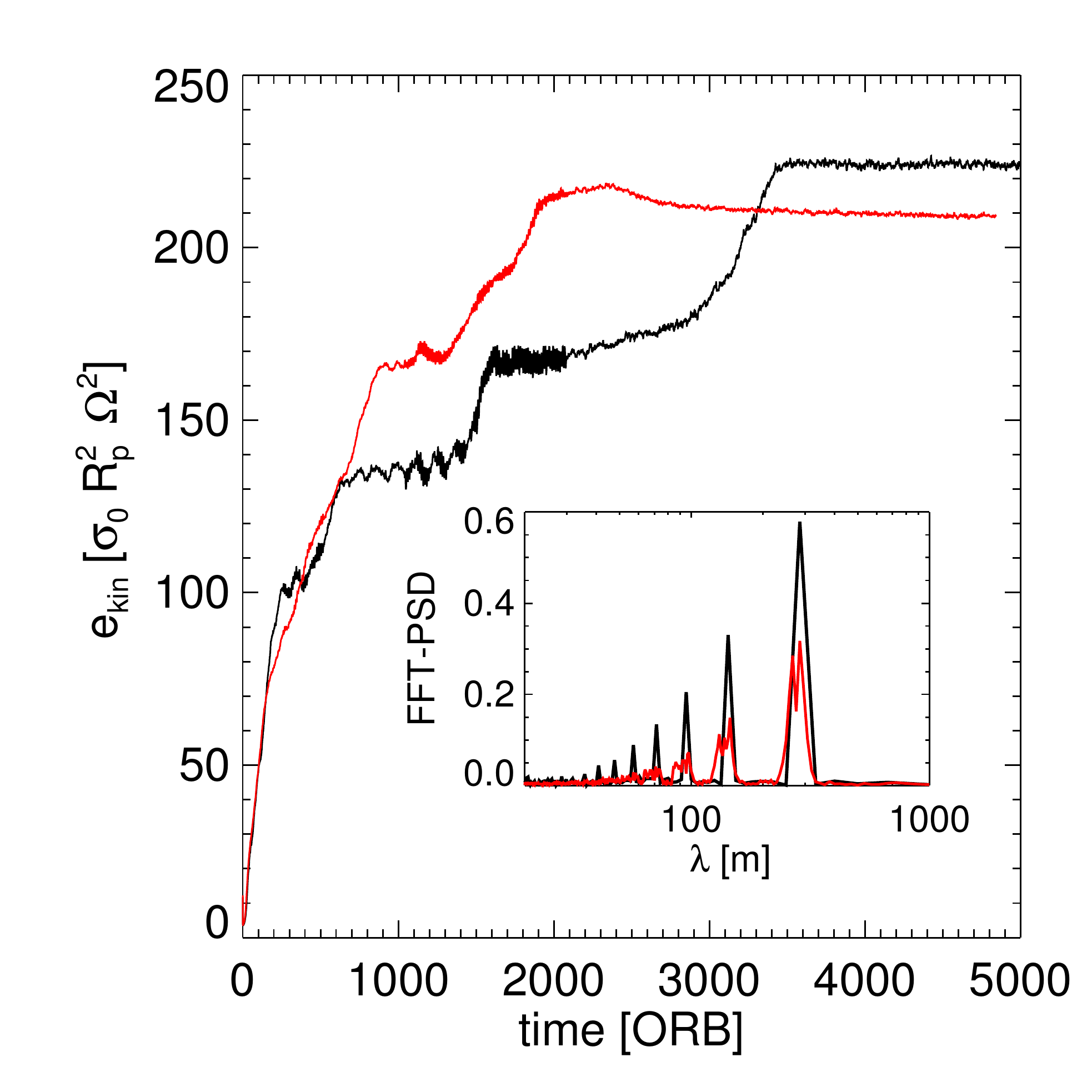}
\caption{Simulations conducted to explore possible effects of the finite box size. \emph{Left}: The curve for $t\leq 5,000\,\text{ORB}$ represents the kinetic 
energy density 
of a simulation with 
$\tau=1.5$, $\Omega_{z}=3.6$ and $\sigma_{0}=150\,\text{kg}\,\text{m}^{-2}$ in a box with $L_{x}=2\,\text{km}$.
The final state of this simulation at $t=5,000 \,\text{ORB}$ is taken as an initial state for three new simulations in larger boxes with $L_{x}=4\,\text{km}$. 
One of these 
simulations keeps the value of $\sigma_{0}=150\,\text{kg}\,\text{m}^{-2}$. The other two continue with $\sigma_{0}=0$ and 
$\sigma_{0}=600\,\text{kg}\,\text{m}^{-2}$, 
respectively.  The curve for $\sigma_{0}=150\,\text{kg}\,\text{m}^{-2}$ for  $t>5,000 \,\text{ORB}$ shows that the box size is not affecting the overstability. 
On the other hand, the other two runs begin to evolve in a manner consistent with the newly imposed surface mass density. The arrows indicate the final values 
of $e_{kin}$ in the simulations with $\sigma_{0}=0$ and $\sigma_{0}=600\,\text{kg}\,\text{m}^{-2}$ 
displayed in Figure \ref{fig:enbody}. \emph{Right}: Comparison of two simulations with $\tau=1.5$, $\Omega_{z}=3.6$ 
and $\sigma_{0}=300\,\text{kg}\, \text{m}^{-2}$ but different box sizes $L_{x}=2\,\text{km}$ and $L_{x}=8\,\text{km}$, respectively.  The insert plot 
represents 
the final power 
spectra of the radially tabulated particle surface number densities. The qualitative similarity of both runs underlines the consistency 
of 
results obtained with the two box sizes.}
\label{fig:boxdouble}
\end{figure}

We performed several tests to assure that the radial box size used for our N-Body simulations is sufficiently large (Figure \ref{fig:boxdouble}). We find that 
the box size is not affecting the outcome of the simulations.

\begin{figure}[h!]
\centering
\includegraphics[width = 0.4 \textwidth]{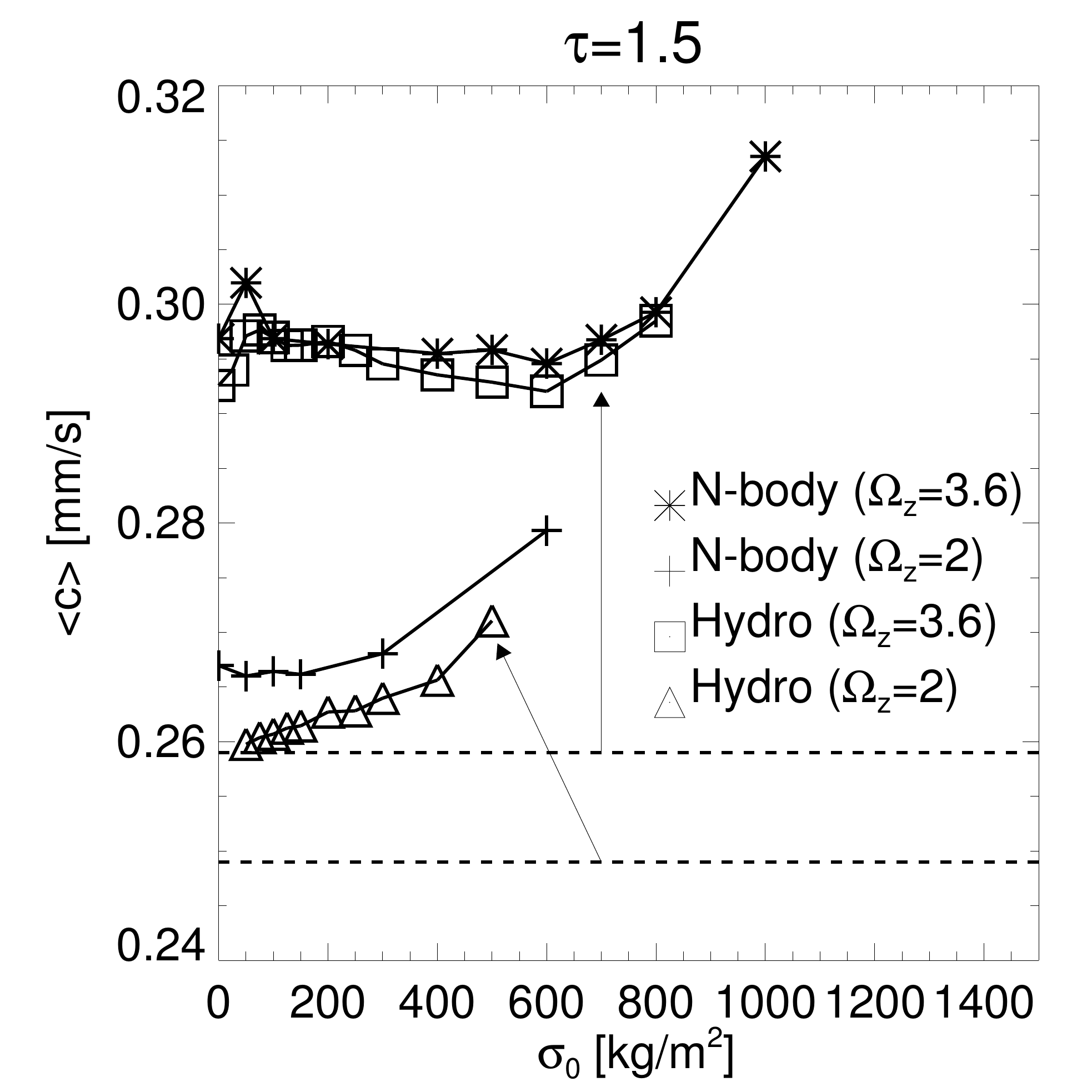}
\includegraphics[width = 0.4 \textwidth]{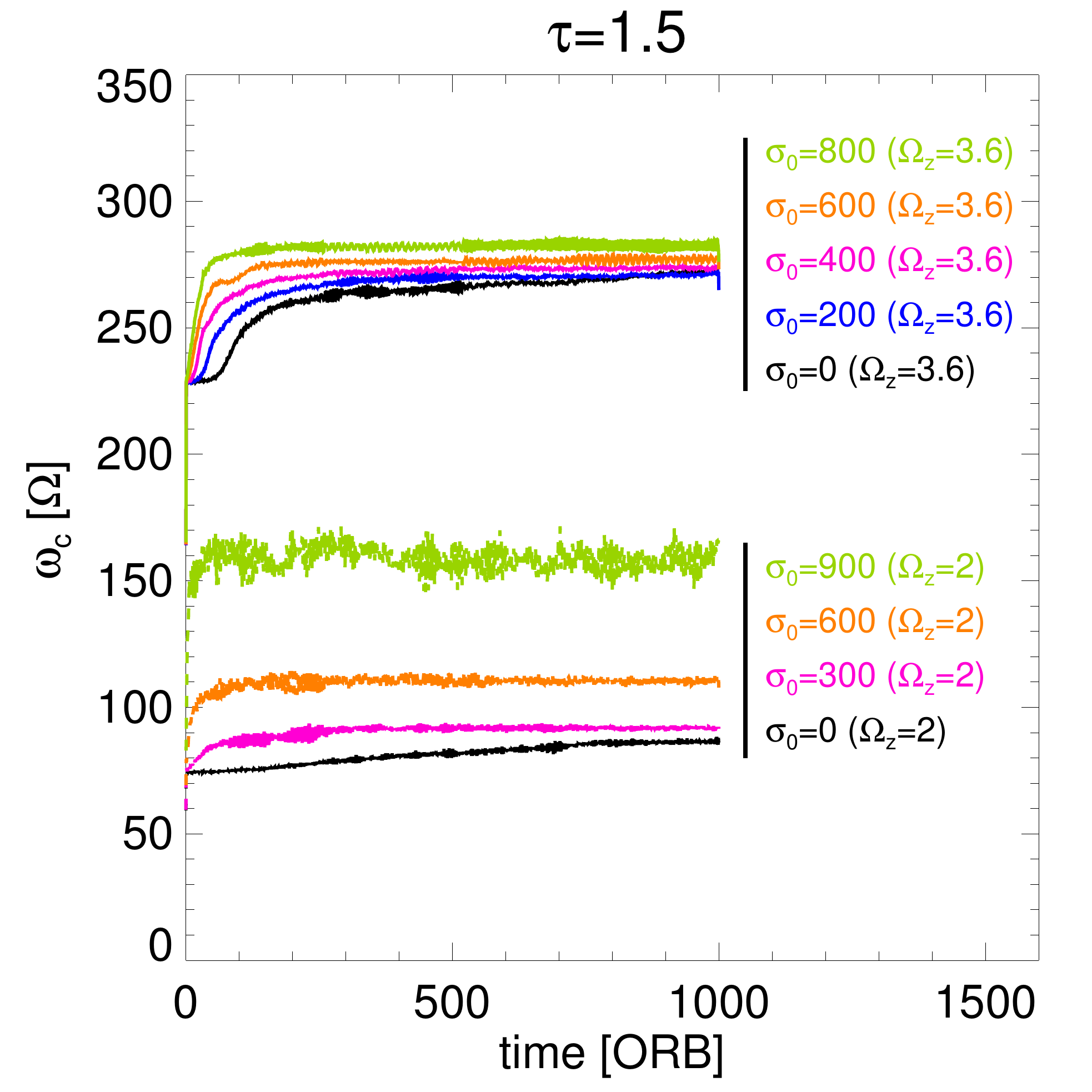}
\caption{\emph{Left}: The velocity dispersion of saturated overstable wave trains, averaged over the simulation box and a time interval of at least 100 orbits, 
as a function of the ground state surface density $\sigma_{0}$ for simulations with optical depth 
$\tau=1.5$ and different vertical frequencies $\Omega_{z}$. The simulations with $\Omega_{z}=3.6$ (asterisks) are the same as in Figure 
\ref{fig:enbody} (left panel), conducted in boxes with 
$L_{x}=5\,\text{km}$ and with a coefficient of 
restitution  $\epsilon=0.5$. For the simulations with $\Omega_{z}=2$ (plus symbols) we utilized $L_{x}=2\,\text{km}$ and $\epsilon=\epsilon_{b}$ [Equation 
(\ref{eq:bridges})]. Also plotted are results from corresponding hydrodynamic computations.
The hydrodynamic results for $\Omega_{z}=3.6$ (squares) are the same as in Figure \ref{fig:etauhydro} (left panel). The hydrodynamic results 
corresponding to $\Omega_{z}=2$ (triangles) are conducted with $L_{x}=10\,\text{km}$. The horizontal dashed 
lines indicate the ground state temperatures of the two parameter sets with $\tau=1.5$ (Table \ref{tab:hydropar}).
\emph{Right}: Evolution of the collision frequencies of the same simulations.}
\label{fig:vdispnb}
\end{figure}

In Figure \ref{fig:vdispnb} we compare final values of the velocity dispersion, averaged over the simulation box and time, denoted by $\langle c \rangle$.
Also shown are collision frequencies of simulations 
with different $\Omega_{z}$ and the same optical depth $\tau=1.5$, as a function of $\sigma_{0}$. 
Results of $\langle c \rangle$ from non-isothermal hydrodynamical computations, drawn for comparison, agree fairly well with the N-body simulation results.
The collision frequencies $\omega_{c}$ of the simulated systems are generally high due to the enhanced vertical frequency $\Omega_{z}$, and 
the growth of overstable modes 
even leads to further enhancement of up to some $20\%$. From the curves one may deduce that it is not the amount of energy ($e_{kin}$) contained in 
overstable oscillations but the magnitude of the radial self-gravity force which dictates the collision frequency. This is evidenced by the fact that in 
contrast to $e_{kin}$, the values of $\omega_{c}$ \emph{increase} with increasing $\sigma_{0}$.
This increase of the collision frequency in the nonlinear wave trains is not accounted for in the hydrodynamic model. 
Therefore, particularly for strong radial self-gravity it can be expected that in the nonlinear state of viscous overstability the hydrodynamic model 
underestimates collisional transport effects, in particular the nonlocal pressure modeled through Equation (\ref{eq:pres}).

Simulations with $\Omega_{z}=2$ exhibit significantly smaller collision frequencies than those with $\Omega_{z}=3.6$. This 
results in a smaller collisional momentum flux and thus a smaller value of the viscous parameter $\beta$ (cf.\ Section \ref{sec:theo}). 
The overstable wave trains found in systems with $\Omega_{z}=2$ have generally smaller amplitudes than those in systems with $\Omega_{z}=3.6$
and are less capable of heating up the system [Figure \ref{fig:vdispnb} (left panel)].

\FloatBarrier

\clearpage


\section{Saturation Wavelength of Viscous Overstability}\label{sec:lambdasat}

\begin{figure}[h!]
\centering
\includegraphics[width = 0.4 \textwidth]{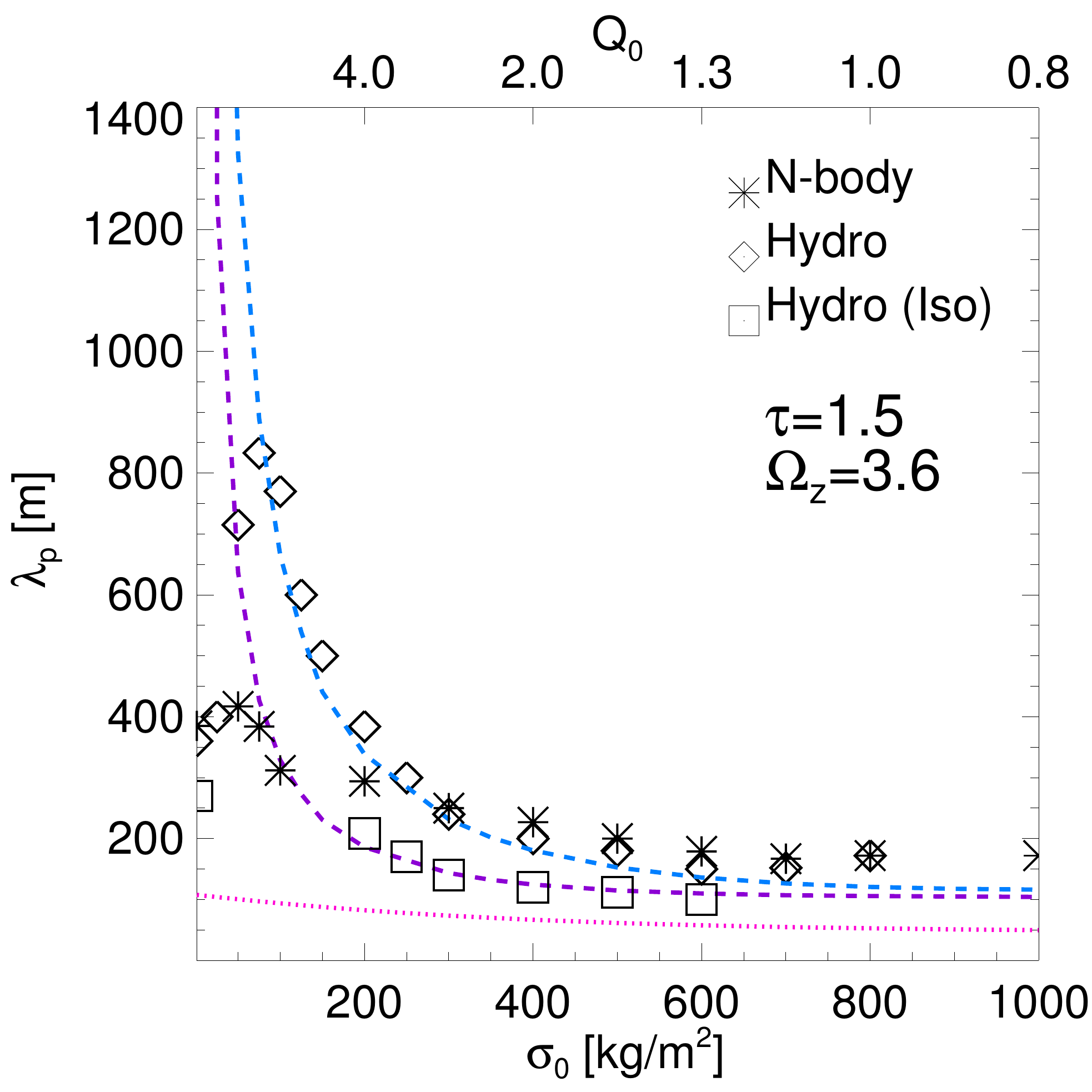}
\includegraphics[width =0.4 \textwidth]{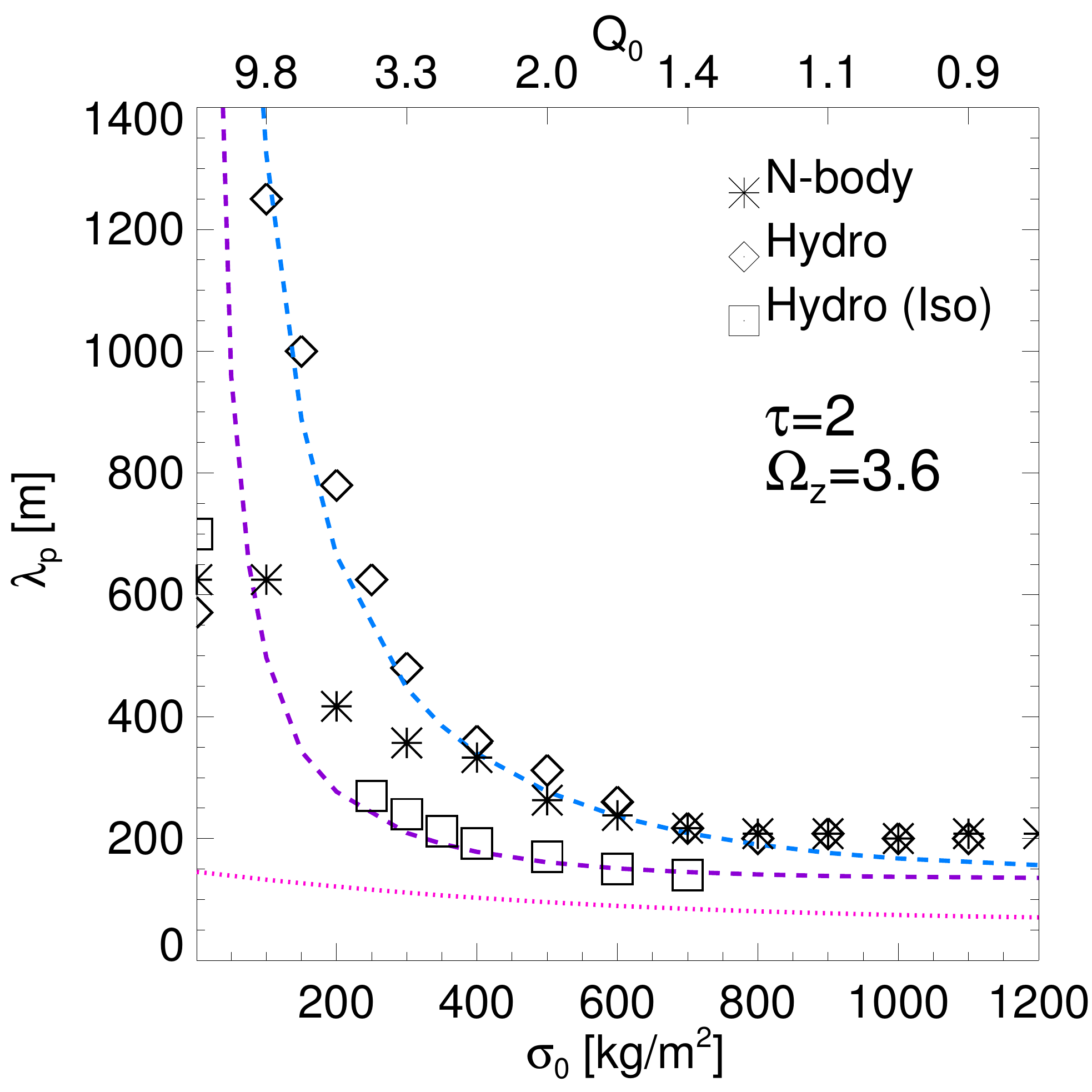}\\
\includegraphics[width = 0.4 
\textwidth]{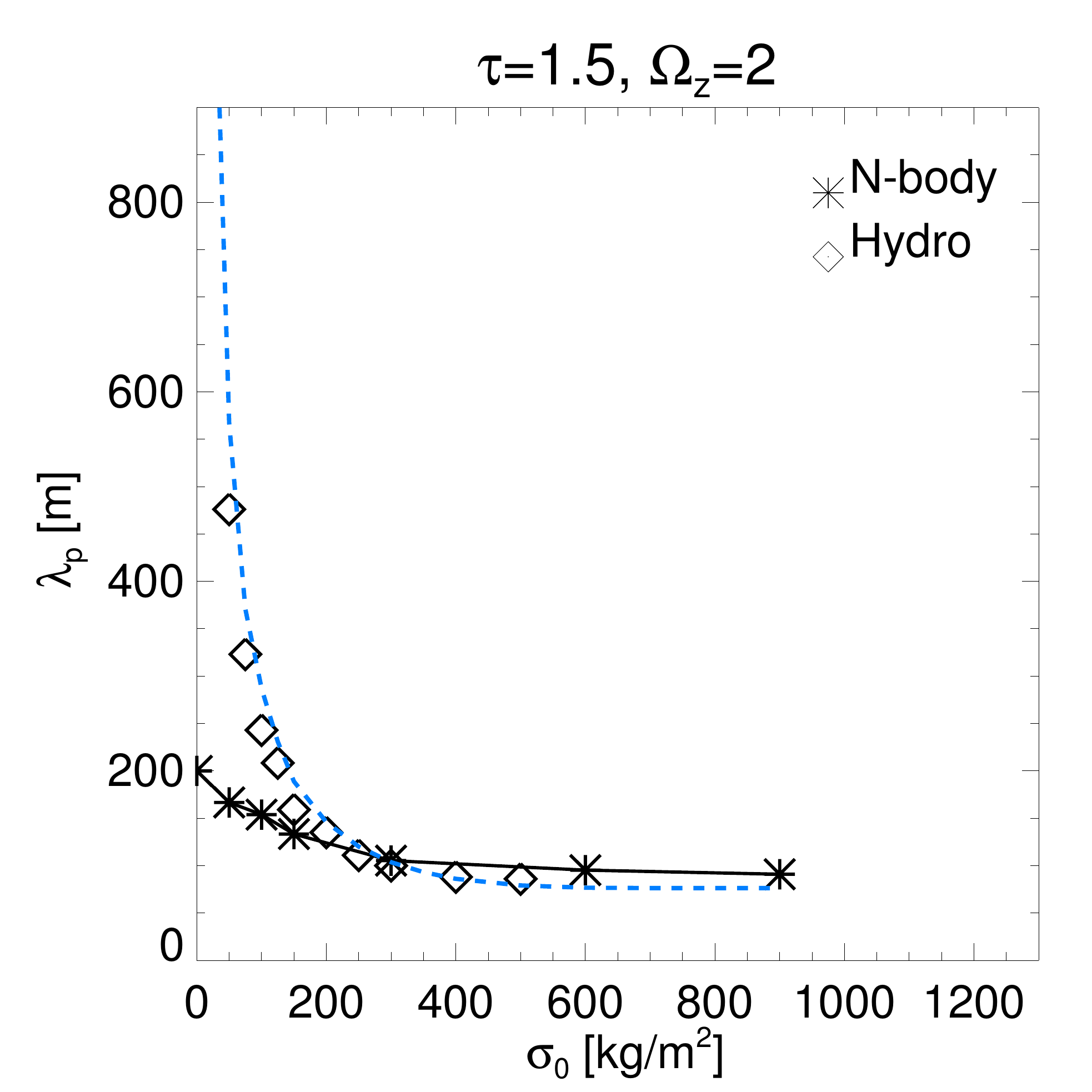}
\includegraphics[width =0.4 \textwidth]{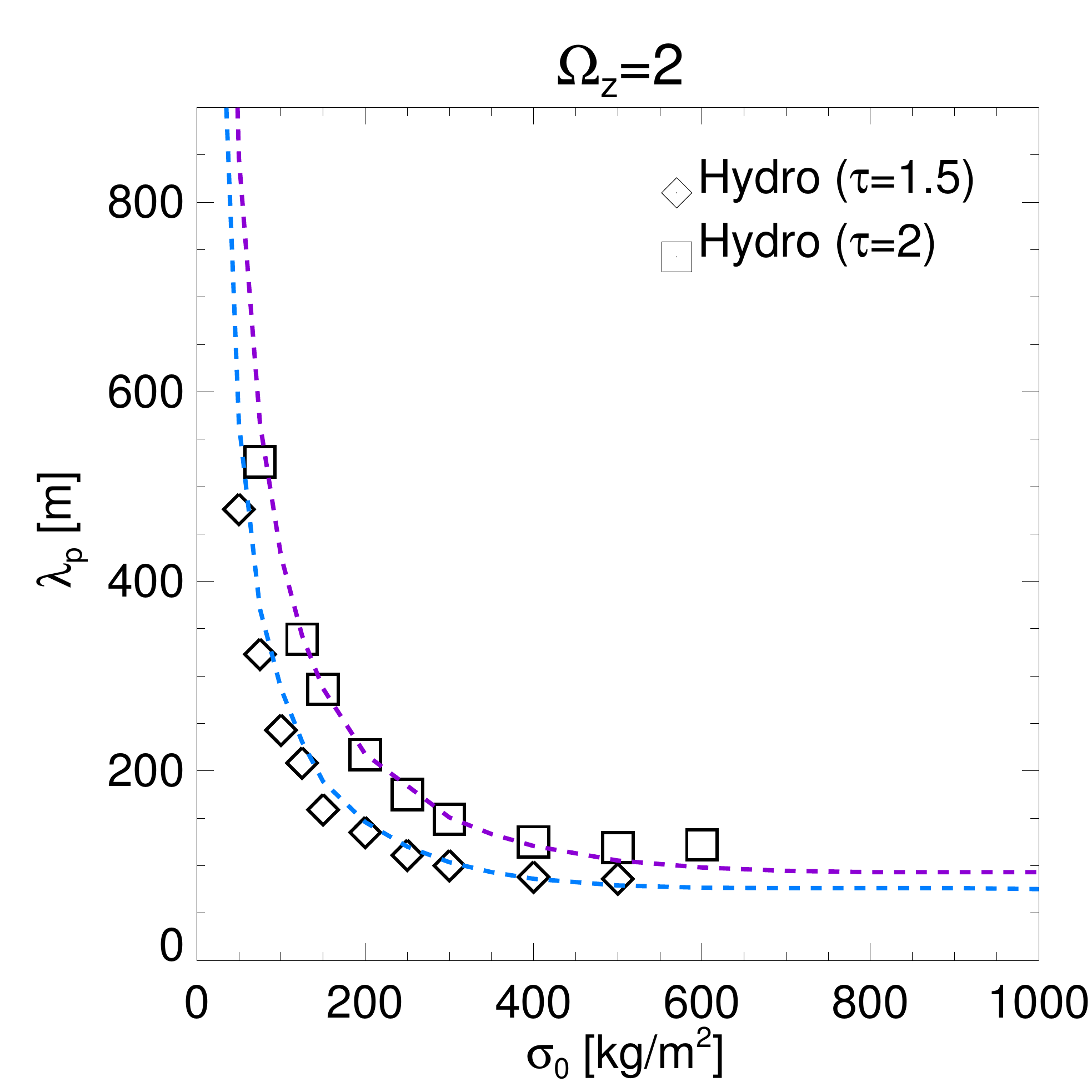}
\caption{Saturated, prevalent overstable wavelengths from N-body 
simulations and hydrodynamic non-isothermal (``Hydro'') and isothermal 
(``Iso'') models. In all frames the dashed curves are the empirical 
relation (\ref{eq:lambdasigma}) with $C=1.4$. For comparison the dotted curves in the 
upper frames represent the wavelengths of highest linear 
(non-isothermal) growth rates, following from Equation (\ref{eq:evp}). The isothermal results in the upper frames correspond to the limit $G_{T}\to \infty$ 
with $p_{s}=1$ (Section \ref{sec:theo}).}
\label{fig:lambdap}
\end{figure}

One important observable quantity is the wavelength of overstable 
oscillations in Saturn's rings, that establishes as the result of the 
long-term nonlinear evolution of the wave pattern. 
From the hydrodynamic 
models and the N-body simulations we define the final, saturated 
wavelength $\lambda_{p}$ (the subscript $p$ denoting \emph{prevalent}), 
as the wavelength with the largest Fourier amplitude in the saturated surface mass density field. Figure \ref{fig:lambdap} summarizes our 
results. Generally, $\lambda_{p}$ decreases with increasing surface mass 
density of the ring, until, for large $\sigma_0$, it settles 
on values that lie around $100-200\,\text{m}$, depending on the precise optical 
depth and the vertical frequency enhancement. 
At small surface densities $\sigma_{0}<200-300\,\text{kg}\,\text{m}^{-2}$ the saturated wavelengths from the N-body 
simulations deviate from the hydrodynamic ones, in that they connect 
smoothly to the wavelength that establishes in non-selfgravitating 
simulations. In the hydrodynamic models, in contrast, the saturated 
wavelengths rise to considerably larger values for small surface mass density, 
exceeding by far the wavelength of non-selfgravitating systems. We will 
return to a discussion of these deviations at small $\sigma_0$, as well as the 
behavior at large $\sigma_0$ in Sections \ref{sec:nldisp}, \ref{sec:per}.

For a wide range of intermediate surface mass densities the hydrodynamic 
prevalent wavelengths follow the simple empirical relation (dashed lines 
in Figure \ref{fig:lambdap})
\begin{equation}\label{eq:lambdasigma}
  \lambda_{p} = C  \lambda_{zero}(\sigma_{0}).
\end{equation}
Here, $\lambda_{zero}(\sigma_{0})$ is the wavelength of vanishing linear 
group velocity (Figure \ref{fig:vgroup}), which in turn corresponds 
to the slowest linear oscillation frequency [cf.\ Equation 
(\ref{eq:osfreq})]. We find that one single factor $C=1.4$ fits quite well all 
results for different optical depths and different enhancements of the 
vertical frequency. Non-isothermal and isothermal models both 
follow this trend with the same factor $C$, if one takes into account the difference in $\lambda_{zero}$ for the two cases. Also, the 
saturation wavelengths from the N-body simulations, in the range of 
surface mass densities $\sigma_{0}\gtrsim 300\,\text{kg}\,\text{m}^{-2}$, settle to attain values that are very similar to the non-isothermal
hydrodynamic ones. 

\FloatBarrier
\subsection{Nonlinear Dispersion Relation}\label{sec:nldisp}

The empirical relation (\ref{eq:lambdasigma}) between the nonlinearly 
saturated wavelengths and the wavelength of vanishing group velocity 
from the \emph{linear} dispersion relations encourages us to determine 
the \emph{nonlinear} dispersion relation of overstable oscillations and 
compare its curve to the saturated $\lambda_{p}$. To this end we extract from the hydrodynamic 
models the nonlinear frequency for a given wavelength 
$\lambda$ from integrations with a calculation region of size $L_{x}=\lambda$. 
To mimick radial self-gravity contributions from distant wave parts, we use an extended force kernel in the self-gravity calculation (Section 
\ref{sec:hydronumsg}).
In contrast to the measurements of the linear 
frequencies (Section \ref{sec:hydronumtests}) now a \emph{large} 
amplitude single wavelength mode is seeded so that only this 
mode saturates. Its evolution can be followed for at least 50 orbital 
periods, until, in cases where $\lambda$ is very large, modes with shorter wavelengths take over before the amplitude of the seeded mode is fully saturated. 
The frequency is then determined with a Lomb normalized periodogram, as 
for the linear frequencies. The nonlinear dispersion relation from 
N-body simulations is determined in a similar manner using a calculation box with $L_x=\lambda$ and applying a large amplitude sinusoidal initial seed for the 
$m=1$ 
oscillation mode (radial velocity amplitudes of the order of 
$10R_{p}\,\Omega$). The radial self-gravity (\ref{eq:sgmodes}) is 
calculated from all modes down to about $10\,\text{m}$, to assure that 
the nonlinear shape of the oscillating wave crests is resolved. As for 
the hydrodynamical model, the initial period during which the prominent 
mode adjusts its oscillation toward the final nonlinearly saturated state is 
excluded.

\begin{figure}[h!]
\vspace{-0.2cm}
\centering
\includegraphics[width = 0.42 
\textwidth]{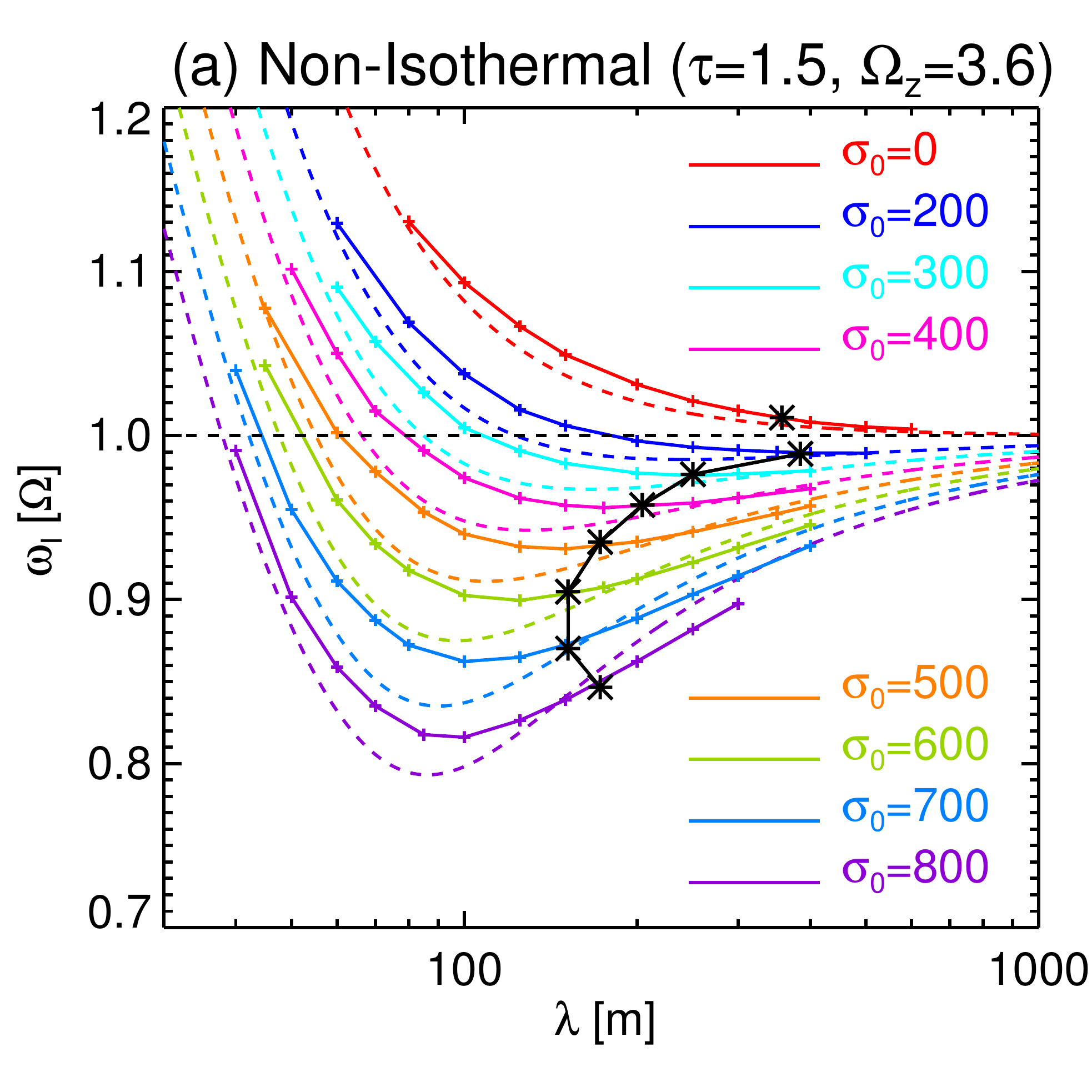}
\includegraphics[width =0.42
\textwidth]{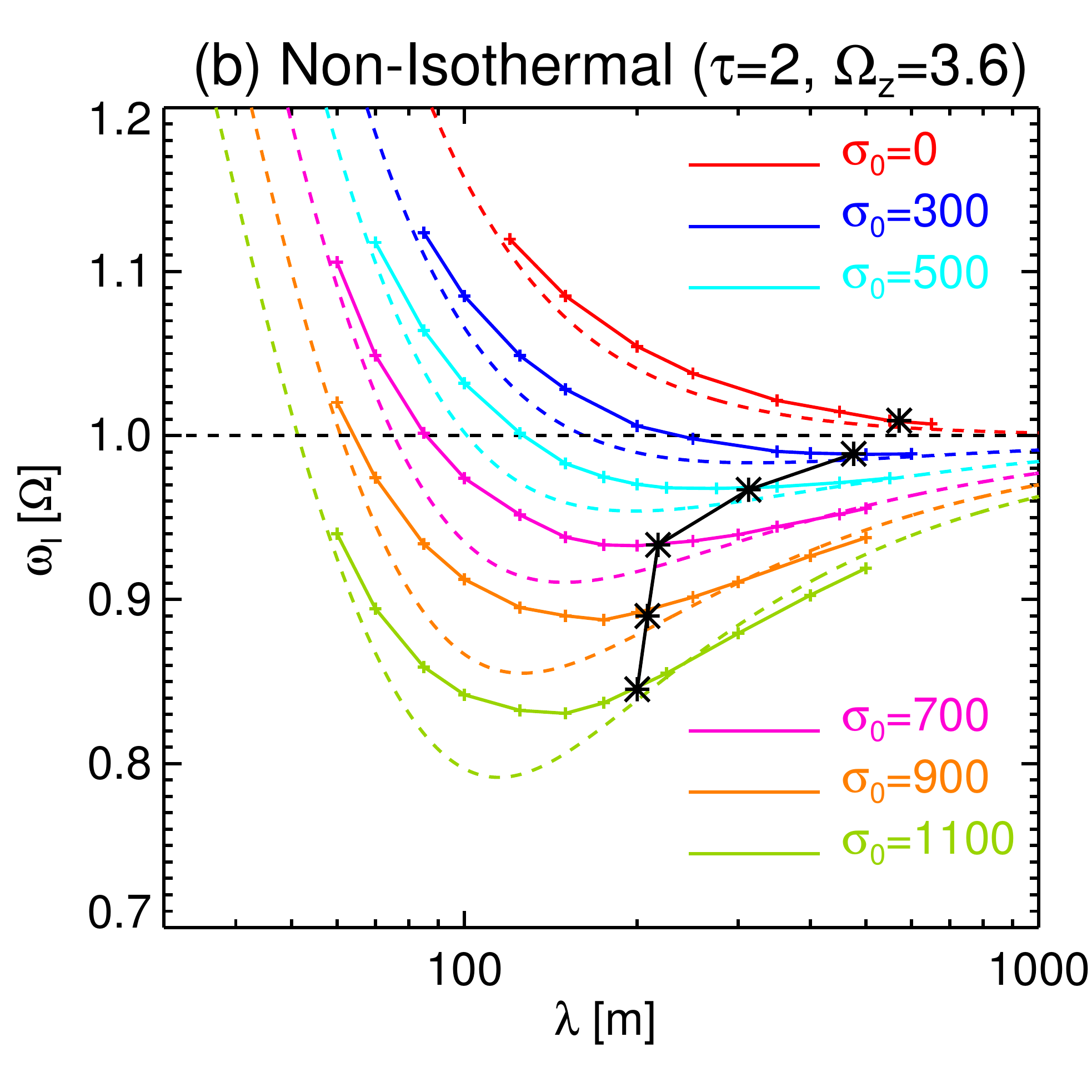}\\
\includegraphics[width = 0.42 
\textwidth]{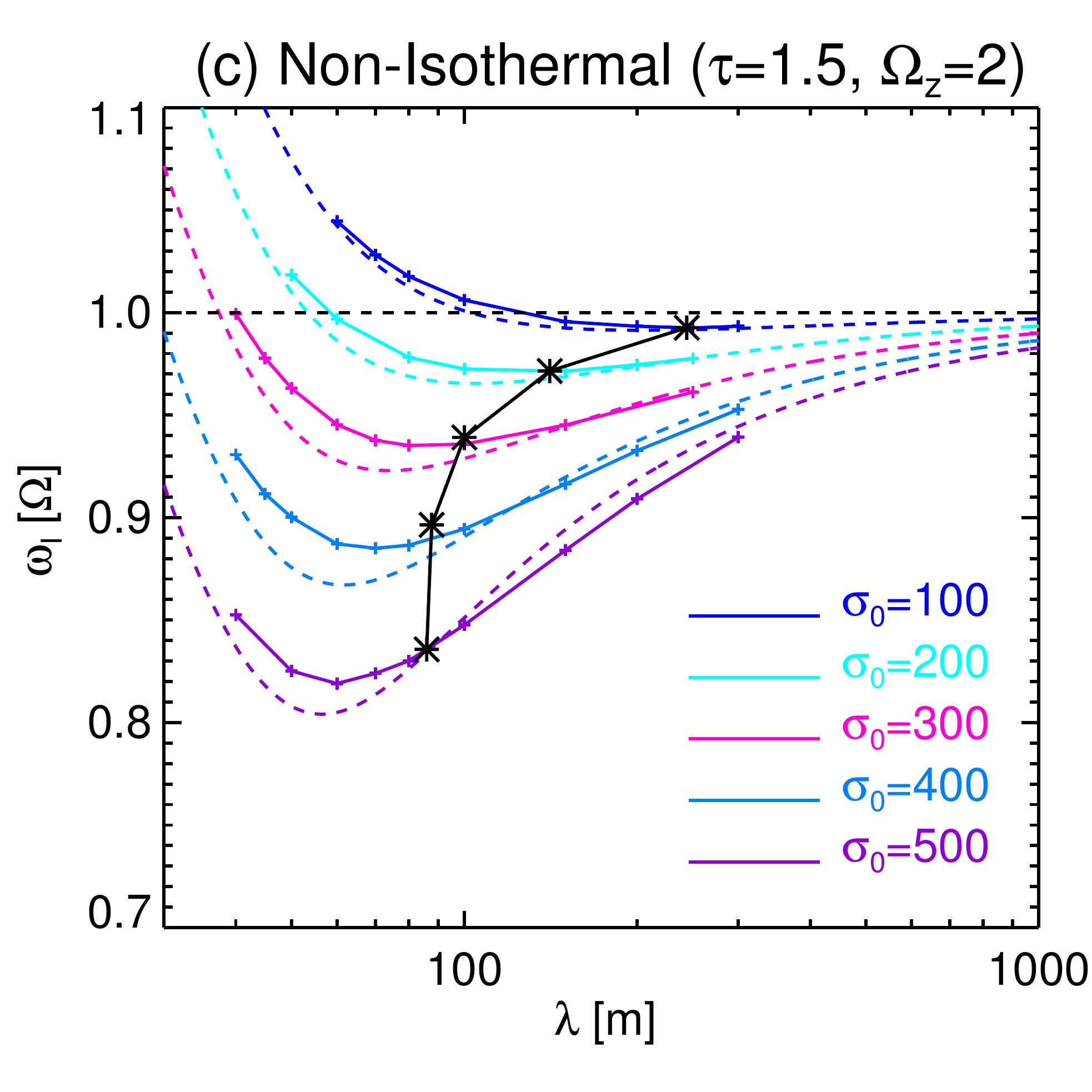}
\includegraphics[width =0.42 
\textwidth]{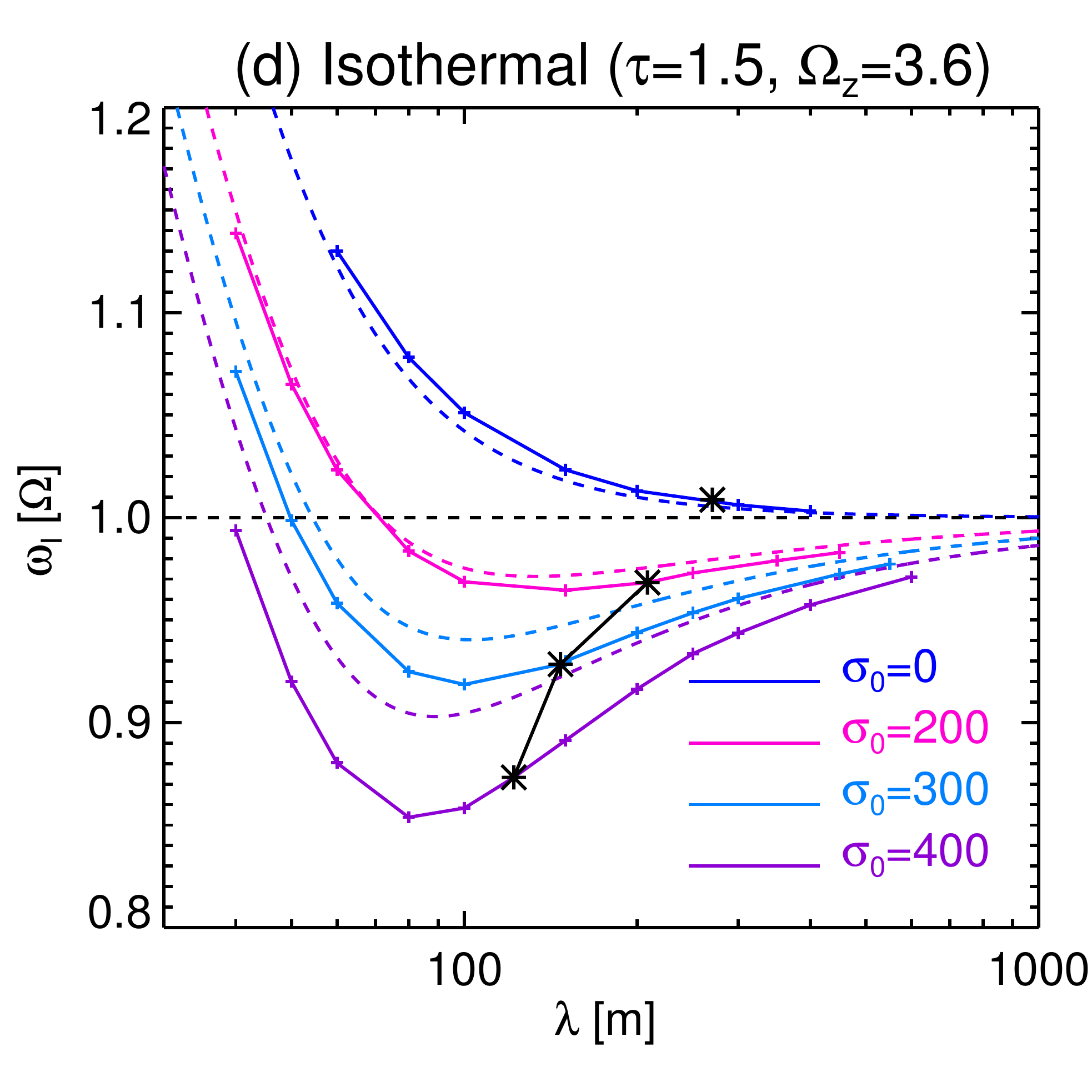}
\caption{Nonlinear hydrodynamic oscillation frequencies extracted from 
computations of saturated traveling waves of different wavelengths and 
for different
parameter sets in periodic domains with radial size corresponding to one wavelength. 
The solid curves represent different surface densities
$\sigma_{0}$.
The dashed curves are the linear oscillation frequencies 
from numerical solution of Equation (\ref{eq:evp}). The over-plotted 
asterisks represent
the
final states of large-scale runs from Figure \ref{fig:lambdap}.}
\label{fig:nlvg}
\end{figure}

The hydrodynamic nonlinear dispersion relations obtained in this manner are shown in Figure 
\ref{fig:nlvg} for the $\tau_{15}$ and the $\tau_{20}$-parameters (upper 
frames) and additionally for a smaller vertical frequency $\Omega_{z}=2$ 
and an isothermal system (lower frames). Generally, we find that for 
self-gravitating systems the wavelength of minimal frequency is shifted by nonlinearity to larger values. 
For the non-isothermal model (panels a-c) also the 
minimal frequency itself is larger, compared to the linear dispersion 
relation (dashed curves). For non-selfgravitating systems such a nonlinear shift, 
attributed to the action of pressure, was already noted by LO2009 (see 
their Equations (33) and (34) as well as their Figure 4). 
Also thermal effects alter the nonlinear frequencies, mitigating the pressure-related increase.
At larger $\lambda$ the curves for the linear and nonlinear dispersion relations 
cross. At these large wavelengths self-gravity begins to dominate the 
deviation of the oscillation frequency from the Keplerian value. 

The nonlinear frequency reduction due to self-gravity is analogous to the 
nonlinear wavenumber reduction found for resonant spiral density waves 
in a dense ring, for which pressure forces play a minor role 
(\citet{shu1985a};~\citet{lehmann2016}). 
In the isothermal model (panel d), with the ideal 
gas equation of state, the presence of any substantial self-gravity 
force results in a nonlinear reduction of the oscillation frequencies. 

One notes that all nonlinear frequency curves converge to the linear 
curves at small $\lambda$, since the saturation amplitudes of the wave 
trains scale linearly with $\lambda$ (SS2003, LO2009) and nonlinear 
effects eventually vanish as the wavelengths approach the linear 
stability boundary.  For sufficiently long wavelengths the effects 
of pressure and self-gravity vanish and the nonlinear curves also approach 
the linear limit.

We find that the prevalent wavelengths in the large-scale, self-gravitating 
hydrodynamic models (asterisk symbols in Figure \ref{fig:nlvg}) depend in a 
systematic manner on the minimum of the corresponding nonlinear 
frequency curve. 
This suggests that nonlinearity accounts for much of the deviation from unity of the factor $C$ in the empirical relation 
(\ref{eq:lambdasigma}). For instance, the saturation wavelengths of integrations with  
 the $\tau_{15}$-parameters and $\sigma_{0}\leq 600\,\text{kg}\,\text{m}^{-2}$ (excluding the case $\sigma_{0}=0$) are very close to the minimum of 
the 
nonlinear dispersion relation [panel (a) in Figure 
\ref{fig:nlvg}]. This implies very small group velocities [Equation (\ref{eq:vg})] of the saturated wave trains (see Figure \ref{fig:vgpert} in Appendix 
\ref{sec:vgroup}). However, with increasing $\sigma_{0}$ the values of $\lambda_{p}$ gradually depart 
from the minimum toward larger wavelengths for all displayed cases in Figure \ref{fig:nlvg}. This apparent inconsistency will be resolved in Section 
\ref{sec:per}.
\FloatBarrier

\FloatBarrier

\begin{figure}[ht]
\vspace{-0.2cm}
\centering
\includegraphics[width = 0.45 \textwidth]{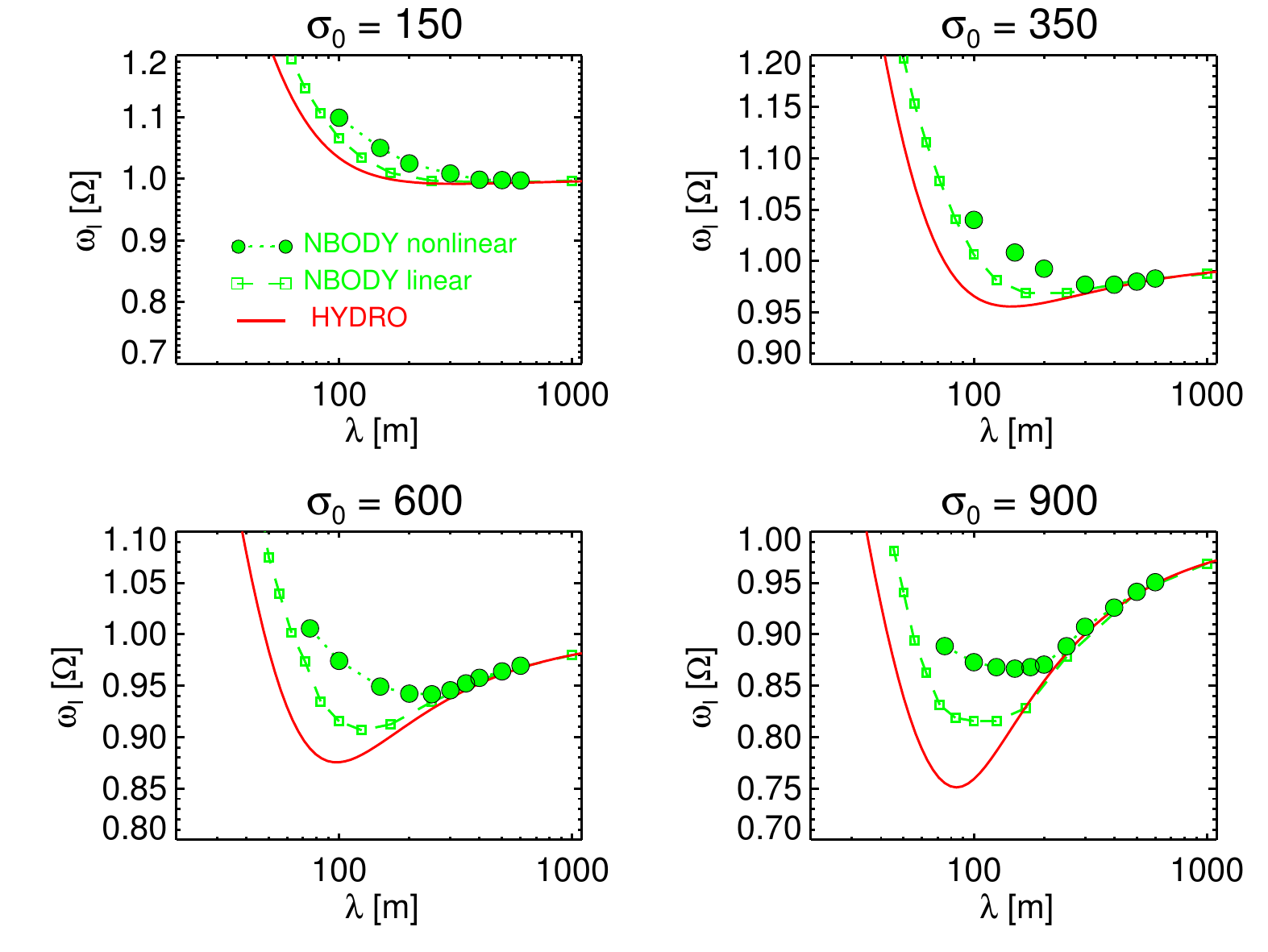}
\includegraphics[width = 0.45 \textwidth]{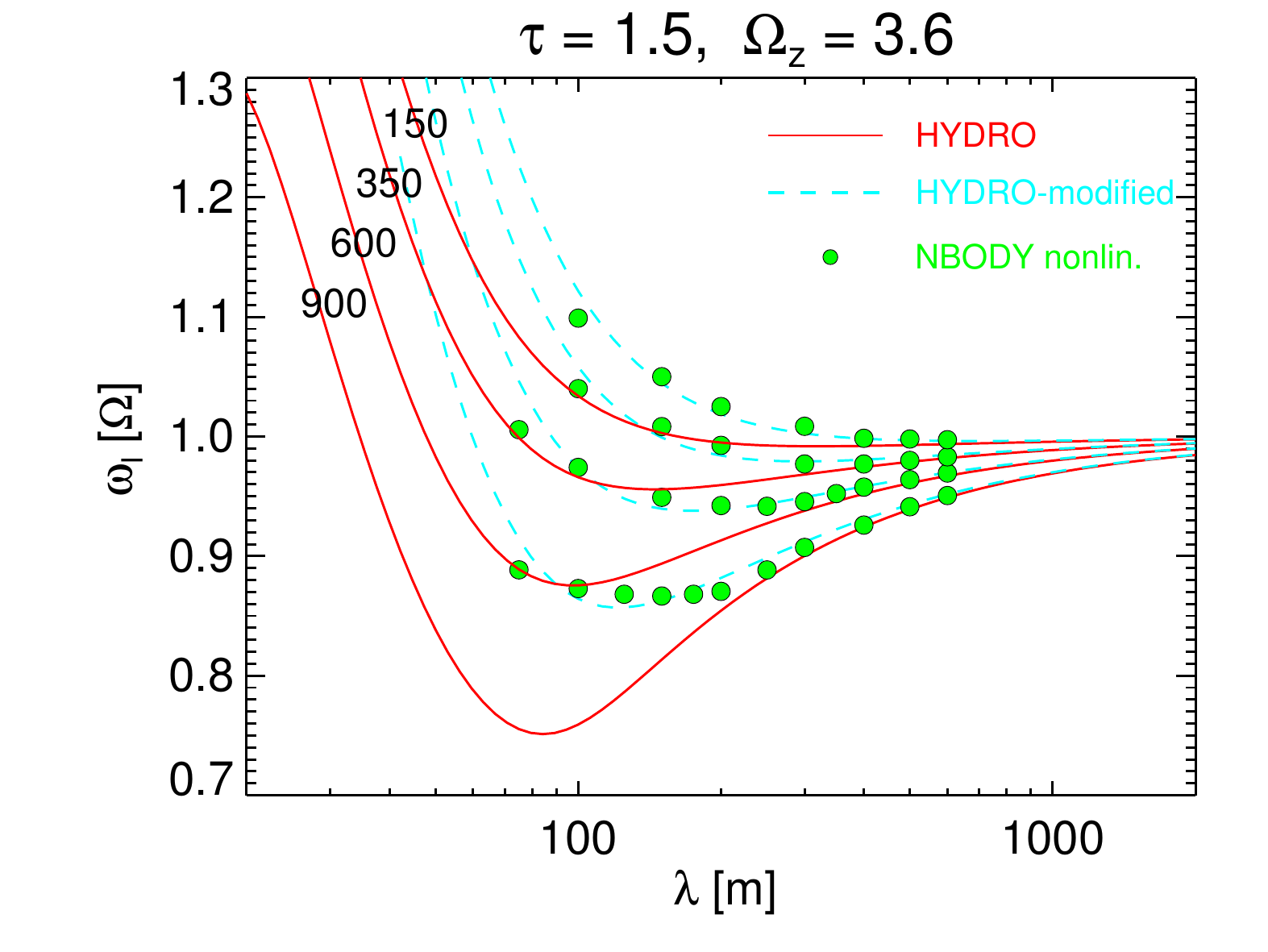}
\caption{Oscillation frequencies in N-body simulations performed with 
four different surface densities ($\sigma_0 = 150, \ 350, \ 600, \ 
900\,\text{kg}\, \text{m}^{-2}$). All simulations used $\tau=1.5$, 
$\Omega_z=3.6$, and relation (\ref{eq:bridges}) for $\epsilon(v)$. 
\textit{Left}: predictions of the linear, non-isothermal hydrodynamical model are 
indicated with solid red curves, while the symbols stand for N-body 
simulations, with frequencies measured both during the linear growth 
period (open boxes), and from the final nonlinear saturated state 
(filled circles). 
\textit{Right}: the nonlinear oscillation frequencies measured from N-body
simulations performed with different surface densities are compared to the
linear hydrodynamical prediction (red solid lines), as well as a
\emph{modified} hydrodynamical model (blue dashed lines), which corresponds
to a 2-fold value of the pressure coefficient $p_s$, keeping all other quantities unchanged.}
\label{fig:nbvga}
\end{figure}
In the N-body simulations the nonlinear dispersion relation assumes 
generally larger frequencies than the linear one (Figure 
\ref{fig:nbvga}, left panel). The latter, determined from simulations with 
small-amplitude overstable waves, has in turn larger frequencies than 
the linear dispersion relation predicted by the hydrodynamic model. Moreover, for the N-body 
simulations the minimum of the nonlinear oscillation frequency is shifted towards larger 
wavelengths, as is the case for the hydrodynamic nonlinear dispersion 
relation (cf.\ Figure \ref{fig:nlvg}).
Much of the difference between the 
hydrodynamic model and the N-body simulations can be attributed to the 
altered action of pressure in the nonlinearly saturated state. That is, 
if we modify the pressure coefficient $p_{s}$ of the hydrodynamic model by a 
factor of two, the hydrodynamic linear dispersion relation matches very well the nonlinear dispersion relation from N-body simulations for a wide range of 
surface mass densities (Figure \ref{fig:nbvga}, right panel). Recall that a modification of $p_{s}$ by a factor of 1.4
led to a similarly good agreement between the \emph{linear} frequencies and growth rates (see Section \ref{sec:nbtest} and Appendix \ref{sec:pmod}).

\begin{figure}[h!]
\centering
\includegraphics[width = 0.5 \textwidth]{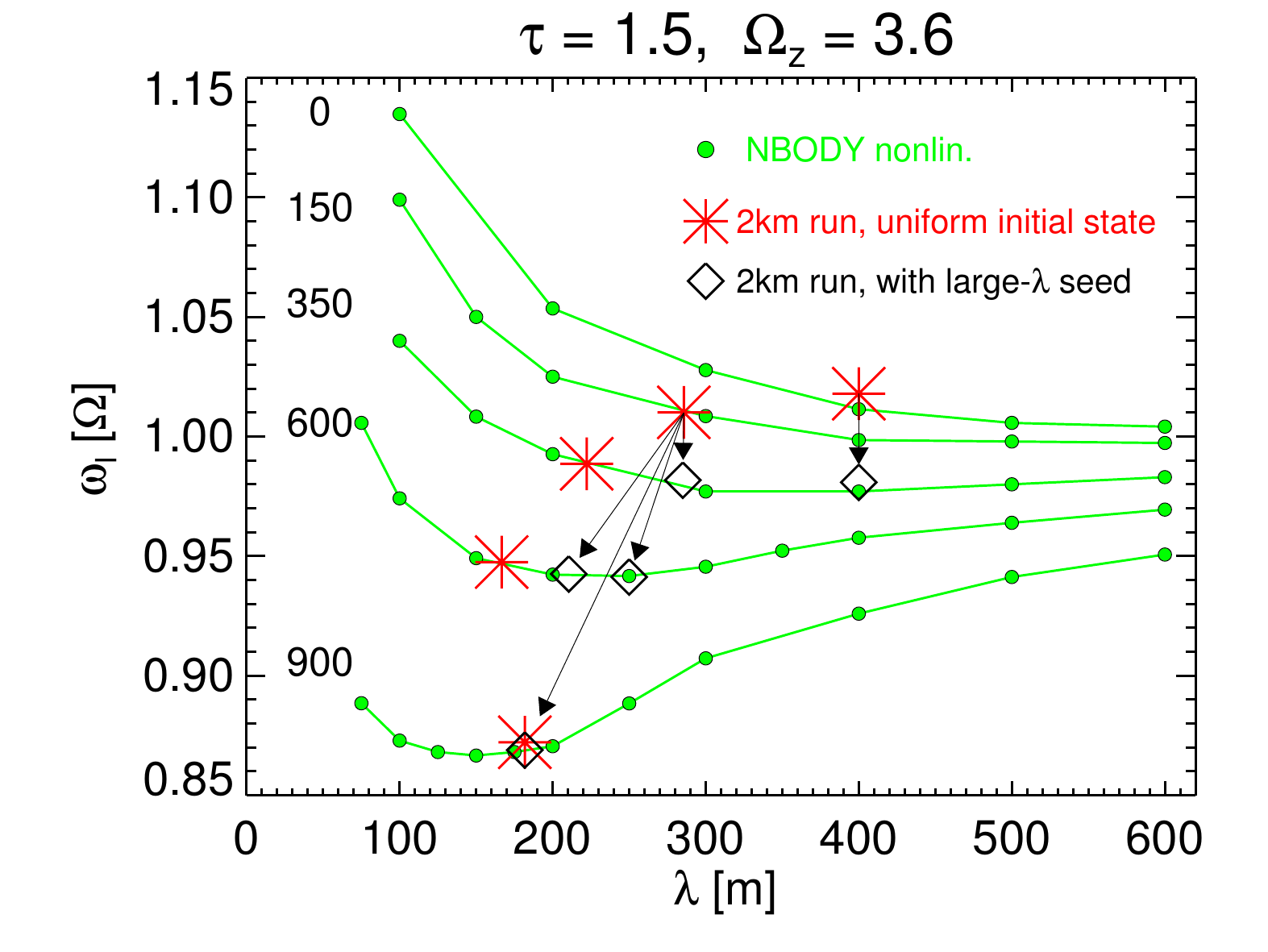}
\caption{Nonlinear oscillation frequencies as a function of wavelength 
from N-body simulations (solid curves with circles). Large asterisk 
symbols mark the wavelength and frequency of the saturated state at the 
end of simulations ($t\sim 5,000\,\text{ORB}$) with 
$L_x=2\,\text{km}$ where overstability evolved from an initially 
uniform state. Although there are no obvious signs of further wavelength 
growth, it cannot be excluded that this might happen on much longer 
timescales. The diamond symbols indicate alternative final saturated 
states, which are stable on the timescales accessible to the 
simulations. These have evolved from different initial states as 
indicated by the arrows (see text). }
\label{fig:nbvgc}
\end{figure}

Also the prevalent wavelengths from large-scale N-body simulations 
($L_{x}\geq 2\text{km}$) lie close to the minimum of the nonlinear oscillation 
frequency in the self-gravitating systems (Figure \ref{fig:nbvgc}), reminiscent of the empirical 
relation (\ref{eq:lambdasigma}). 
The group velocity of the waves 
vanishes at this frequency minimum and therefore any spacial variations 
in the wave pattern will not propagate anymore. For this reason the 
interaction of spatially separated wave states, with slightly different 
properties, will become weaker, and ultimately vanish, when the 
wavelength approaches the value of zero group velocity. We expect that 
in the limit $t\rightarrow\infty$ the simulations evolve towards this 
wavelength. 

On the timescales 
accessible to the numerical exploration (thousands to ten thousands of 
orbits) we expect a belt of quasi-stable wavelengths 
around this minimum, which practically do not evolve. To explore this possibility (see also Appendix \ref{sec:altsat}) we perform simulations where a 
non-sinusoidal large-$\lambda$ seed is employed, with a wavelength very close 
to the frequency minimum of the nonlinear dispersion relation. In 
practice the initial positions and velocities of the simulation 
particles for this seeded state are taken from the final state of a simulation with smaller surface mass density and a larger 
saturated $\lambda$ (the starting points of the arrows in Figure 
\ref{fig:nbvgc}). The diamond symbols mark the alternative final 
states reached in these new simulations (end points of the arrows). 
In this way arrows in the figure indicate the evolution of the prevalent wavelengths in the new, 
large-scale simulations. For $\sigma_{0}=600\,\text{kg}\,\text{m}^{-2}$ 
we find that two simulations with different size ($L_x=2$ and 
$4\,\text{km}$) reach slightly different final states with 
$\lambda_{p}=250\,\text{m}$ and $\lambda_{p}=210\,\text{m}$, respectively, both 
being very close to the minimum of the overstable oscillation frequency. 

In the limit of vanishing surface mass density the 
wavelength of the minimum oscillation frequency tends to very large values, theoretically 
approaching infinity. But especially for small surface mass 
density the minimum becomes very shallow. 
The particle flow in N-body simulations is subject to a variety 
of fluctuations (small non-axisymmetries, low contrast particle 
clumping, variations in the scale height). Consequently, there 
exists a certain threshold surface mass density below which the system does 
not feel the effect of the minimum anymore. We believe that this is the 
reason for the deviation in the saturation wavelength of the N-body 
simulations and the hydrodynamic systems for small surface densities (Figure \ref{fig:lambdap}). In 
the idealized hydrodynamic models the wavelength follows even in the 
small $\sigma_0$ limit the curve implied by relation 
(\ref{eq:lambdasigma}), formally diverging for vanishing surface mass 
density, with a non-smooth transition to the finite saturated wavelength 
of the non-selfgravitating case. The prevalent wavelengths of the N-body 
simulations, in contrast, converge monotonically to the 
non-selfgravitating value. The deviations of the hydrodynamic models 
from relation (\ref{eq:lambdasigma}) at very small (but non-zero) 
surface mass densities we attribute to the influence of the finite size 
of the computational domain ($L_{x}$), which becomes important for these very large 
wavelengths (see  Appendix \ref{sec:boxsize}). The integration with $\sigma_{0}=0$ in Appendix \ref{sec:boxsize} demonstrates that in the 
non-selfgravitating case stable uniform traveling wave 
solutions exist for all wavelengths larger than some critical value, as has been shown by LO2009 for the isothermal limit.

We note that for small surface densities $\sigma_{0}\lesssim 
150\,\text{kg}\,\text{m}^{-2}$ the applied frequencies of vertical motions 
$\Omega_{z}= 2-3.6$ imply unrealistic disk scale heights smaller than 
one particle radius [cf.\ (\ref{eq:wtfac})]. When using a 
self-consistent self-gravity implementation, systems with such small 
surface densities would not be viscously overstable.
Furthermore, for sufficiently large $\sigma_{0}$ ($Q_{0}$ well below unity), the 
N-body simulations exhibit a gravitational instability, while the 
hydrodynamical scheme runs into numerical instability.

 \FloatBarrier

\subsection{Hydrodynamical Integrations with a Buffer-Zone}\label{sec:per}

In Section \ref{sec:nldisp} we found that the prevalent wavelength $\lambda_{p}$ of large-scale hydrodynamic integrations and N-body simulations closely 
follows the wavelength corresponding to the frequency minimum of the nonlinear dispersion relation of overstable waves. However, with 
increasing surface mass density $\sigma_{0}$ the hydrodynamic values of $\lambda_{p}$ are increasingly displaced from the minimum towards larger wavelengths 
(asterisk symbols in Figure 
\ref{fig:nlvg}).

\begin{figure}[h!]
\vspace{-0.2cm}
\centering
\includegraphics[width = 0.45 
\textwidth]{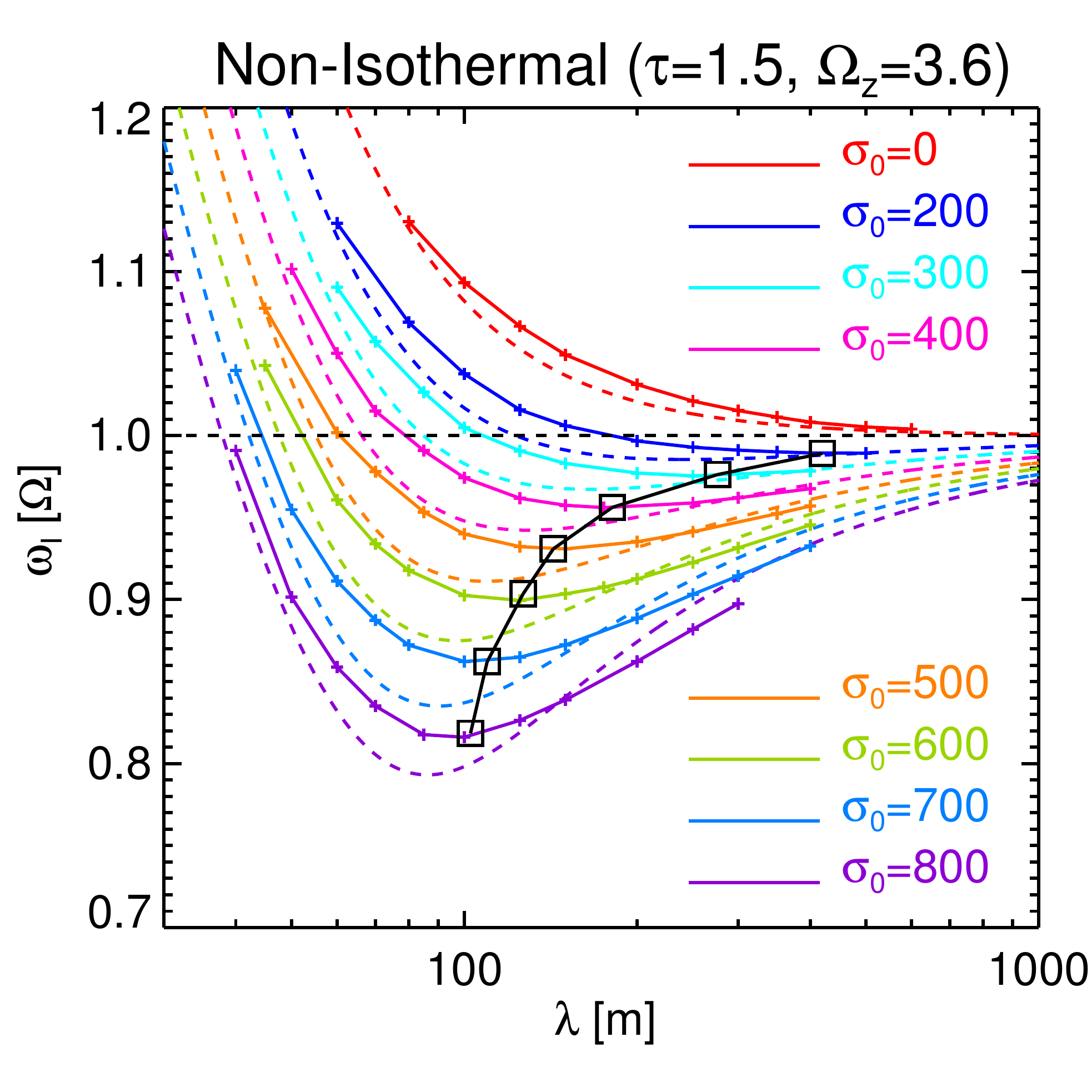}
\caption{The same as figure \ref{fig:nlvg}a with the difference that now the symbols (open squares) mark the saturation wavelengths obtained from hydrodynamic 
integrations which started from low amplitude white noise and contained a buffer-zone in the calculation region (see also Figure \ref{fig:osfrat}).}
\label{fig:nlvgbuff}
\end{figure}

Here we check for a possible artificial influence of the periodic boundary conditions on the nonlinear saturation mechanism and the resulting 
wavelength $\lambda_{p}$. Therefore we perform hydrodynamic integrations, starting from spectral white noise with the $\tau_{15}$-parameters and 
various surface densities $200\, \text{kg}\,\text{m}^{-2} \leq \sigma_{0}\leq 800\, \text{kg}\,\text{m}^{-2}$, where we include a buffer-zone in the 
computational 
domain ($L_{x}=6-8\,\text{km}$). For an example of such an integration see Figure \ref{fig:sig500buff} in Appendix \ref{sec:altsat}. As 
outlined before (Section \ref{sec:isonosg}), the buffer-zone prevents the system from settling on a uniform nonlinear wave train and should eliminate possible 
spurious effects on the long term evolution, provoked by the 
periodic boundary conditions (\citet{latter2010}).
The saturation wavelengths we find for these integrations are for all values of $\sigma_{0}$ close to the 
nonlinear frequency minimum (Figure \ref{fig:nlvgbuff}), in a more 
consistent manner than the values found in homogeneous periodic integration domains.
In conclusion, the presence of a buffer-zone shifts the saturation wavelength closer to the frequency minimum of the nonlinear dispersion relation.

\begin{figure}[h!]
\centering
\includegraphics[width = 0.5 \textwidth]{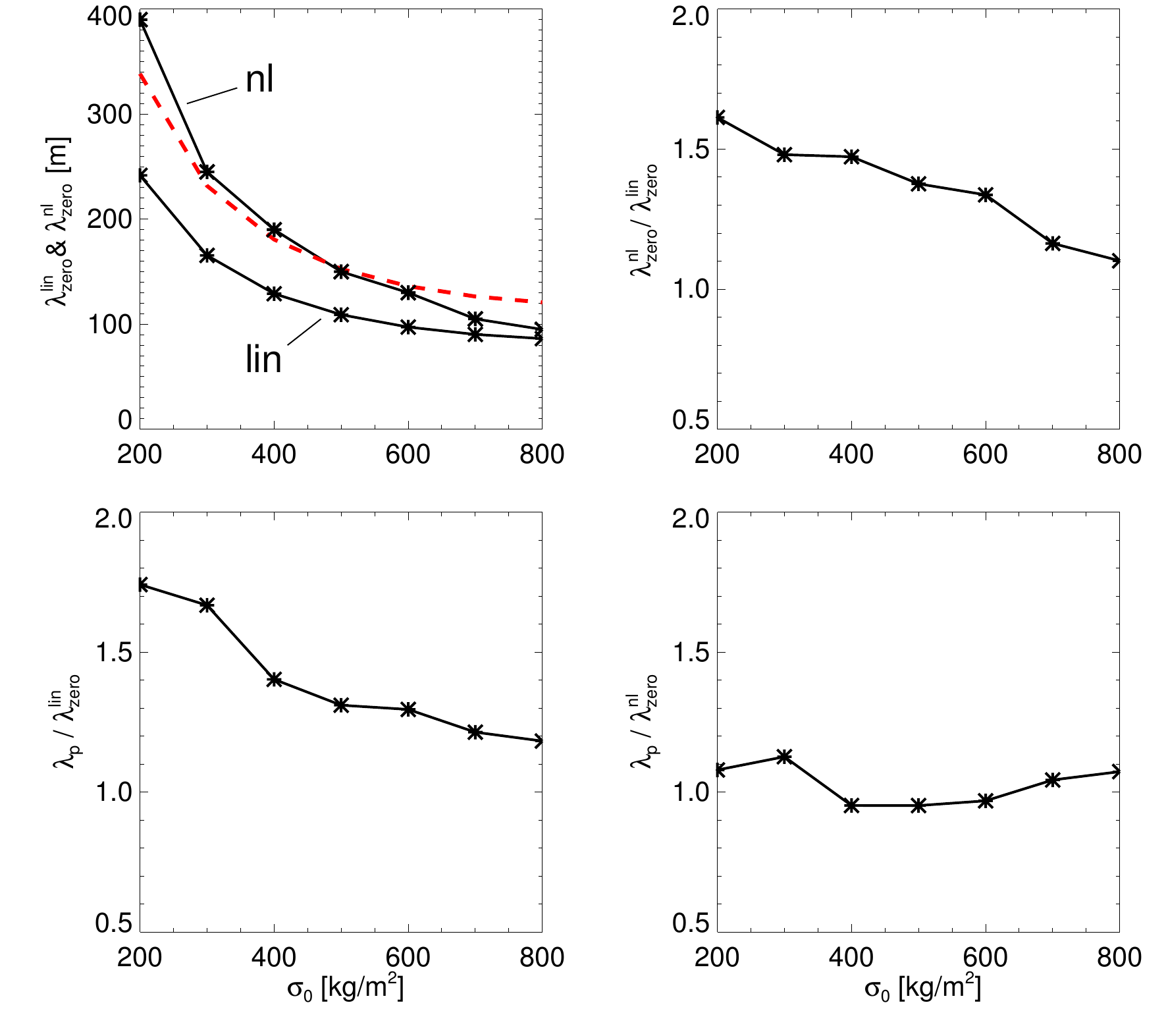}
\caption{Plots involving the hydrodynamic ($\tau_{15}$-parameters) wavelengths 
corresponding to the linear frequency minimum ($\lambda_{zero}^{lin}$), the nonlinear frequency minimum 
($\lambda_{zero}^{nl}$) and the saturation of overstability ($\lambda_{p}$), the latter being extracted from large-scale integrations which 
include a buffer-zone. In the upper left panel the labels ``nl'' and 
``lin'' denote the curves of $\lambda_{zero}^{nl}$ and $\lambda_{zero}^{lin}$, respectively. The red dashed line is relation 
(\ref{eq:lambdasigma}). Due to the buffer-zone, the prevalent wavelength in the final state can in some cases fluctuate within a given narrow range 
(about $5\%$).
Here, and also in Figure \ref{fig:nlvgbuff} we display the smallest value of $\lambda_{p}$ for each surface mass density. }
\label{fig:osfrat}
\end{figure}

%
In Figure \ref{fig:osfrat} we display for the same integrations different ratios of 
$\lambda_{zero}^{lin}$ (the wavelength of the linear frequency minimum), $\lambda_{zero}^{nl}$ (the wavelength of the nonlinear frequency 
minimum) and $\lambda_{p}$ (the prevalent wavelengths of the final saturated states). 
Also these plots show that, albeit with some scatter, the saturation wavelength $\lambda_{p}$ tends to follow the \emph{nonlinear} frequency minimum 
rather than the linear one.

It should be noted that it is difficult to obtain very accurate values of the involved nonlinear wavelengths for different reasons. 
On the one hand, the minima of the nonlinear frequency curves, which we can only probe with integrations as 
described in Section \ref{sec:nldisp}, are very mild for surface densities $\sigma_{0}\leq 400\,\text{kg}\,\text{m}^{-2}$.
On the other hand, as already outlined in the discussion of Figure \ref{fig:nbvgc}, the timescale of nonlinear evolution is prolonged if $\lambda_{p}$ is 
close to the nonlinear frequency minimum, requiring very long integrations.

\FloatBarrier

Most integrations presented in Figure \ref{fig:osfrat} (and Figure \ref{fig:nlvgbuff}) develop a persistent source/sink pair as a consequence of the 
buffer-zone, where the latter serves as 
the sink. In the cases with small surface densities ($\sigma_{0}=200-300\,\text{kg}\,\text{m}^{-2}$) all source and sink structures outside of 
the buffer-zone annihilate within a few thousand orbits so that the buffer-zone contains the remaining source/sink pair. 
In the case of very large surface density ($\sigma_{0}=800\,\text{kg}\,\text{m}^{-2}$) the pattern is more complicated and disturbed and it is 
not possible to identify source and sink structures.
Nevertheless, in all integrations of Figure \ref{fig:osfrat} the presence of the buffer-zone leads to the formation of nonlinear modes with wavelengths in  
direct vicinity of the nonlinear frequency minimum (cf.\ Appendix \ref{sec:altsat}). 

As already outlined in Section \ref{sec:intro}, in Saturn's rings one expects large-scale variations in the background parameters so that patches of 
overstable modes are permanently subject to 
perturbations. This situation might prevent the development of large uniform wave trains. The inclusion of a buffer-zone that damps overstability is a 
possible way to model this situation and should therefore provide a more realistic description than a (periodic) homogeneous integration region.

\subsection{Variation of the Particle Radius}\label{sec:radius}

All simulations and integrations discussed thus far assume a mono-disperse ring consisting of particles with radius $R_{p}=1\text{m}$.
A particle radius on the order of $1\text{m}$ follows from the formula for the geometric optical depth of a system of uni-sized spheres 
 $\tau = \pi R_{p}^2 \, \sigma /m_{p}$ (particle mass $m_{p}$) if one uses plausible parameters for Saturn's A-ring [$\tau\sim 0.5-0.8$ (\citet{colwell2009}), 
$\rho_{p}\sim 450\, \text{kg}\, \text{m}^{-3}$ (\citet{french2007}), $\sigma\sim 300 \, \text{kg}\, \text{m}^{-2}$ (\citet{tiscareno2007b})].

A thorough investigation of the effects of a wider size distribution on viscous overstability in terms of N-body simulations is computationally not feasible 
at present, due to the very high 
particle collision rates.
It can, however be expected that the presence of a particle size distribution has a mitigating effect on overstability (see Figure 5 in 
\citet{salo2001b}). 
Instead, we perform simulations with varying particle size, but keeping all other quantities (optical depth, elasticity law, surface mass density) 
unchanged. Note that this restriction can imply very unrealistic particle internal densities.

The effect of changing the particle radius on the saturation wavelength of overstability can be estimated as follows.
For a large range of Toomre-parameters 
$Q_{0}>1$
relation (\ref{eq:lambdasigma}) follows the estimate 
\begin{equation}\label{eq:lambdasigmaap}
  \lambda_{p} \approx  2 C\left(p_{s}-F_{2}\right)  \lambda_{J}
\end{equation}
 with the Jeans-wavelength 
\begin{equation}\label{eq:jeans}
 \lambda_{J}= \frac{c_{0}^2 }{G \sigma_{0}}.
\end{equation}
This approximation follows directly from Equation (\ref{eq:osfreq}) for $Q_{0}\gg 1$.
In a dense ring the effective velocity dispersion [Equation (\ref{eq:ptot})] scales roughly linearly with the particle radius, $c_{0} \sim R_{p}\Omega$, on 
account of the dominance of nonlocal pressure.
Thus, we expect a roughly quadratic dependence of the saturation wavelength on the particle radius, $\lambda_{p}\sim R_{p}^2$, at least for a range of values 
$R_{p}$.
For sufficiently small $R_{p}$, nonlocal effects would eventually diminish so that the condition for viscous overstability
is not fulfilled anymore. Furthermore, for large values of the surface density $\sigma_{0}$ the system develops a gravitational instability for particle radii 
in the range $0.1-1\,\text{m}$. 

\begin{figure}[h!]
\centering
\includegraphics[width = 0.4 \textwidth]{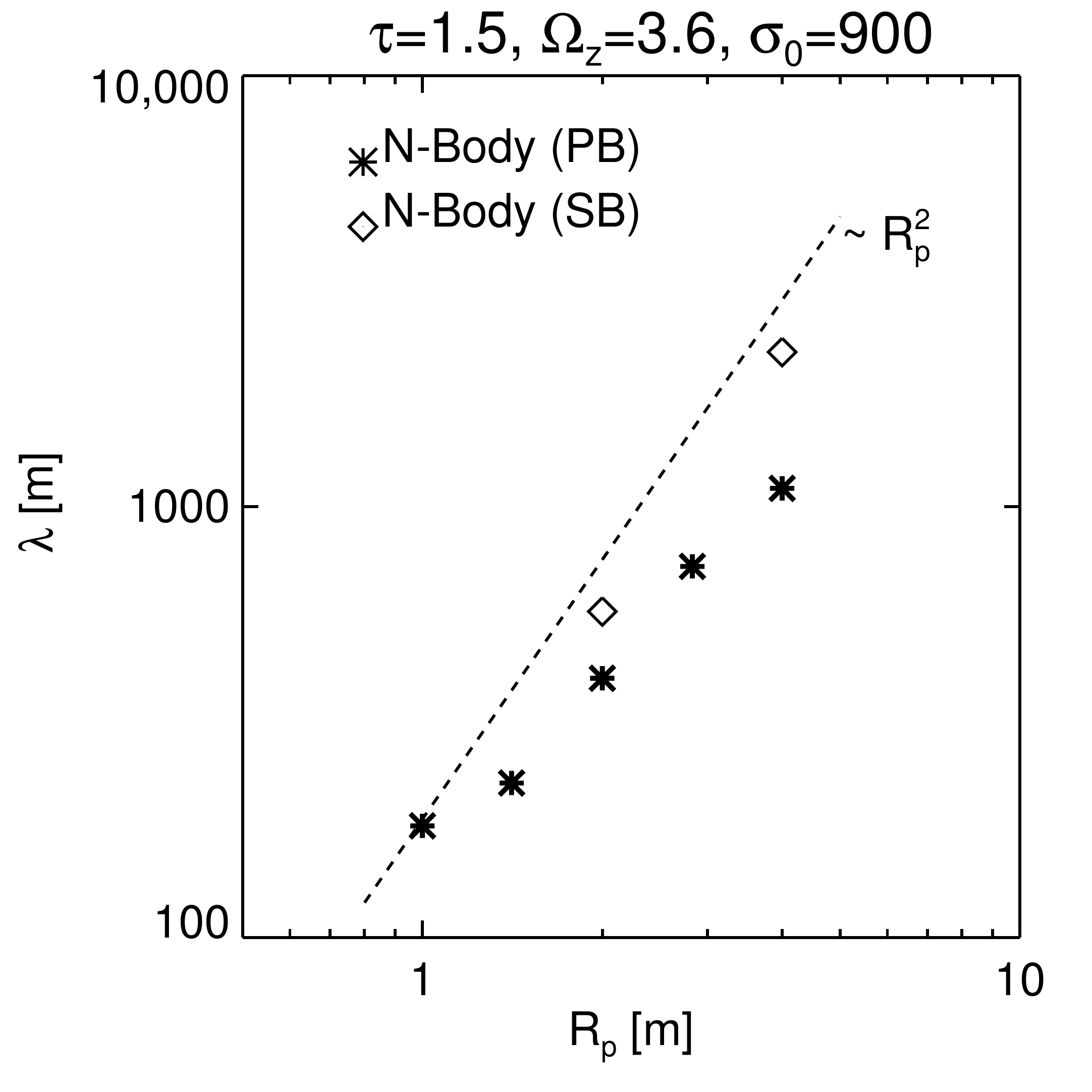}
\caption{Saturation wavelengths of viscous overstability in N-Body simulations with varying particle radius and fixed optical depth $\tau=1.5$, vertical 
frequency $\Omega_{z}=3.6$ and surface density $\sigma_{0}=900\,\text{kg}\,\text{m}^{-2}$. The asterisks represent results from simulations with 
periodic boundaries 
('PB') while the diamonds correspond to simulations with spreading boundaries ('SB'). (See text).}
\label{fig:radvar}
\end{figure}

Figure \ref{fig:radvar} shows the results of N-Body simulations performed with varying particle radius in the range $1-4\,\text{m}$ with a surface density 
$\sigma_{0}=900\,\text{kg}\,\text{m}^{-2}$ and the Bridges-type elasticity law (\ref{eq:bridges}). The radial width of the simulation region for these runs is 
chosen to depend quadratically on the particle radius, i.e.\ $L_{x}=2\,\text{km} \times (R_{p}/1\text{m})^2 $ to accommodate for the expected behavior of the 
saturation wavelengths in these runs. The asterisk symbols in Figure \ref{fig:radvar}
represent the resulting saturation wavelengths found for runs with periodic boundary conditions (as applied in all other N-Body simulations presented so far).
The diamond symbols represent two runs where we adopt ``spreading boundary'' conditions. In these runs the initial particle positions do not fill out the 
complete radial extent of the simulation region. These simulations resemble the hydrodynamic integrations with a buffer-zone, since the nonlinear wave trains 
are damped out as they enter the rarefied region. This also implies the emergence of a sink structure since the boundaries represent a buffer region and 
are found to act as source. Eventually, viscous spreading causes the particles to fill the entire simulation box and the system becomes similar to our standard 
periodic boundary case. When this happens the system is able to sustain the larger prevalent wavelength attained during the expanding phase. The saturation 
wavelengths found in these runs are larger, and closer to the corresponding nonlinear frequency minima. Thus, in agreement with the hydrodynamic integrations 
presented in Section \ref{sec:per} as well as Appendix \ref{sec:altsat}, the presence of a buffer-zone shifts the prevalent 
wavelength of the saturated pattern 
closer to the nonlinear frequency minimum. 
Furthermore, in the absence of radial self-gravity ($\sigma_{0}=0$) we find a nearly linear dependence $\lambda_{p}\sim R_{p}$, which is expected as in this 
case the particle radius is the only physical scale parameter of the system.

\subsection{Comparison to Previous Studies}\label{sec:comp}
Our results indicate that with increasing optical depth $\tau$, as well as with increasing vertical frequency enhancement $\Omega_{z}$ the saturation 
wavelength of the viscous overstabiliy increases.
This is in agreement with LO2009 and LO2010, who have shown (in the absence of radial self-gravity) that the wavelength $\lambda_{st}$ of nonlinear 
traveling waves, which are the preferred saturation of the viscous overstability (SS2003), is a steeply increasing function of the viscous 
parameter $\beta$ (see their Table 3). This parameter in turn is an increasing function of the ground state optical depth $\tau$ (\citet{salo2001}). A 
positive correlation between the overstable saturation wavelength of final state traveling waves and the equilibrium optical depth was later found in 
non-gravitating N-body 
simulations (RL2013). 
In N-body simulations, an increase of the parameters $\tau$ and $\Omega_{z}$ results in both cases in an increased collision frequency $\omega_{c}$.
The latter is the quantitity which eventually affects the value of $\beta$.
Nevertheless, a linear stability analysis of nonlinear wave train solutions of the system (\ref{eq:nleq}), including radial self-gravity, should be performed
to verify our hydrodynamic results. A method similar to that used in LO2009 for the isothermal limit without radial self-gravity, 
might be suitable in the present case as well.

The results of \citet{schmit1999} agree with ours in that their single isothermal integration with radial self-gravity resulted in
a saturation of 
overstability with prevalent wavelengths of a few times the corresponding Jeans-wavelength.
Indeed, the wavelength of vanishing linear group velocity, $\lambda_{zero}$, appearing in our relation (\ref{eq:lambdasigma}) for the saturation wavelength, 
reads in the isothermal limit $\lambda_{zero}\approx 2 p_{s} \lambda_{J}$ with the Jeans-wavelength (\ref{eq:jeans}).

\section{Conclusion}\label{sec:conc}
We investigate the influence of self-gravity on the long term and large-scale evolution of axisymmetric waves induced by viscous overstability in a dense 
planetary 
ring. In our approach we use hydrodynamic models and N-body simulations. These take into account the effect of collective radial self-gravity, exerted by the 
wave pattern on the ring matter. Aspects of the vertical component of self-gravity are incorporated in terms of the overall enhancement of the 
vertical frequency of oscillations. The effect of direct particle-particle gravity is not included.

We find a reasonably good agreement between N-body simulations and the hydrodynamic treatment for the nonlinear saturation of the viscous 
overstability in a dense ring. For the majority of surface mass densities the main effect of the radial self-gravity force is a reduction 
of the saturation wavelength of viscous overstability. 
In particular, the agreement of both modeling approaches is good for surface densities $\sigma_{0} \gtrsim 300 \,\text{kg}\,\text{m}^{-2}$, which are relevant 
for Saturn's dense rings (\citet{tiscareno2007b,hedman2016}), where overstability has been detected by Cassini instruments \citep{colwell2007, Thomson2007, 
sremcevic2009,hedman2014a}. The range of observed wavelengths $\lambda\sim 150\,\text{m}-250\,\text{m}$ compares well with the prevalent wavelengths
we find in our models.
Our results show that this length scale of saturation is closely related to the wavelength of minimal oscillation frequency of the 
nonlinear dispersion relation of oscillatory ring modes. This minimum exists only for a non-vanishing radial self-gravity force and it shifts to shorter 
wavelengths with increasing 
strength of radial self-gravity. Precisely at this minimum the group velocity of waves vanishes, so that the characteristic timescale for the nonlinear mode 
interaction diverges. 
Apparent deviations of the saturation wavelength from the nonlinear frequency minimum which we encounter in our results can have several reasons.  
Most importantly, we find that influences resulting from the application of periodic boundary conditions in a homogeneous model ring generally lead to an 
increase of the saturation wavelength in hydrodynamic integrations. But also the timescale of nonlinear evolution can prevent a proper determination of the 
saturation wavelength, particularly in N-body simulations. Moreover, the details of the numerical scheme to solve the hydrodynamic equations can have small but 
notable effects. 

In our hydrodynamical integrations and N-body simulations with vanishing radial self-gravity we find, in agreement with previous studies 
(SS2003, LO2009, LO2010, RL2013), that viscous overstability saturates in form of nonlinear traveling waves.
The same holds true if the radial self-gravity force is sufficiently weak. The 
nonlinear evolution toward this saturated state generally comprises source and sink structures at some point, separating counter-propagating wave trains. 
With increasing strength of self-gravity, saturated wave trains generally become more distorted, eventually showing persistent complex, 
standing-wave like interactions.

Generally, our hydrodynamic description that includes the energy equation yields a better match with N-body simulations than the isothermal approximation. We 
find that a good representation for the equation of state is essential to obtain an adequate description of the nonlinear saturation behavior of viscous 
overstability. In this vein, 
one route of future investigation could be the use of more suitable constitutive relations for dense particulate systems from the theory of granular matter 
(e.g. \citet{haff1983b,hutter1995}) in place of the power-law parameterizations (\ref{eq:pres})-(\ref{eq:coolfunc}) employed in this study. However, this will 
require also a self-consistent modeling of the disk's vertical thickness (\citet{borderies1985}). Another direction for future research could be an extension of the 
kinetic treatment by \citet{latter2008}, so that it can be applied to dense systems, employing an Enskog collision term (\citet{chapman1970,araki1986}). 
The considerable mathematical complexity of this approach could potentially be overcome by use of suitable approximations to solve the collision integrals 
(\citet{hameen1993}). Such a treatment would allow to model the effect of additional modes in the components of the pressure tensor \citep{latter2008}, 
which are not contained in the Newtonian approximation used in our study. 

A major challenge will be the inclusion of direct particle-particle self-gravity and its effect on the large-scale, long-term evolution of viscous 
overstability 
in Saturn's rings. In principle this can be achieved in terms of N-body simulations. But an investigation of the long radial scales we have 
studied in this paper is at present not feasible with this method, because of the prohibitively high CPU demand. It is known that the self-gravity of ring 
particles leads in 
large parts of Saturn's rings to the formation of gravitational wakes \citep{salo1992a,daisaka2001,hedman2007,colwell2007,french2007}, non-axisymmetric 
structures of wavelengths below 
$100\,\text{m}$. N-body simulations with full 
self-gravity show that these wakes interact with overstable modes in a complex manner \citep{salo2001,salo2018}. We believe that the overall 
saturation 
behavior of viscous overstability, like the prevalent wavelength of overstable modes, and the occurrence of source/sink patterns, is captured by the 
axisymmetric gravity model we have employed in this paper. But 
gravitational wakes will have additional effects. For instance, the heating induced by the wakes, as well as their typical 
non-axisymmetric pattern superimposed 
to the overstable modes, will affect the stability boundary of viscous overstability, i.e.\ the affinity of the ring to produce spontaneous 
axisymmetric overstable waves. The wakes and overstability generally depend differently on the local properties of the ring, such as particle size, internal 
density and elasticity, as well as optical depth (\citet{ballouz2017}). Therefore a detailed study of overstability in a fully self-gravitating system has the 
potential to constrain these still poorly known parameters by comparison to the precise pattern of occurrence and non-occurrence of overstable waves and 
self-gravity wakes, observed in Saturn's rings.

\FloatBarrier
\clearpage
\section*{Acknowledgments}

We acknowledge support from the Academy of Finland and the University of Oulu Graduate School.
We thank an anonymous reviewer for a constructive report that helped us to improve the paper.

\newpage

\appendix

\section{Stroboscopic Space-Time Diagram}\label{sec:strob}

\begin{figure}[h!]
\vspace{-0.2cm}
\centering
\includegraphics[width = 0.4 \textwidth]{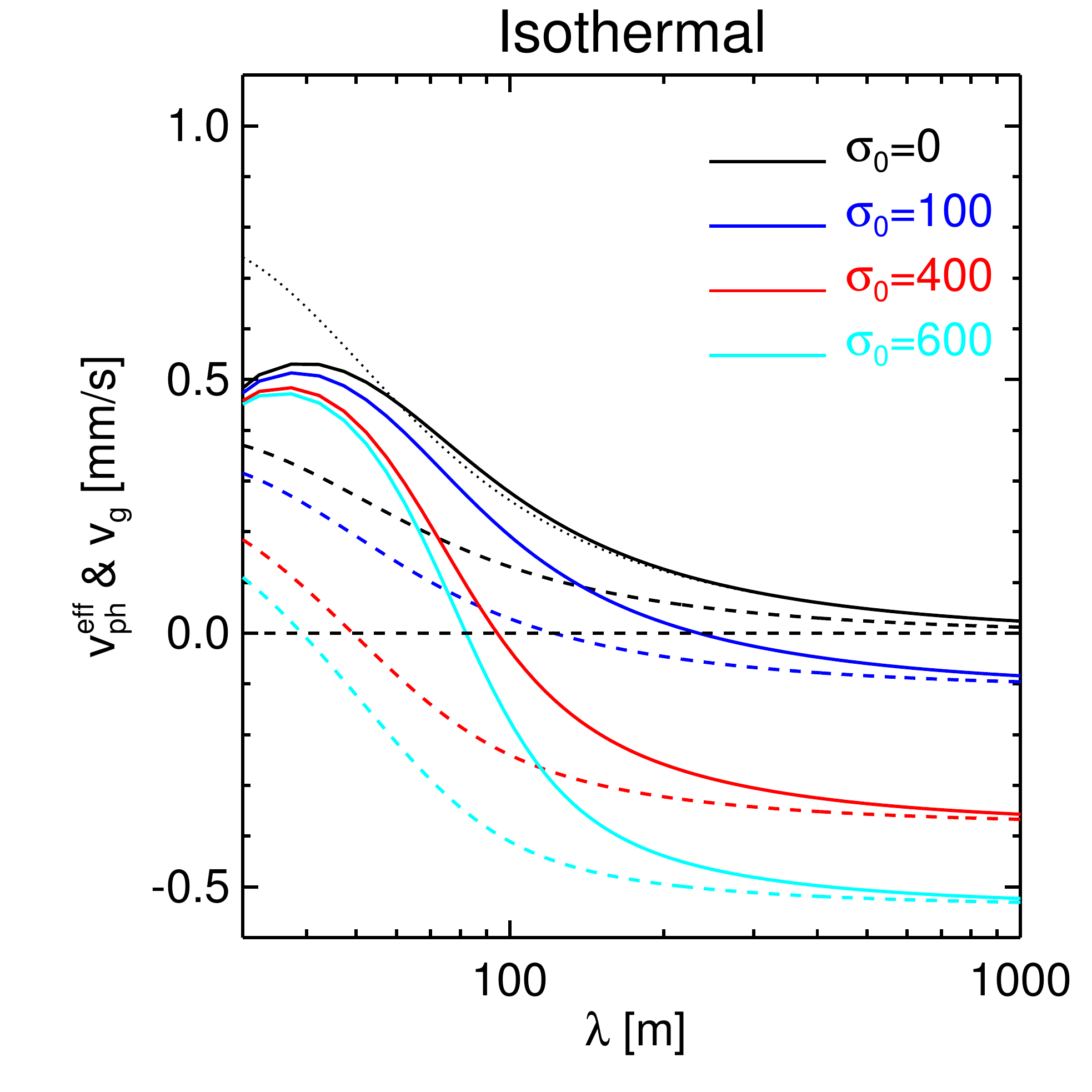}
\vspace{-0.5cm}
\includegraphics[width = 0.4 \textwidth]{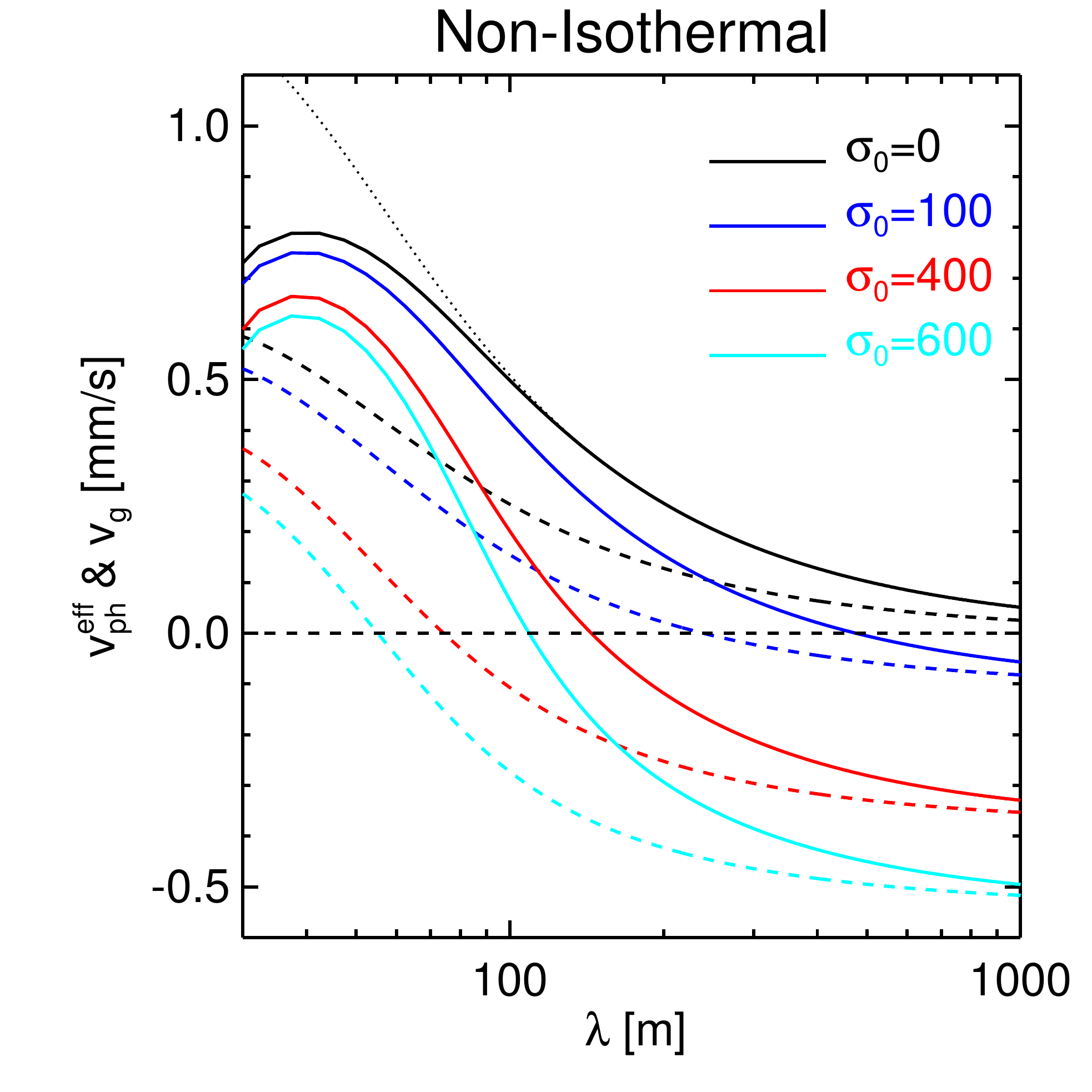}
\vspace{0.2cm}
\caption{Linear group velocity $v_{g}$ (solid curves) and the corresponding \emph{effective} linear phase velocity $v_{ph}-\Omega/k$ (dashed 
curves) 
for the $\tau_{15}$-parameters with different surface densities $\sigma_{0}$(in units $\text{kg}\, \text{m}^{-2}$). The black 
dotted curve equals twice the effective phase velocity (black dashed curve) for the case $\sigma_{0}=0$. In a stroboscopic space-time diagram with a sampling 
of 1/orbit  $v_{ph}^{\mbox{eff}}$ is the apparent phase velocity of wave structures.}
\label{fig:vgph}
\end{figure}

In a \emph{stroboscopic} space-time diagram, such as Figure \ref{fig:tau15nga} (left panel), the waves posses an \emph{effective} (unscaled) phase 
velocity $v_{ph}^{\mbox{eff}}=(\omega_{I}-\Omega)/k \approx  c_{0}\, p_{s}\, \Omega k/2$, following from (\ref{eq:osfreq}) as long as self-gravity is 
negligible. The 
group velocity (\ref{eq:vg}) in the absence of self-gravity yields $v_{g}\approx c_{0}\, p_{s}\, \Omega k$. Thus, in Figure \ref{fig:tau15nga} (and also Figure 
\ref{fig:tau20nonisonosga}) the 
identification of sources and sinks is straightforward, because the group velocity has the same sign as the apparent phase velocity for all wavelengths.
As an illustration, Figure \ref{fig:vgph} displays the linear group velocities $v_{g}$ (solid lines) and effective phase velocities $v_{ph}^{\mbox{eff}}$ 
(dashed lines) for the $\tau_{15}$-parameters and different surface densities $\sigma_{0}$. The left panel shows the isothermal model whereas the right panel 
corresponds to the non-isothermal model. The black dotted curves in both frames represent twice the phase velocity for the case $\sigma_{0}=0$, and agree in 
both cases with the group velocity (black solid curves) for all wavelengths larger than some $100\,\text{m}$. 
For large surface densities, however, 
$v_{ph}^{\mbox{eff}}$ and $v_{g}$ do not follow this relation anymore and the two quantities can even have opposite signs in the relevant wavelength range. 
Additionally, we find that \emph{nonlinear effects} alter both the phase and the group velocities of overstable waves (Section \ref{sec:nldisp}). Therefore 
one 
needs to be careful when interpreting structures in stroboscopic space-time plots.

 \section{Additional Hydrodynamical Integrations}

In this appendix we briefly describe hydrodynamical integrations which address specific topics that were pointed out in Section \ref{sec:nldisp}. 

\subsection{Influence of the Radial Domain Size on the Saturation Wavelength}\label{sec:boxsize}

\begin{figure}[h!]
\centering
\includegraphics[width = 0.3 \textwidth]{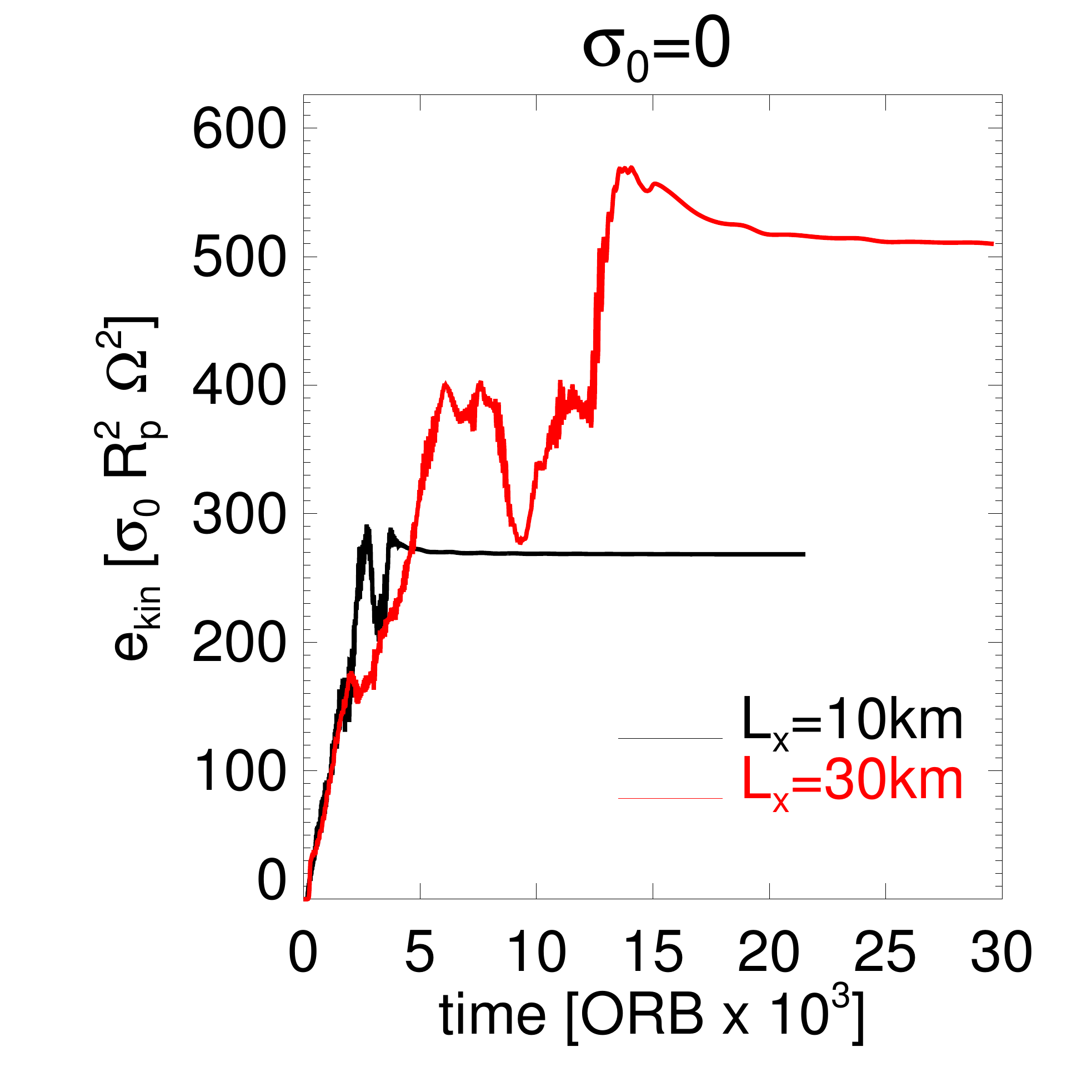}
\includegraphics[width = 0.3 \textwidth]{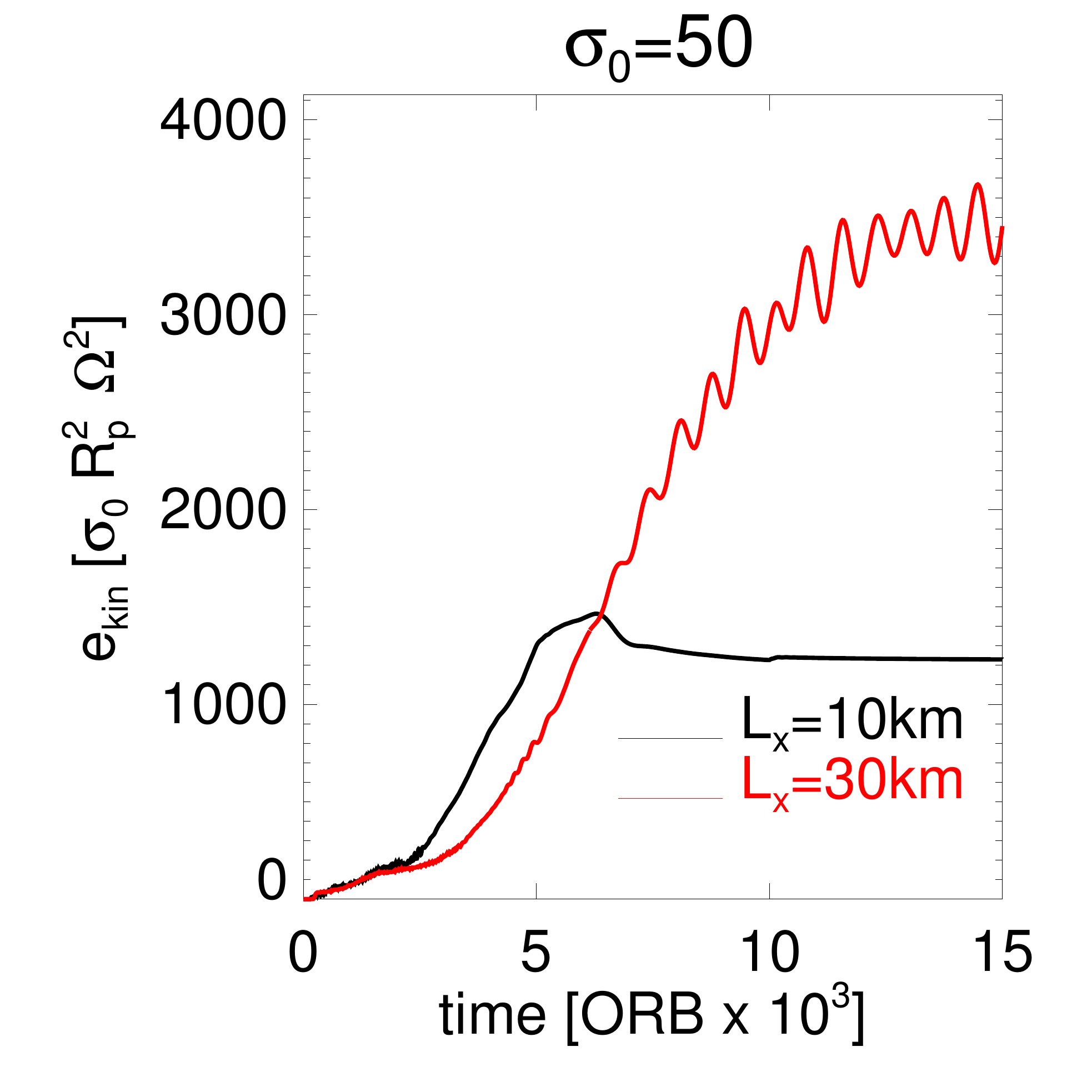}
\includegraphics[width = 0.3 \textwidth]{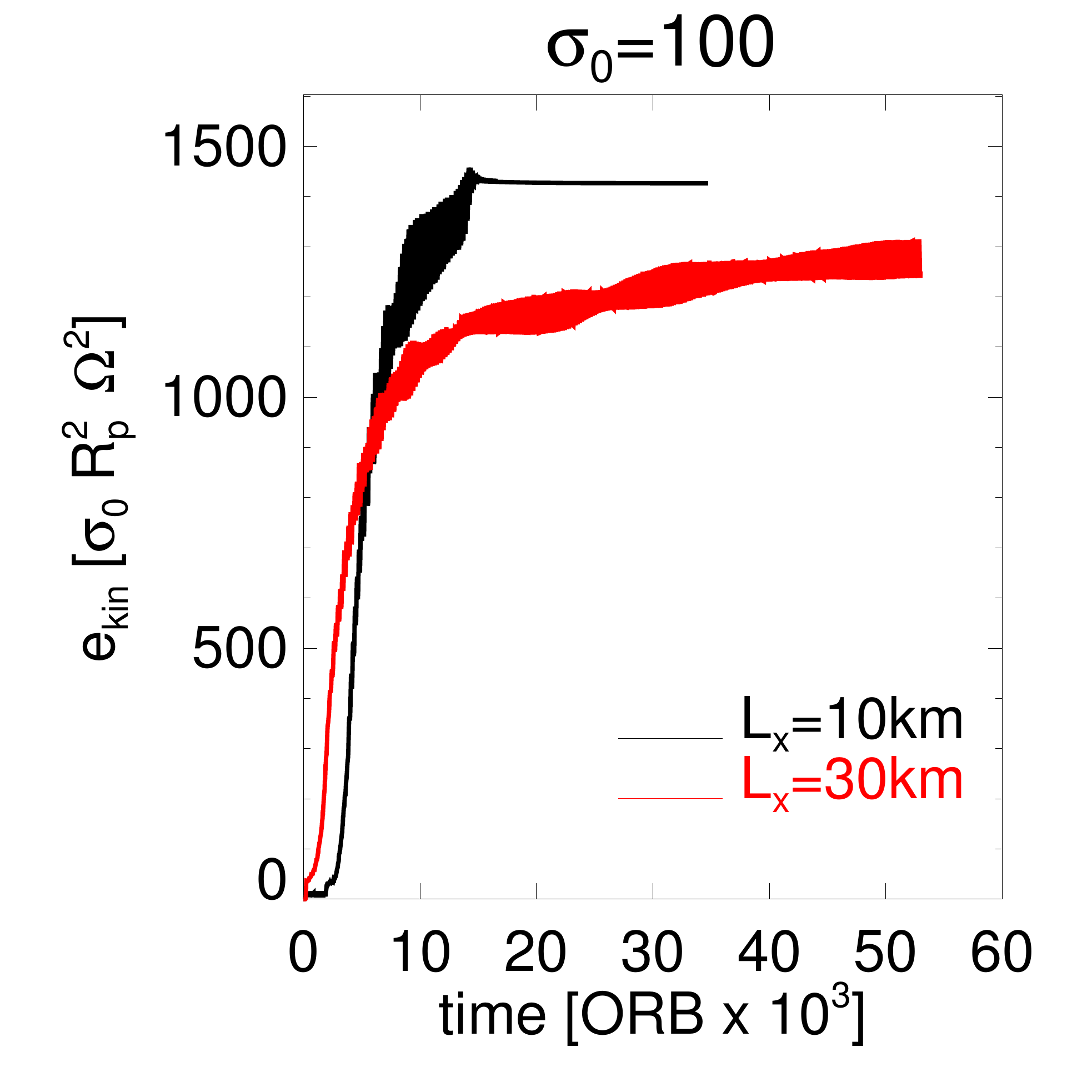}\\
\includegraphics[width = 0.3 \textwidth]{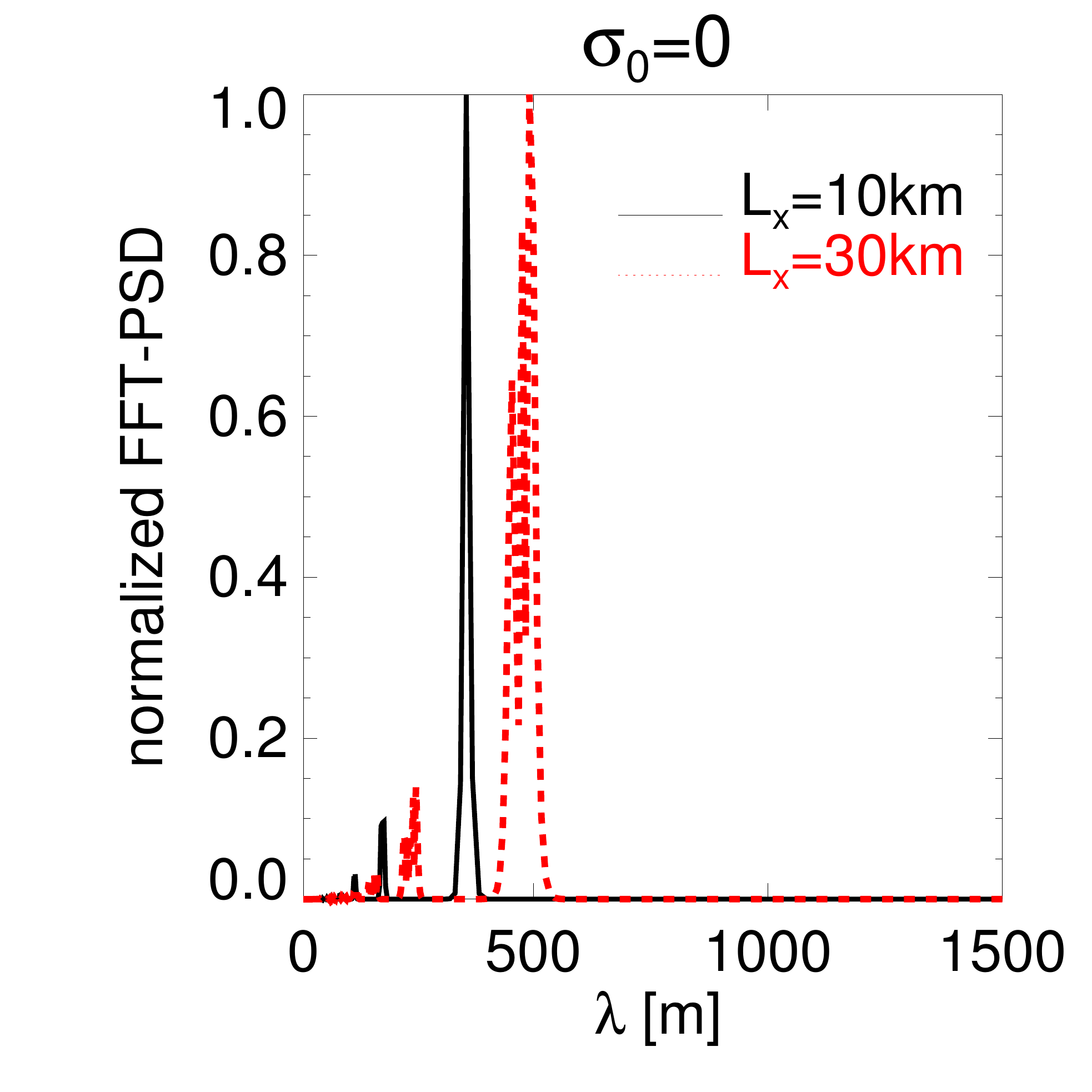}
\includegraphics[width = 0.3 \textwidth]{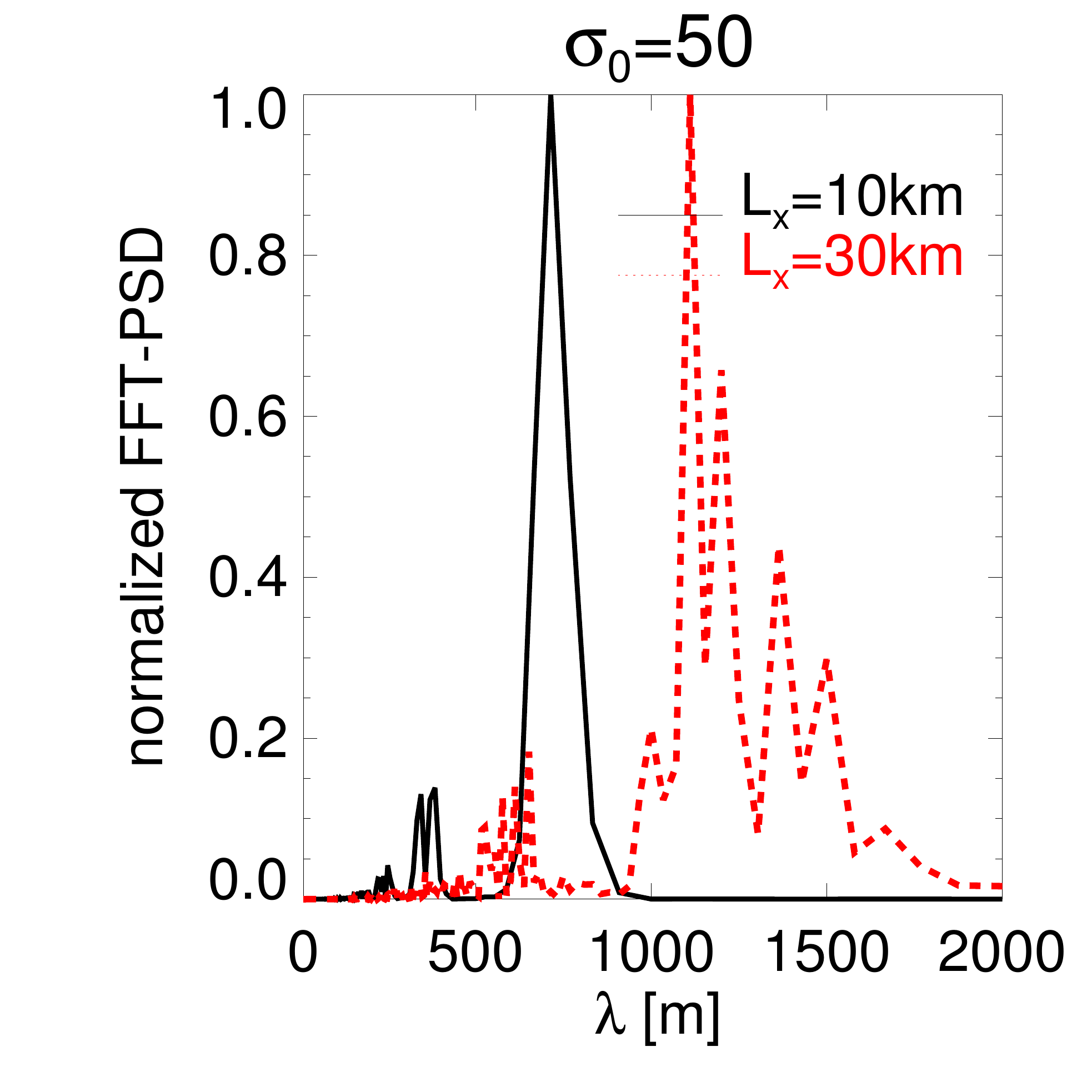}
\includegraphics[width = 0.3 \textwidth]{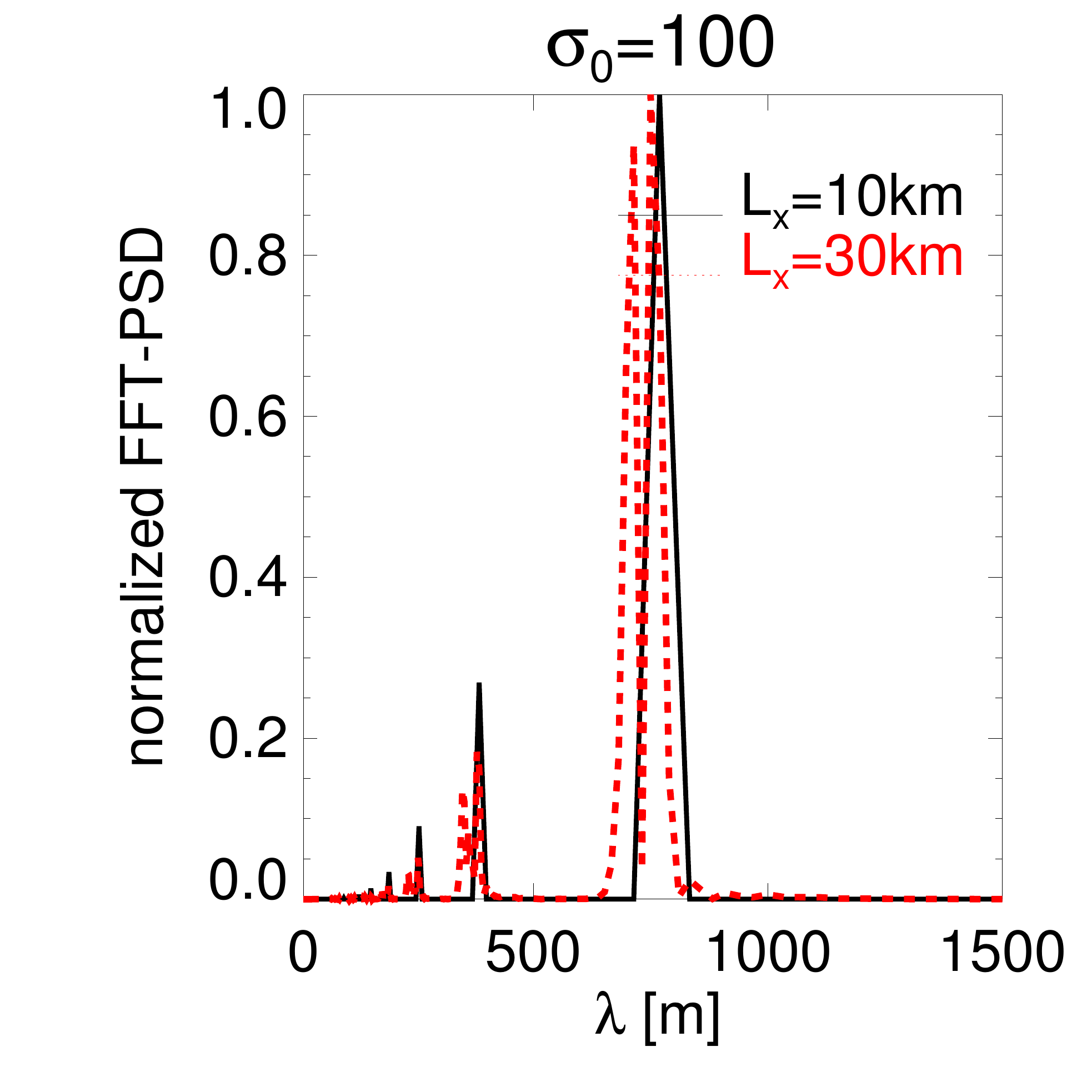}
\caption{Illustration of the limiting effect of the size of calculation region $L_{x}$ on the evolution of the viscous overstability in hydrodynamic 
computations for small surface densities $\sigma_{0}$ ($\tau_{15}$-parameters). The upper three frames show the evolution of the kinetic energy densities and 
the lower frames 
the final power spectra for integrations with considerably different $L_{x}$. For the cases $\sigma_{0}=0$ and $\sigma_{0}=50\,\text{kg}\,\text{m}^{-2}$, the 
initial 
seed was 
equal (spectral white 
noise) for the two integrations with different $L_{x}$. For the integrations with $\sigma_{0}=100\,\text{kg}\,\text{m}^{-2}$ the initial conditions were a 
single 
wavelength mode $\lambda=100\,\text{m}$ ($L_{x}=10\,\text{km}$) and spectral white noise ($L_{x}=30\,\text{km}$).}
\label{fig:boxsizecomp01}
\end{figure}

In Section \ref{sec:nldisp} we noted that for very small surface mass densities $\sigma_{0}$ the 
finite size of the computational domain limits the growth of large-scale structures in hydrodynamical integrations, thereby affecting the saturation 
wavelength $\lambda_{p}$. This limiting effect, demonstrated in Figure \ref{fig:boxsizecomp01}, becomes weaker with increasing $\sigma_{0}$. 
That is, for the $\tau_{15}$-parameters, a domain size of \emph{at least} $L_{x}=30\,\text{km}$ is required for a surface density 
$\sigma_{0}=50\,\text{kg}\,\text{m}^{-2}$ (Figure \ref{fig:boxsizecomp01} second column). Whereas, a size of $L_{x}=10\,\text{km}$ is sufficient for 
$\sigma_{0}=100 \,\text{kg} \, 
\text{m}^{-2}$ (Figure \ref{fig:boxsizecomp01} third column).  Furthermore, for  $\sigma_{0}\gtrsim 300 \,\text{kg} \, \text{m}^{-2}$ a size 
$L_{x}=5\,\text{km}$ suffices to obtain a consistent value of 
the saturation wavelength [i.e.\ close to the value predicted by the empirical relation (\ref{eq:lambdasigma})].

 \subsection{Integrations with Modified Equation of State}\label{sec:pmod}

Related to the discussion of Figures \ref{fig:nblin2} and \ref{fig:nbvga} (right panel), we performed several non-isothermal hydrodynamic 
integrations with the $\tau_{15}$-parameters and $L_{x}=10\,\text{km}$, employing the 
increased values of $p_{s}$ 
by factors of 1.4 and 2, mentioned in sections \ref{sec:nbtest} and \ref{sec:nldisp} (see also Figures \ref{fig:nblin2} and \ref{fig:nbvga}). The result is 
that these modifications bring the saturation wavelength $\lambda_{p}$ for small surface densities $\sigma_{0}\leq 150 \, \text{kg}\,\text{m}^{-2}$ very close 
to 
the value of $\lambda_{p}$ obtained for the case $\sigma_{0}=0$ with the original $p_{s}$ (with $L_{x}=10\,\text{km}$).
Thus, a sufficiently large value of $p_{s}$ effectively removes the influence of the frequency minimum for small $\sigma_{0}$, by shifting it to very 
large wavelengths. Consequently, its approach is hindered by the size of the calculation region (Figure \ref{fig:boxsizecomp01}). However, the modified 
$p_{s}$ 
also leads to considerably increased values of $\lambda_{p}$ for larger surface densities $\sigma_{0}\gtrsim 300\, \text{kg}\,\text{m}^{-2}$, which makes the 
agreement with the N-body simulations worse.

 \subsection{Influence of the Initial State on the Saturation Wavelength}\label{sec:altsat}

The saturation wavelengths of the hydrodynamic model (the asterisks in Figure \ref{fig:nlvg}) are obtained from integrations where the 
initial 
state consists of spectral 
white noise.
In order to investigate possible saturation on alternative wavelengths (cf.\ Figure \ref{fig:nbvgc}) we perform a series of integrations in a radial domain 
with 
$L_{x}=5-10\,\text{km}$, 
 employing the
$\tau_{15}$-parameters, where we seed a single wavelength large amplitude mode. As the seeded mode saturates, we follow the subsequent evolution for 10,000 
orbits.
In actual fact, we find for each surface density a whole range of wavelengths supporting stable traveling 
waves, not showing any signs of wavelength change for at least 10,000 orbits.
For instance, for wavelengths $150\,\text{m} \lesssim \lambda \lesssim 250\,\text{m}$, all surface densities $\sigma_{0}\gtrsim 
200\,\text{kg}\,\text{m}^{-2}$ 
support such  stable traveling waves.

As speculated in Section \ref{sec:nldisp}, in some cases the smallness of the group velocity might explain the absence of changes of the wave train on 
the considered timescale.
In other cases, however, the (quasi-)stability of single wavelength modes 
substantially different from those over-plotted in Figure \ref{fig:nlvg}, seems to contradict with N-body simulations, such as those in Figure 
\ref{fig:boxdouble} (left panel). 
As outlined before (Section \ref{sec:nldisp}), in N-body simulations numerous fluctuations due to the discrete nature of the particle flow are present at all 
times. These perturbations do not exist in the 
hydrodynamic model system, which might explain the resistance to 
change the dominant wavelength.
When adding a buffer-zone in the computational domain though (cf.\ Sections \ref{sec:isonosg} and \ref{sec:per}), the system immediately responds. That is, it 
immediately 
excites power on a range of wavelengths located in close vicinity to the nonlinear frequency minimum. The power on these wavelengths subsequently increases and 
the 
prevalent wavelength approaches asymptotically a value close to the nonlinear frequency minimum (Section \ref{sec:per}).

\begin{figure}[h!]
\vspace{-0.2cm}
\centering
\includegraphics[width = 0.4 \textwidth]{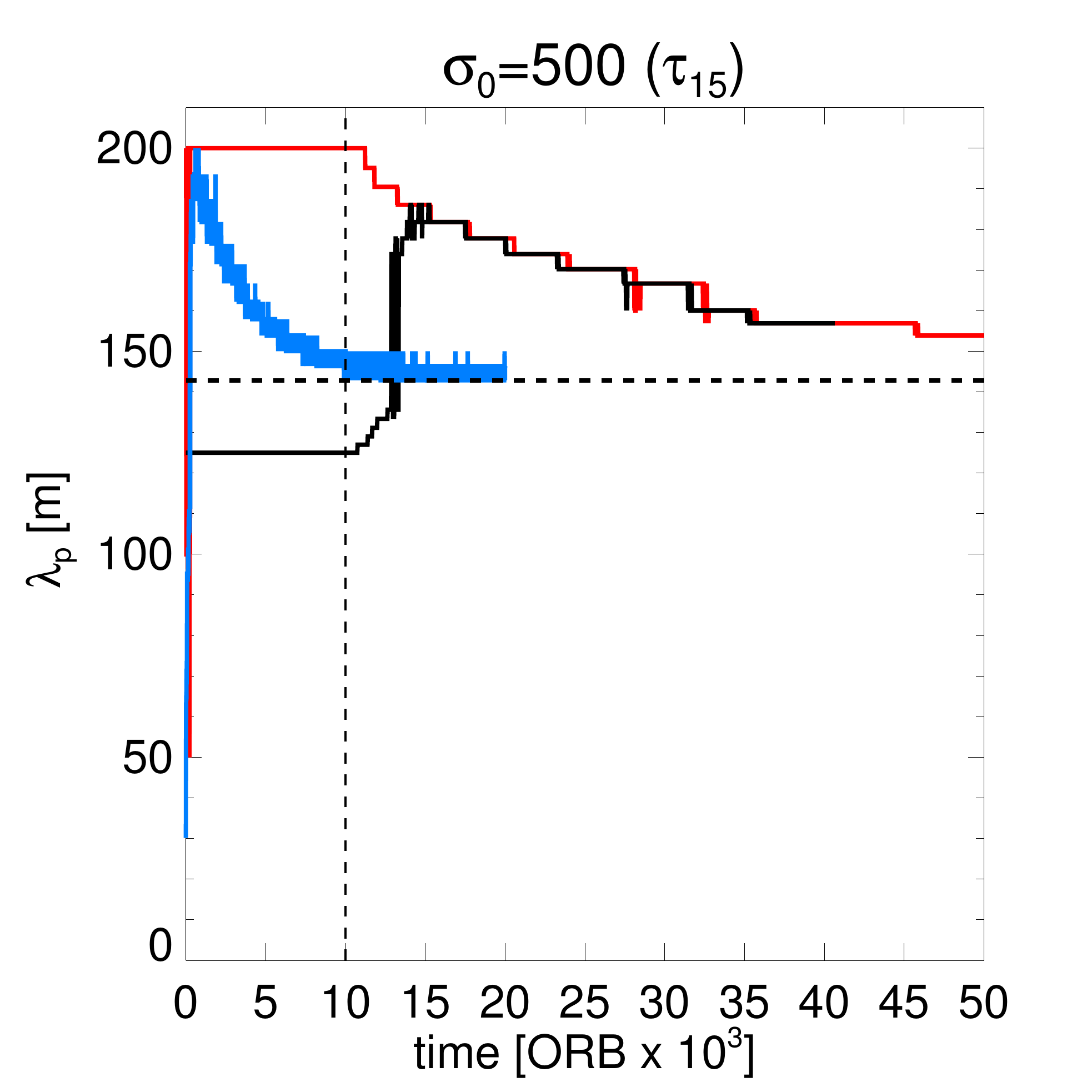}
\includegraphics[width = 0.4 \textwidth]{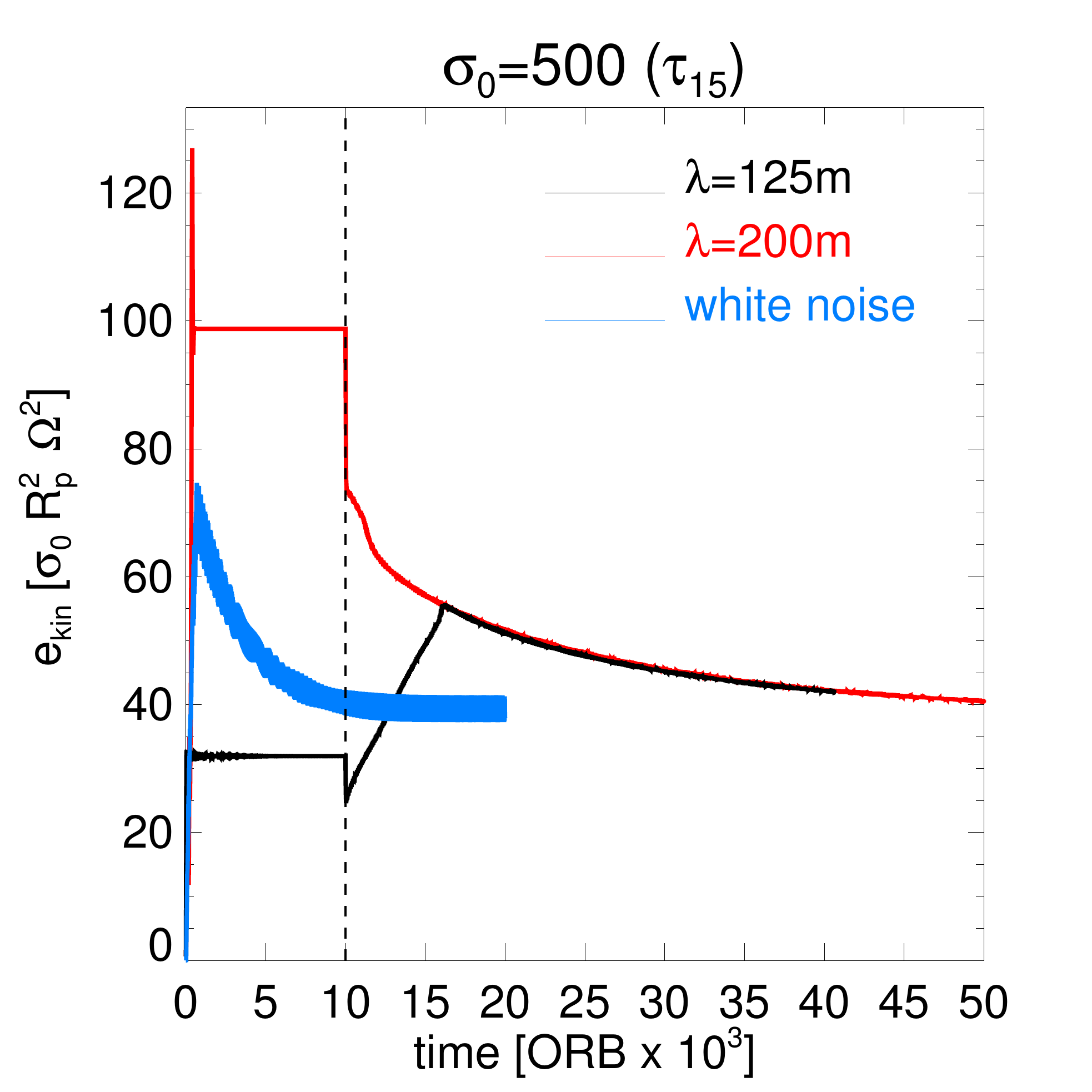}
\caption{Two hydrodynamic integrations with $\sigma_{0}=500\,\text{kg}\, \text{m}^{-2}$ ($\tau_{15}$-parameters) started with different large amplitude single 
wavelength modes ($\lambda=125\,\text{m}$ and $\lambda=200\,\text{m}$, respectively) in a periodic domain with $L_{x}=8\,\text{km}$. Left and right panels show 
the prevalent wavelengths and kinetic energy densities for both integrations. The wave trains saturate 
rapidly and remain stable for at least 10,000 orbits.
At time $t=10,000\,\text{ORB}$ (marked by the vertical dashed line) the integration regions are superimposed by a buffer-zone ($\beta=-0.5$ in the region 
$[-0.5\,\text{km};+0.5\,\text{km}]$). 
Note the sudden drop of $e_{kin}$ in these integrations at 10,000 orbits due to the buffer-zone. 
Also shown are $\lambda_{p}$ and $e_{kin}$ for an integration that started from white noise with a buffer-zone present 
at all times.}
\label{fig:sig500buff}
\end{figure}

To illustrate this behavior, Figure \ref{fig:sig500buff} shows the evolution of the prevalent wavelengths and the kinetic energy densities of two integrations 
with $L_{x}=8\,\text{km}$ where modes with respective wavelengths 
$\lambda=125\,\text{m}$ and $\lambda=200\,\text{m}$ saturate and their evolution is followed for 10,000 orbits. At time 
$t=10,000\,\text{ORB}$ in both cases a buffer-zone is superimposed to the integration region such that $\beta<\beta_{c}$ for 
$x=[-0.5\,\text{km};0.5\,\text{km}]$.
Also displayed is the evolution of an integration which started from low amplitude white noise and that included a buffer-zone from the beginning (cf.\ 
Section \ref{sec:per}).
As a result of the buffer-zone, in all of the three integrations source/sink structures form and the prevalent wavelength asymptotically approaches the 
nonlinear frequency minimum, which is marked by a horizontal dashed line in the left frame.
In the integration which started from white noise the timescale for this approach is significantly shorter. Also, in the same integration, there are stronger 
fluctuations in $e_{kin}$ and $\lambda_{p}$, indicating stronger perturbations emitted by the source. These differences might be a 
consequence of the different source/sink configuration.
In this integration a sink persists exterior to the buffer-zone so that the buffer-zone as a whole serves as a source.
In contrast, the integrations starting from nonlinear unidirectional wave trains form source and sink structures within the buffer-zone so that the overall 
wave pattern effectively ``tunnels'' through the latter.  Whether this difference is the reason for the different time 
scales is, however, speculative and should be addressed in future work.
In this regard we like to note that the precise asymptotic behavior of these hydrodynamic integrations can depend slightly on the details of the applied 
numerical scheme, such as the reconstruction method used for the numerical flux vector (Section \ref{sec:numerics}). 

\FloatBarrier

\section{Group Velocities of Saturated Wave Trains}\label{sec:vgroup}

\begin{figure}[h!]
\centering
\includegraphics[width = 0.33 \textwidth]{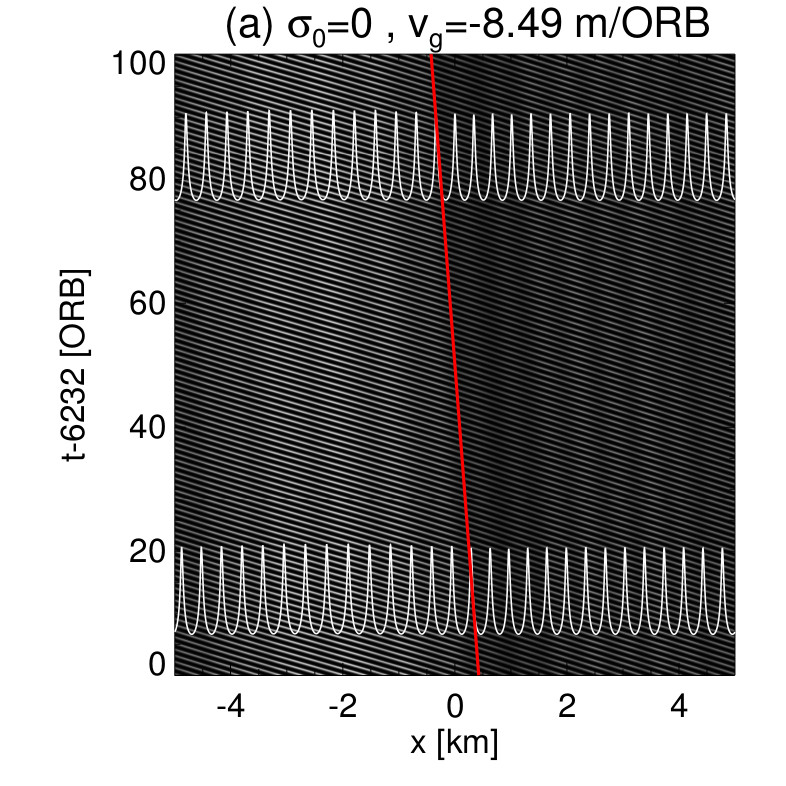}
\includegraphics[width = 0.33 \textwidth]{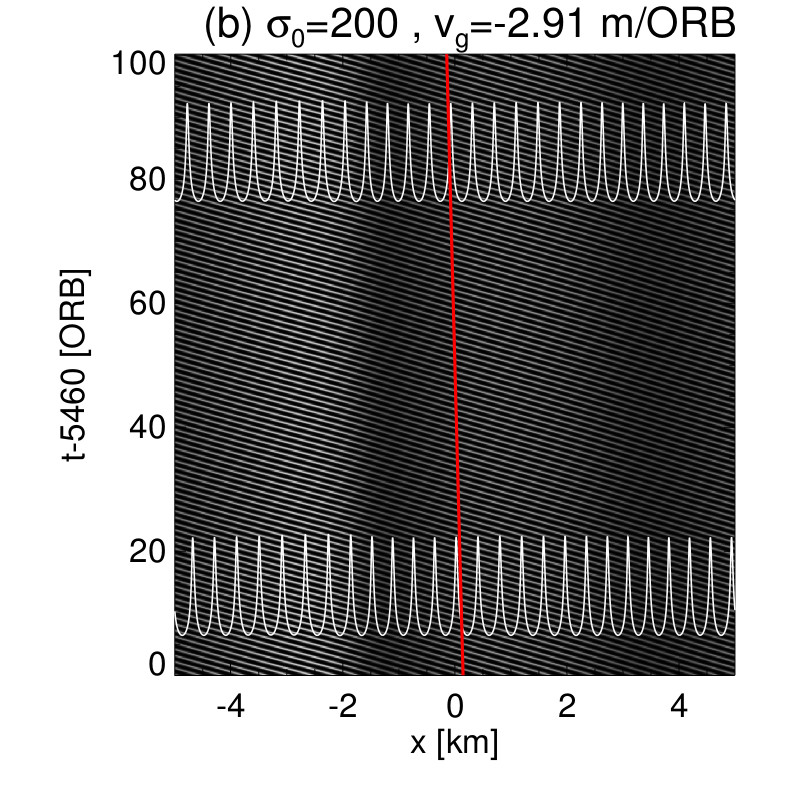}\\
\includegraphics[width = 0.33 \textwidth]{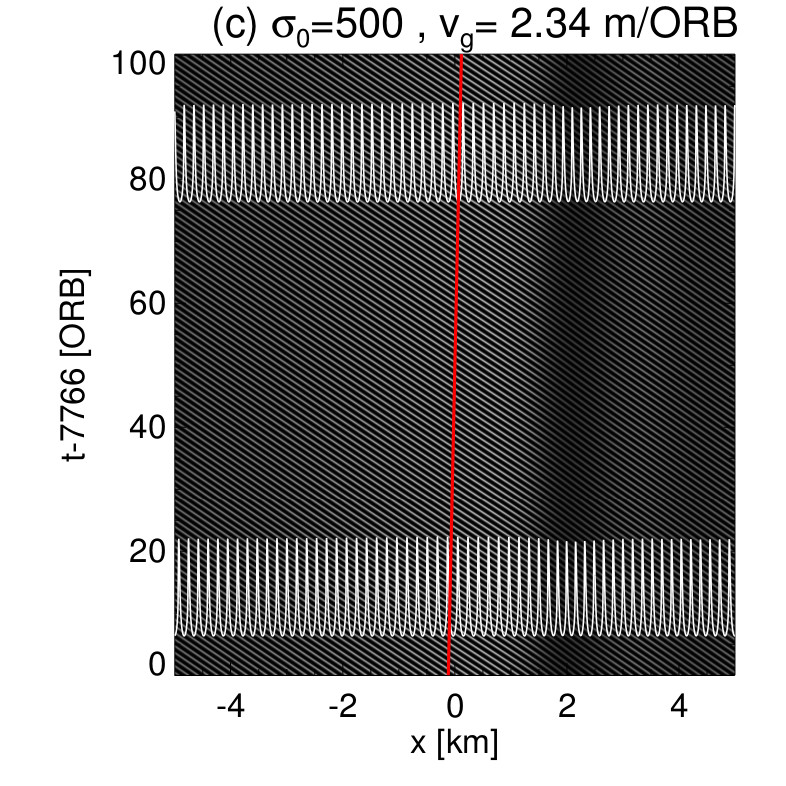}\\
\includegraphics[width = 0.33 \textwidth]{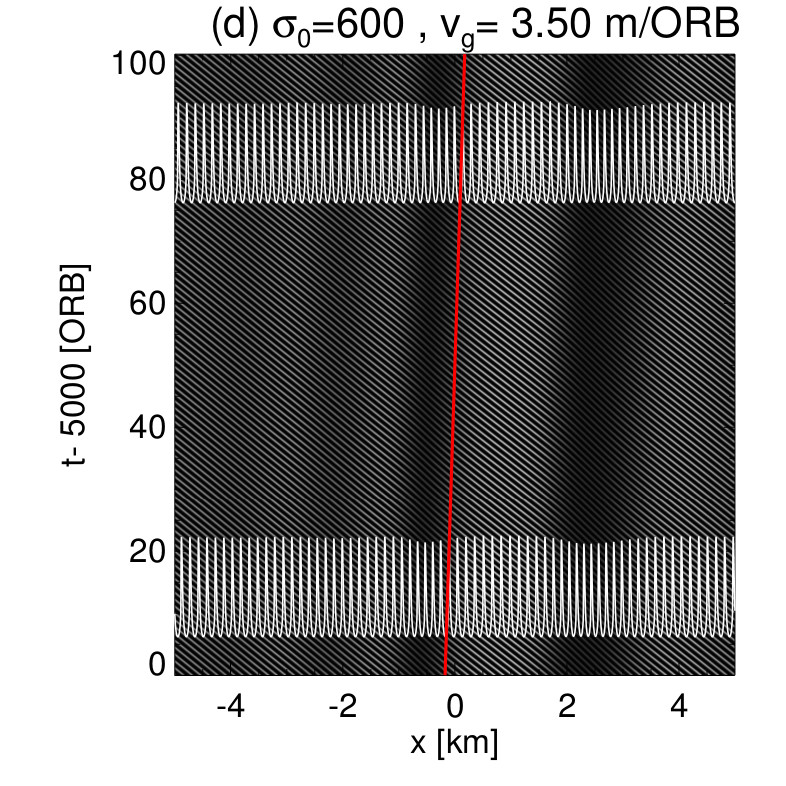}
\includegraphics[width = 0.33 \textwidth]{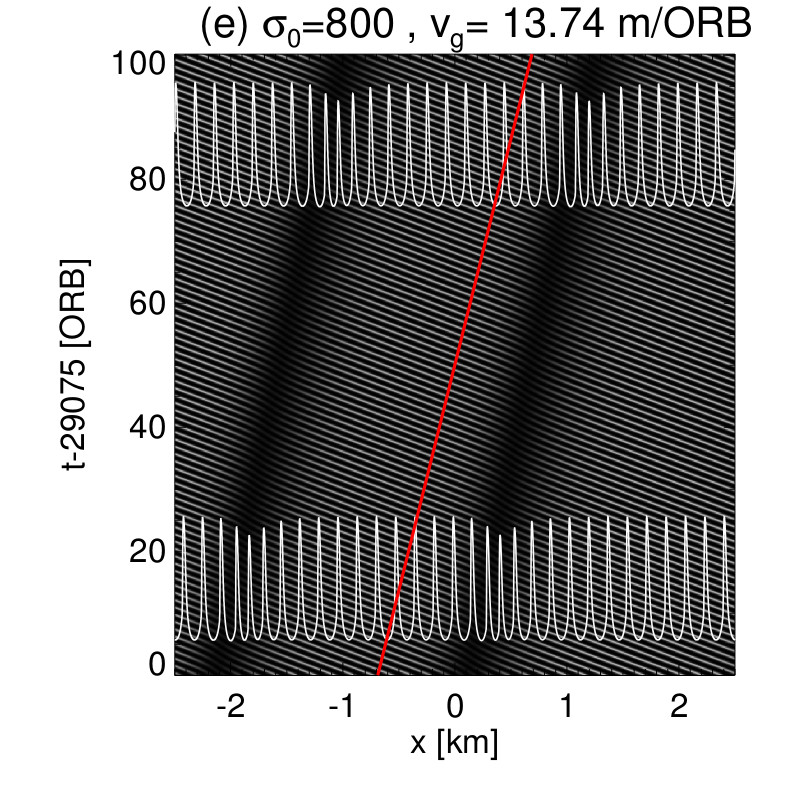}
\vspace{-0.3cm}
\caption{Space-time diagrams of saturated traveling waves resulting from integrations with the $\tau_{15}$-parameters.
The movement of long wavelength undulations occurs with the group velocity of the wave train which is indicated by red lines.
Also shown in each frame are the profiles of surface mass density (enhanced by a factor of 10) corresponding to early and late stages of the displayed 
evolution. See the text for more explanation.}
\label{fig:vgpert}
\end{figure}

In order to verify the computed nonlinear dispersion relations $\omega_{I}^{nl}(k)$ (Figure \ref{fig:nlvg} in Section \ref{sec:nldisp}) of overstable 
waves, we
can compare the propagation speed of small perturbations imposed to saturated wave trains with the group velocity 
$\mathrm{d}\omega_{I}^{nl}(k)/\mathrm{d}k$, obtained by numerical differentiation.
Figure \ref{fig:vgpert} presents space-time diagrams of saturated (left) traveling waves resulting from integrations with the $\tau_{15}$-parameters which 
started from white noise. 
Frame (a) shows the same integration as in Figure \ref{fig:tau20nonisonosga}. Frames (b) and (e) are the same integrations as in Figure \ref{fig:etauhydro}. 
The primary waves in each frame correspond to the small scale structure. Perturbations in the wave amplitudes are visible on much larger scale and develop in 
all
cases in the course of the nonlinear evolution. The red lines represent the 
group velocities computed by numerical differentiation (using 3-point Lagrangian interpolation) of the measured nonlinear frequency curves in Figure 
\ref{fig:nlvg}a at the corresponding saturation wavelengths (the asterisks) for each $\sigma_{0}$ (units $\text{kg}\,\text{m}^{-2}$). The values of $v_{g}$ are 
indicated for all cases, matching well the propagation of the long wavelength undulations.

\vspace{-0.5cm}


\begin{thebibliography}{}

\bibitem[{Araki} and {Tremaine}, 1986]{araki1986}
{Araki}, S. and {Tremaine}, S. (1986).
\newblock {The dynamics of dense particle disks.}
\newblock {\em Icarus}, 65:83--109.

\bibitem[{Ballouz} et~al., 2017]{ballouz2017}
{Ballouz}, R.-L., {Richardson}, D.~C., and {Morishima}, R. (2017).
\newblock {Numerical Simulations of Saturn's B Ring: Granular Friction as a
  Mediator between Self-gravity Wakes and Viscous Overstability}.
\newblock {\em \aj}, 153:146.

\bibitem[Binney and Tremaine, 1987]{binney1987}
Binney, J. and Tremaine, S. (1987).
\newblock {\em Galactic Dynamics}.
\newblock Princeton University Press.

\bibitem[Borderies et~al., 1985]{borderies1985}
Borderies, N., Goldreich, P., and Tremaine, S. (1985).
\newblock A granular flow model for dense planetary rings.
\newblock {\em Icarus}, 63:406--420.

\bibitem[Bridges et~al., 1984]{bridges1984}
Bridges, F., Hatzes, A., and Lin, D. (1984).
\newblock {Structure, stability and evolution of Saturn's rings}.
\newblock {\em Nature}, 309:333--338.

\bibitem[Chapman and Cowling, 1970]{chapman1970}
Chapman, S. and Cowling, T. (1970).
\newblock {\em The mathematical theory of non-uniform gases}.
\newblock Cambridge University Press, Cambridge.

\bibitem[{Colwell} et~al., 2007]{colwell2007}
{Colwell}, J.~E., {Esposito}, L.~W., {Srem{\v c}evi{\'c}}, M., {Stewart},
  G.~R., and {McClintock}, W.~E. (2007).
\newblock {Self-gravity wakes and radial structure of Saturn's B ring}.
\newblock {\em Icarus}, 190:127--144.

\bibitem[{Colwell} et~al., 2009]{colwell2009}
{Colwell}, J.~E., {Nicholson}, P.~D., {Tiscareno}, M.~S., {Murray}, C.~D.,
  {French}, R.~G., and {Marouf}, E.~A. (2009).
\newblock {\em {The Structure of Saturn's Rings}}, page 375.

\bibitem[Daisaka et~al., 2001]{daisaka2001}
Daisaka, H., Tanaka, H., and Ida, S. (2001).
\newblock {Viscosity in a dense planetary ring with self-gravitating particles}.
\newblock {\em Icarus}.

\bibitem[{Dilley}, 1993]{dilley1993}
{Dilley}, J.~P. (1993).
\newblock {Energy loss in collision of icy spheres: Loss mechanism and
  size-mass dependence.}
\newblock {\em Icarus}, 105:225--234.

\bibitem[{French} et~al., 2007]{french2007}
{French}, R.~G., {Salo}, H., {McGhee}, C.~A., and {Dones}, L. (2007).
\newblock {HST observations of azimuthal asymmetry in Saturn's rings}.
\newblock {\em Icarus}, 189:493--522.

\bibitem[{Goldreich} and {Lynden-Bell}, 1965]{goldreich1965}
{Goldreich}, P. and {Lynden-Bell}, D. (1965).
\newblock {II. Spiral arms as sheared gravitational instabilities}.
\newblock {\em mnras}, 130:125.

\bibitem[Goldreich and Tremaine, 1978]{goldreich1978a}
Goldreich, P. and Tremaine, S. (1978).
\newblock {The velocity dispersion in Saturn's rings.}
\newblock {\em Icarus}, 34:227--239.

\bibitem[{Gottlieb} et~al., 2001]{gottlieb2001}
{Gottlieb}, S., {Shu}, C.-W., and {Tadmor}, E. (2001).
\newblock {Strong Stability-Preserving High-Order Time Discretization Methods}.
\newblock {\em SIAM Review}, 43:89--112.

\bibitem[{Haff}, 1983]{haff1983b}
{Haff}, P.~K. (1983).
\newblock {Grain flow as a fluid--mechanical phenomenon}.
\newblock {\em J. Fluid Mech.}, 134:401--430.

\bibitem[{H{\"a}meen-Anttila} and {Salo}, 1993]{hameen1993}
{H{\"a}meen-Anttila}, K.~A. and {Salo}, H. (1993).
\newblock {Generalized Theory of Impacts in Particulate Systems}.
\newblock {\em Earth Moon and Planets}, 62:47--84.

\bibitem[{Hedman} and {Nicholson}, 2016]{hedman2016}
{Hedman}, M.~M. and {Nicholson}, P.~D. (2016).
\newblock {The B-ring's surface mass density from hidden density waves: Less
  than meets the eye?}
\newblock {\em Icarus}, 279:109--124.

\bibitem[{Hedman} et~al., 2014]{hedman2014a}
{Hedman}, M.~M., {Nicholson}, P.~D., and {Salo}, H. (2014).
\newblock {Exploring Overstabilities in Saturn's A Ring Using Two Stellar
  Occultations}.
\newblock {\em AJ}, 148:15.

\bibitem[{Hedman} et~al., 2007]{hedman2007}
{Hedman}, M.~M., {Nicholson}, P.~D., {Salo}, H., {Wallis}, B.~D., {Buratti},
  B.~J., {Baines}, K.~H., {Brown}, R.~H., and {Clark}, R.~N. (2007).
\newblock {Self-Gravity Wake Structures in Saturn's A Ring Revealed by Cassini
  VIMS}.
\newblock {\em \aj}, 133:2624--2629.

\bibitem[{Hwang} and {Hutter}, 1995]{hutter1995}
{Hwang}, H. and {Hutter}, K. (1995).
\newblock {A new kinetic model for rapid granular flow}.
\newblock {\em Continuum Mechanics and Thermodynamics}, 7:357--384.

\bibitem[{Latter} and {Ogilvie}, 2006]{latter2006}
{Latter}, H.~N. and {Ogilvie}, G.~I. (2006).
\newblock {The linear stability of dilute particulate rings}.
\newblock {\em Icarus}, 184:498--516.

\bibitem[{Latter} and {Ogilvie}, 2008]{latter2008}
{Latter}, H.~N. and {Ogilvie}, G.~I. (2008).
\newblock {Dense planetary rings and the viscous overstability}.
\newblock {\em Icarus}, 195:725--751.

\bibitem[{Latter} and {Ogilvie}, 2009]{latter2009}
{Latter}, H.~N. and {Ogilvie}, G.~I. (2009).
\newblock {The viscous overstability, nonlinear wavetrains, and finescale
  structure in dense planetary rings}.
\newblock {\em Icarus}, 202:565--583.

\bibitem[{Latter} and {Ogilvie}, 2010]{latter2010}
{Latter}, H.~N. and {Ogilvie}, G.~I. (2010).
\newblock {Hydrodynamical simulations of viscous overstability in Saturn's
  rings}.
\newblock {\em Icarus}, 210:318--329.

\bibitem[{Lehmann} et~al., 2016]{lehmann2016}
{Lehmann}, M., {Schmidt}, J., and {Salo}, H. (2016).
\newblock {A Weakly Nonlinear Model for the Damping of Resonantly Forced
  Density Waves in Dense Planetary Rings}.
\newblock {\em \apj}, 829:75.

\bibitem[{Lin} and {Bodenheimer}, 1981]{lin1981}
{Lin}, D. N.~C. and {Bodenheimer}, P. (1981).
\newblock {On the stability of Saturn's rings}.
\newblock 248:L83--L86.

\bibitem[{Liou} and {Steffen}, 1993]{liou1993}
{Liou}, M.-S. and {Steffen}, C.~J. (1993).
\newblock {A New Flux Splitting Scheme}.
\newblock {\em Journal of Computational Physics}, 107:23--39.

\bibitem[Lukkari, 1981]{lukkari1981}
Lukkari, J. (1981).
\newblock Collisional amplification of density fluctuations in {S}aturn's
  rings.
\newblock {\em Nature}, 292:433--435.

\bibitem[{Press} et~al., 1992]{press1992}
{Press}, W.~H., {Teukolsky}, S.~A., {Vetterling}, W.~T., and {Flannery}, B.~P.
  (1992).
\newblock {\em {Numerical recipes in FORTRAN. The art of scientific
  computing}}.

\bibitem[{Rein} and {Latter}, 2013]{latter2013}
{Rein}, H. and {Latter}, H.~N. (2013).
\newblock {Large-scale N-body simulations of the viscous overstability in
  Saturn's rings}.
\newblock {\em \mnras}, 431:145--158.

\bibitem[Ruuth, 2006]{ruuth2006}
Ruuth, S.~J. (2006).
\newblock Global optimization of explicit strong-stability-preserving
  {R}unge-{K}utta methods.
\newblock {\em Math. Comp.}, 75(253):183--207 (electronic).

\bibitem[Salo, 1992]{salo1992a}
Salo, H. (1992).
\newblock {Numerical simulations of dense collisional systems: II. Extended
  distribution of particle size}.
\newblock {\em Icarus}, 96:85--106.

\bibitem[{Salo}, 1995]{salo1995}
{Salo}, H. (1995).
\newblock {Simulations of dense planetary rings. III. Self-gravitating
  identical particles.}
\newblock {\em Icarus}, 117:287--312.

\bibitem[{Salo}, 2001]{salo2001b}
{Salo}, H. (2001).
\newblock {Numerical Simulations of the Collisional Dynamics of Planetary
  Rings}.
\newblock In {P{\"o}schel}, T. and {Luding}, S., editors, {\em Granular Gases},
  volume 564 of {\em Lecture Notes in Physics, Berlin Springer Verlag}, page
  330.

\bibitem[{Salo} et~al., 2018]{salo2018}
{Salo}, H., {Ohtsuki}, K., and {Lewis}, M.~C. (2018).
\newblock {\em {Computer Simulations of Planetary Rings, to appear in the book
  "Planetary Ring Systems", eds. M. Tiscareno and C. Murray }}.

\bibitem[{Salo} and {Schmidt}, 2010]{salo2009}
{Salo}, H. and {Schmidt}, J. (2010).
\newblock {N-body simulations of viscous instability of planetary rings}.
\newblock {\em Icarus}, 206:390--409.

\bibitem[Salo et~al., 2001]{salo2001}
Salo, H., Schmidt, J., and Spahn, F. (2001).
\newblock Viscous overstability in {S}aturn's {B} ring: {I}. {D}irect
  simulations and mesurement of transport coefficients.
\newblock {\em Icarus}, 153:295--315.

\bibitem[{Schmidt} et~al., 2009]{schmidt2009}
{Schmidt}, J., {Ohtsuki}, K., {Rappaport}, N., {Salo}, H., and {Spahn}, F.
  (2009).
\newblock {\em {Dynamics of Saturn's Dense Rings}}, pages 413--458.

\bibitem[Schmidt and Salo, 2003]{schmidt2003}
Schmidt, J. and Salo, H. (2003).
\newblock A weakly nonlinear model for viscous overstability in {S}aturn's
  dense rings.
\newblock {\em Physical Review Letters}, 90(6):061102.

\bibitem[Schmidt et~al., 2001]{schmidt2001b}
Schmidt, J., Salo, H., Spahn, F., and Petzschmann, O. (2001).
\newblock Viscous overstability in {S}aturn's {B} ring: {II}. {H}ydrodynamic
  theory and comparison to simulations.
\newblock {\em Icarus}, 153:316--331.

\bibitem[{Schmit} and {Tscharnuter}, 1995]{schmit1995}
{Schmit}, U. and {Tscharnuter}, W. (1995).
\newblock {A fluid dynamical treatment of the common action of
  self-gravitation, collisions, and rotation in {S}aturn's B-ring}.
\newblock {\em Icarus}, 115:304--319.

\bibitem[{Schmit} and {Tscharnuter}, 1999]{schmit1999}
{Schmit}, U. and {Tscharnuter}, W. (1999).
\newblock {On the formation of the fine--scale structure in {S}aturn's B ring}.
\newblock {\em Icarus}, 138:173--187.

\bibitem[{Shu}, 2009]{shu2009}
{Shu}, C.-W. (2009).
\newblock {High Order Weighted Essentially Nonoscillatory Schemes for
  Convection Dominated Problems}.
\newblock {\em SIAM Review}, 51:82--126.

\bibitem[{Shu} and {Osher}, 1988]{shu1988}
{Shu}, C.-W. and {Osher}, S. (1988).
\newblock {Efficient Implementation of Essentially Non-oscillatory
  Shock-Capturing Schemes}.
\newblock {\em Journal of Computational Physics}, 77:439--471.

\bibitem[Shu et~al., 1985]{shu1985a}
Shu, F., Yuan, C., and Lissauer, J. (1985).
\newblock Nonlinear spiral density waves: an inviscid theory.
\newblock {\em Astrophysical Journal}, 291:356--376.

\bibitem[{Shu} and {Stewart}, 1985]{shu1985c}
{Shu}, F.~H. and {Stewart}, G.~R. (1985).
\newblock {The collisional dynamics of particulate disks}.
\newblock {\em Icarus}, 62:360--383.

\bibitem[Spahn et~al., 2000]{spahn2000a}
Spahn, F., Schmidt, J., Petzschmann, O., and Salo, H. (2000).
\newblock Stability analysis of a {K}eplerian disk of granular grains:
  influence of thermal diffusion.
\newblock {\em Icarus}, 145:657--660.

\bibitem[{Sremcevic} et~al., 2009]{sremcevic2009}
{Sremcevic}, M., {Colwell}, J.~E., and {Esposito}, L.~W. (2009).
\newblock {Small-scale ring structure observed in Cassini UVIS occultations}.
\newblock {\em AGU Fall Meeting Abstracts}.

\bibitem[{Stewart} et~al., 1984]{stewart1984}
{Stewart}, G.~R., {Lin}, D. N.~C., and {Bodenheimer}, P. (1984).
\newblock Collision-induced transport processes in planetary rings.
\newblock In Greenberg, R. and Brahic, A., editors, {\em Planetary Rings},
  pages 447--512, Tucson Arizona. Univ. of Arizona Press.

\bibitem[{Suresh} and {Huynh}, 1997]{suresh1997}
{Suresh}, A. and {Huynh}, H.~T. (1997).
\newblock {Accurate Monotonicity-Preserving Schemes with Runge Kutta Time
  Stepping}.
\newblock {\em Journal of Computational Physics}, 136:83--99.

\bibitem[{Thomson} et~al., 2007]{Thomson2007}
{Thomson}, F.~S., {Marouf}, E.~A., {Tyler}, G.~L., {French}, R.~G., and
  {Rappoport}, N.~J. (2007).
\newblock {Periodic microstructure in Saturn's rings A and B}.
\newblock {\em GRL}, 34:24203--+.

\bibitem[{Tiscareno} et~al., 2007]{tiscareno2007b}
{Tiscareno}, M.~S., {Burns}, J.~A., {Nicholson}, P.~D., {Hedman}, M.~M., and
  {Porco}, C.~C. (2007).
\newblock {Cassini imaging of Saturn's rings II: A wavelet technique for
  analysis of density waves and other radial structure in the rings}.
\newblock {\em {Icarus}}.

\bibitem[Toomre, 1964]{Toomre1964}
Toomre, A. (1964).
\newblock On the gravitational stability of a disk of stars.
\newblock {\em Astrophysical Journal}, 139:1217--1238.

\bibitem[{van Hecke} et~al., 1999]{vanhecke1999}
{van Hecke}, M., {Storm}, C., and {van Saarloos}, W. (1999).
\newblock {Sources, sinks and wavenumber selection in coupled CGL equations and
  experimental implications for counter-propagating wave systems}.
\newblock {\em Physica D: Nonlinear Phenomena}, 134:1-47.

\bibitem[Ward, 1981]{ward1981}
Ward, W.~R. (1981).
\newblock {On the radial structure of Saturn's rings}.
\newblock 8:641--643.

\bibitem[{Wisdom} and {Tremaine}, 1988]{wisdom1988}
{Wisdom}, J. and {Tremaine}, S. (1988).
\newblock {Local simulations of planetary rings}.
\newblock {\em Astron. J.}, 95:925--940.

\end{thebibliography}
\end{document}